\documentclass[superscriptaddress,aps,preprintnumbers,amsmath,amssymb,sort&compress,nofootinbib,10pt]{revtex4-2}

\usepackage{amsfonts,amsmath,amssymb,ascmac,bm,tensor}
\usepackage{fnpct} 
\usepackage{comment}
\usepackage{ifpdf}
\usepackage{slashed}
\usepackage{color}
\usepackage[mathscr]{eucal}
\usepackage[utf8]{inputenc}
\usepackage{physics}
\usepackage{cancel}
\usepackage{soul}
\usepackage{simpler-wick}

\ifpdf
  \usepackage{graphicx}     %   usepackage without driver option
  \usepackage[bookmarksopen,colorlinks=true,linkcolor=bblue,citecolor=bblue,urlcolor=ppink]{hyperref}
\else     % For (p)LaTeX + dvipdfmx
\fi

\definecolor{red}{rgb}{1,0,0}
\definecolor{darkred}{rgb}{0.6,0,0}
\definecolor{darkgreen}{rgb}{0.992447,0.623778,0.034597}
\definecolor{ppink}{rgb}{1,0.4,0.4} 
\definecolor{bblue}{rgb}{0.284602,0.317763,0.963947}
\definecolor{purple}{rgb}{0.5 ,0, 0.7}

\newcommand{\fo}{{(1)}}
\renewcommand{\so}{{(2)}}
\newcommand{\tho}{{(3)}}

\newcommand{\Pl}{\text{Pl} }
\newcommand{\ee}{\text{e}}
\newcommand{\UV}{\text{UV} }
\newcommand{\IR}{\text{IR} }

\newcommand{\vx}{\text{vx} }
\newcommand{\tre}{\text{tr} }
\newcommand{\pe}{\text{peak}}
\newcommand{\lmax}{\text{lmax}}
\newcommand{\hc}{\text{h.c.}}

\renewcommand{\Re}{\text{Re}}
\renewcommand{\Im}{\text{Im}}

\makeatletter
\newcommand\footnoteref[1]{\protected@xdef\@thefnmark{\ref{#1}}\@footnotemark}
\makeatother

\allowdisplaybreaks[1]

\begin{document}

%%%%%%%%%%%%%%%%%%%%%%%%%%%
%%%%%%%%%%% Title %%%%%%%%%%%
%%%%%%%%%%%%%%%%%%%%%%%%%%%

\title{
Questions on calculation of primordial power spectrum with large spikes: \\the resonance model case\\ 
}

\author{Keisuke Inomata
}
\affiliation{Kavli Institute for Cosmological Physics, The University of Chicago, Chicago, IL 60637, USA}

\author{Matteo Braglia
}
\affiliation{Center for Cosmology and Particle Physics, New York University, 726 Broadway, New York, NY 10003, USA}
\affiliation{INAF/OAS Bologna, via Gobetti 101, I-40129 Bologna, Italy}

\author{Xingang Chen
}
\affiliation{
Institute for Theory and Computation, Harvard-Smithsonian Center for Astrophysics, 60
Garden Street, Cambridge, MA 02138, USA}

\author{Sébastien Renaux-Petel
}
\affiliation{
Institut d’Astrophysique de Paris, UMR 7095 du CNRS et de Sorbonne Université, 98 bis bd Arago, 75014 Paris, France}

\begin{abstract} 
\noindent
Inflationary models predicting a scale-dependent large amplification of the density perturbations have recently attracted a lot of attention because the amplified perturbations can seed a sizable amount of primordial black holes (PBHs) and stochastic background of gravitational waves (GWs).
While the power spectra in these models are computed based on the linear equation of motion,
it is not obvious whether loop corrections are negligible when such a large amplification occurs during inflation. 
In this paper, as a first step to discuss the loop corrections in such models, we use the in-in formalism and calculate the one-loop scalar power spectrum numerically and analytically in an illustrative model where the density perturbations are resonantly amplified due to oscillatory features in the inflaton potential. 
Our calculation is technically new in that the amplified perturbations are numerically taken into account in the in-in formalism for the first time.
In arriving at our analytical estimates, we highlight the role that the Wronskian condition of perturbations, automatically satisfied in our model, plays in obtaining the correct estimates.
We also discuss the necessary conditions for subdominant loop corrections in this model.
We find that, for the typical parameter space leading to the $\mathcal O(10^7)$ amplification of the power spectrum required for a sufficient PBH production, the one-loop power spectrum dominates over the tree-level one, indicating the breakdown of the perturbation theory.
\end{abstract}

%\date{\today}
\maketitle

\tableofcontents

%%%%%%%%%%%%%%%%%%%%%%%%%%%%%%%%
\section{Introduction}
%%%%%%%%%%%%%%%%%%%%%%%%%%%%%%%%

The inflationary theory is the leading candidate for the explanations of the initial condition of the Big Bang model~\cite{Starobinsky:1980te,Sato:1980yn,Guth:1980zm}. 
The inflation not only solves the flatness, the horizon, and the monopole problems, but also provides the production mechanism of the cosmological perturbations, which seed the cosmic microwave background (CMB) anisotropies and the large scale structure (LSS) observed in our Universe~\cite{Mukhanov:1981xt,Mukhanov:1982nu,Starobinsky:1982ee,Guth:1982ec,Panagiotakopoulos:1982rn,Bardeen:1983qw}.
The power spectrum of such primordial perturbations is tightly constrained by the latest cosmological observations on the large scales $k\in[10^{-3},\,1]\, {\rm Mpc}^{-1}$~\cite{Nicholson:2009new,Nicholson:2010new,bird2011minimally}. On those scales, observations of CMB anisotropies and LSS suggest that the power spectrum takes the form of a nearly scale-invariant power law, with the spectral index $n_s-1$ constrained to $n_s=0.9649\pm0.0042$ at 68\% CL \cite{Planck:2018jri}. 
Analyses of the Planck temperature and polarization maps also provide the most stringent constraints to date on primordial Non-Gaussianities and are consistent with the simple picture where the perturbations are Gaussian~\cite{Planck:2019kim}. 
These observational results are consistent with the slow-roll inflation, which is described by a slowly-rolling scalar field (inflaton)~\cite{Linde:1981mu,Albrecht:1982wi,Linde:1983gd}.
On smaller scales ($k > 1$\,Mpc$^{-1}$), however, the primordial perturbations are not investigated well due to the diffusion damping of the CMB anisotropies~\cite{Silk:1967kq} and the resolution limit of the Lyman-$\alpha$ forest for the LSS observations~\cite{Nicholson:2009new,Nicholson:2010new,bird2011minimally}, while future observations of the redshifted 21cm emission line from neutral hydrogen~\cite{Loeb:2003ya,Chen:2016zuu,Cole:2019zhu,Balaji:2022zur}
 and spectral distortions of the CMB~\cite{Unal:2020mts,Schoneberg:2020nyg} offer promising prospects to test the power spectrum up to much smaller scales $k\lesssim 10^4\,{\rm Mpc}^{-1}$.

One of the current observational limits on the small-scale power spectrum is the requirement that primordial black holes (PBHs) are not overproduced.
While deriving a precise limit relies on the modeling of the PBH formation and their accretion, $\mathcal{P}_\zeta \lesssim\mathcal{O}(10^{-2})$ is widely accepted as a conservative upper bound~\cite{Josan:2009qn,Cole:2017gle,Sasaki:2018dmp,Sato-Polito:2019hws}. 
In fact, PBHs have recently seen a surge of interest from many areas of astrophysics and cosmology after it was proposed in Refs.~\cite{Bird:2016dcv,Sasaki:2016jop,Clesse:2016vqa} that they could explain the first direct detection of gravitational waves (GWs) by the LIGO-Virgo collaboration~\cite{LIGOScientific:2016aoc}. 
In some mass ranges, PBHs may also constitute a (significant) fraction of the cold dark matter (DM), which permeates our universe~\cite{Carr:2016drx}.
A quite general requirement for PBHs to form is the significant enhancement of the amplitude of curvature perturbations with respect to the COBE normalization so that, when they re-enter the horizon (typically) during the radiation dominated era, they can efficiently overcome the radiation pressure and collapse into black holes~\cite{Zeldovich:1967lct,Hawking:1971ei,Carr:1974nx}.
The PBH scenarios for the LIGO/Virgo events and DM typically require $\mathcal P_\zeta \sim \mathcal O(10^{-2})$ on the small scales~\cite{Sasaki:2018dmp,Wang:2022nml}, realized by $\mathcal O(10^7)$ enhancement from that on the CMB/LSS observation scales.
The quest for inflationary model builders is therefore to engineer mechanisms to realize such a large amplification of the primordial power spectrum. Over the past years, many such models have been proposed in the context of hybrid inflation~\cite{GarciaBellido:1996qt,Clesse:2015wea,Kallosh:2022ggf}, double inflation~\cite{Kawasaki:1997ju,Kawasaki:1998vx,Kawasaki:2006zv,Kawasaki:2016pql,Inomata:2016rbd,Inomata:2017okj,Inomata:2017vxo,Pi:2017gih}, single-field chaotic new inflation~\cite{Yokoyama:1998pt,Saito:2008em}, axion-like curvaton model~\cite{Kasuya:2009up,Kawasaki:2012wr,Ando:2017veq,Ando:2018nge}, ultra-slow-roll inflation~\cite{Ivanov:1994pa,Kinney:1997ne,Inoue:2001zt,Kinney:2005vj,Martin:2012pe,Garcia-Bellido:2017mdw,Motohashi:2017kbs,Germani:2017bcs,Ballesteros:2017fsr}, turns in the field space~\cite{Palma:2020ejf,Fumagalli:2020adf,Braglia:2020eai,Fumagalli:2020nvq,Braglia:2020taf,Iacconi:2021ltm}, resonant amplification with oscillatory features~\cite{Cai:2019jah,Cai:2019bmk,Zhou:2020kkf,Peng:2021zon,Cai:2021wzd}, or other types of sharp features
\cite{Ozsoy:2018flq,Mishra:2019pzq,Kefala:2020xsx,Inomata:2021tpx,Dalianis:2021iig,Inomata:2021uqj}. 
On the phenomenological side, another interesting prediction of the models with large-amplitude curvature perturbations is the generation of a stochastic gravitational wave background (SGWB) both during \cite{Cai:2021yvq,Inomata:2022ydj,Fumagalli:2021mpc} and after inflation~\cite{Ananda:2006af,Baumann:2007zm,Saito:2008jc,Saito:2008jc} (see Ref.~\cite{Domenech:2021ztg} for a review). As opposed to the abundance of PBHs, which is exponentially sensitive to the amplitude of the primordial power spectrum, the energy density of the SGWB is quadratically dependent on it.
Conversely, the required amplitude of the power spectrum is more sensitive to the amount of SGWB than the PBH abundance, which allows future GW detectors to test a large portion of the parameter space of these models down to $\mathcal{P}_\zeta \sim \mathcal{O}(10^{-6})$~\cite{Assadullahi:2009jc,Byrnes:2018txb,Inomata:2018epa}.  
It is therefore of utmost importance to assess the robustness of theoretical predictions from models involving the amplification of curvature perturbations if we want to test them in the near future.

In this paper, we pay attention to the fact that many previous works on the amplification models perform the calculation of the PBH abundance or the SGWB based on the tree-level power spectrum.\footnote{
The effects beyond the tree-level power spectrum are discussed in some specific situations.
In the ultra-slow-roll model, the effects can be taken into account with the use of the stochastic-$\delta N$ formalism~\cite{Pattison:2017mbe,Biagetti:2018pjj,Ezquiaga:2018gbw}. 
The inflation model with a step feature is discussed with lattice simulations, which can solve the perturbations in a non-perturbative way~\cite{Caravano:2021pgc}.
Also, in Ref.~\cite{Fumagalli:2020nvq,Inomata:2021tpx}, the conditions for the non-linear perturbation effects to be negligible are discussed by comparing the quadratic and the higher order Lagrangian.
The authors of Ref.~\cite{Celoria:2021vjw} propose another non-perturbative way of calculation focusing on the tail of the probability function of the perturbations.
In Ref.~\cite{Meng:2022ixx}, the non-linear effects on the tree power spectrum are discussed with the local type ansatz of the non-Gaussianity of curvature perturbations.}
The ``tree-level power spectrum'' here means the power spectrum calculated with the linear equation of motion for the perturbations.
The linear equation of motion is typically solved numerically and non-perturbatively, which amounts to a non-perturbative sum of all tree-level diagrams \cite{Chen:2015dga,Werth:2023pfl}.
However, the linear equation is modified by non-linear perturbations in general, whose effects can be seen as loop corrections~\cite{Jordan:1986ug,Calzetta:1987abcd,Weinberg:2005vy,Sloth:2006az,Sloth:2006nu,Seery:2007we,Seery:2007wf,Adshead:2008gk,Senatore:2009cf} to the tree-level power spectrum in the in-in formalism (see Refs.~\cite{Weinberg:2005vy,Chen:2010xka,Wang:2013zva} for reviews).
When the perturbations are amplified by some mechanism, it is not obvious whether the loop corrections are subdominant to the tree power spectrum.
In this work, as a first step towards this direction, using the in-in formalism that is necessary for the models we study, we explicitly check the one-loop corrections to the tree power spectrum in a single-field model with oscillatory features, where the amplification mechanism consists in the parametric resonance of the field fluctuations with the background oscillation. 
We first numerically calculate the loop power spectrum, which is technically new in that we take into account the resonantly amplified perturbations all numerically in the in-in formalism for the first time for these models with large scale-dependent spikes.\footnote{
Recently, GWs induced by the amplified perturbations during inflation have been discussed in the in-in formalism with an analytical ansatz of the perturbation evolution~\cite{Ota:2022hvh}.
Also, the authors of Ref.~\cite{Kristiano:2021urj} have focused on the case with a small sound speed of the inflaton and mentioned the possible importance of the loop corrections to the power spectrum in the PBH scenarios with the use of an analytical ansatz.
}
 Then, we obtain analytical estimates for the peak amplitude, which show that the Wronskian condition of the perturbations plays an important role in obtaining the correct order of the estimates.
With the analytical estimates, we also obtain the necessary conditions for the loop power spectrum to be subdominant.
One of the main results we will show is that, to realize the subdominant one-loop corrections in the typical oscillatory feature model, the amplification of the tree power spectrum should be $< \mathcal O(10^7)$, whereas $\mathcal O(10^7)$ amplification is often considered in the PBH scenarios.
Note that the oscillatory features of the potential lead to large potential derivatives and enable the loop power spectrum to be dominant even though the tree power spectrum of curvature perturbations is much smaller than unity.
At this stage, we would like to highlight that the conventional computational method for the power spectrum could fail if the large amplification of the perturbations occurs during inflation, rather than providing the ultimate technique to correctly compute the leading order power spectrum.
Quantitative predictions of the power spectrum in such models based on the linear equation of motion are in question, and a completely new computational tool is in need and remains an open question.

Our paper is organized as follows.
In Sec.~\ref{sec:in_in}, we summarize the equations for the loop calculation in the in-in formalism.
Then, we introduce our fiducial setup and show the numerical results in Sec.~\ref{sec:num_result}.
In Sec.~\ref{sec:analytical_estimates}, we obtain analytical estimates of the loop power spectrum on its peak scale and the one-loop correction to the background. 
Using the estimates, we discuss in Sec.~\ref{sec:necessary_conditions} the necessary conditions for the loop power spectrum to be subdominant, which are not satisfied in the models of interest in the PBH scenarios.
Finally, we conclude our paper in Sec.~\ref{sec:conclusion}.

Throughout this paper, we use a natural unit in which $c=\hbar =1$ and the mostly plus metric convention, $(-,+,+,+)$.

%%%%%%%%%%%%%%%%%%%%%%%%%%%%%%%%
\section{Equations for loop calculation} 
\label{sec:in_in}
%%%%%%%%%%%%%%%%%%%%%%%%%%%%%%%%

In this section, we summarize the equations for the calculation of the loop power spectrum in the in-in formalism.
In many models where features generate spikes in the power spectrum, the relevant modes are those inside or crossing the horizon instead of those long after the horizon exit. For those models, the local type non-Gaussianity ansatz~\cite{Meng:2022ixx} or the $\delta N$ formalism~\cite{Pattison:2017mbe,Biagetti:2018pjj,Ezquiaga:2018gbw} generally does not apply, and we need to use the first principle in-in formalism.

In this work, we focus on a single-field inflation model, whose action is given by 
\begin{align}
	S = \int \dd \eta\, \dd^3 x\, \sqrt{-g} \mathcal L = \int \dd \eta \, \dd^3 x\, a^4 \mathcal L,
	\label{eq:action}
\end{align}
where $\eta$ is the conformal time and the Lagrangian is given by 
\begin{align}
	\mathcal L =& -\frac{1}{2} \partial^\mu \phi \partial_\mu \phi - V(\phi).
\end{align}
Note that, throughout this work, we take the spatially flat gauge and neglect the metric perturbations because we focus on the inflaton potential with sharp features that resonantly amplify the field fluctuations. 
In Appendix~\ref{app:met_pertb}, we show that, in our fiducial potential, which is introduced in the next section, the order of metric perturbation contributions are suppressed by the slow-roll parameter $\epsilon \equiv -\dot H/H^2$ compared to the dominant potential derivative terms, where the dot denotes the physical time derivative.

To use the in-in formalism, we expand the Lagrangian with respect to the order of perturbations as 
\begin{align}
	\mathcal L_2 &= \frac{1}{2}\delta {\dot \phi}^2 - \frac{1}{2a^2}(\partial_i \delta \phi)^2 - \frac{1}{2} V^{(2)}(\phi) \delta \phi^2 \nonumber \\
	&= \frac{1}{a^2} \left[\frac{1}{2}\delta {\phi'}^2 - \frac{1}{2}(\partial_i \delta \phi)^2 - \frac{a^2}{2} V^{(2)}(\phi) \delta \phi^2 \right], \\
  \mathcal L_{n(>2)} &=  -\frac{1}{n!} V^{(n)}(\phi) \delta\phi^n, 
\end{align}
where the prime denotes the conformal time derivative, the subscripts of Lagrangian indicate the order of field fluctuations, $(\partial_i \delta \phi)^2 = \delta^{ij} \partial_i \delta \phi \partial_j \delta \phi$, and $V^{(n)} \equiv \partial^n V(\phi)/\partial \phi^n$.
From the quadratic Lagrangian, we can derive the linear equation of motion for the field fluctuation:
\begin{align}
		\delta \phi'' + 2 \mathcal H \delta \phi' - \nabla^2 \delta \phi + a^2 V^{(2)}(\phi) \delta \phi = 0.
	\label{eq:delta_phi_eom}
\end{align}

The Hamiltonian density for $\delta \phi$ is given by 
\begin{align}
	a^4 \mathcal H = \Pi_{\delta \phi} \delta \phi' - a^4 \mathcal L,
\end{align}
where $\Pi_{\delta \phi}$ is the momentum conjugate to $\delta \phi$, given by 
\begin{align}
	\Pi_{\delta \phi} = \frac{\partial (a^4 \mathcal L)}{\partial {\delta \phi'}} = a^2 \delta \phi'.
\end{align}
Similar to the Lagrangian, we expand the Hamiltonian density as  
\begin{align}
	\mathcal H_2 &= \frac{1}{a^2} \left[ \frac{1}{2}\delta {\phi'}^2 + \frac{1}{2}(\partial_i \delta \phi)^2 + \frac{a^2}{2} V^{(2)}(\phi) \delta \phi^2 \right], \\
	\mathcal H_{n(>2)}  &=  \frac{1}{n!} V^{(n)}(\phi) \delta\phi^n.
\end{align}
We here choose to group all the quadratic terms in the free Hamiltonian, which includes the tree-level resonance effect (see Appendix~\ref{app:P_tree_pertb}), and all the other terms in the interaction Hamiltonian:
\begin{align}
	H_{\text{int}} &= \sum_{n>2} H_{\text{int},n}, \\
	H_{\text{int},n}& \equiv \int \dd^3 x\,\, a^4 \mathcal H_{n}.
\end{align}
In the in-in formalism, the expectation value of an arbitrary quantity, $W(\eta)$, is given by~\cite{Weinberg:2005vy}
\begin{align}
	\expval{W(\eta)} \equiv \vev{ \left(T \ee^{-i \int^\eta_{-\infty} \dd \eta' H_{\text{int}}(\eta')}\right)^\dagger W(\eta) \left(T \ee^{-i \int^\eta_{-\infty} \dd \eta'' H_{\text{int}}(\eta'')}\right)}.
	\label{eq:in_in_form}
\end{align}
With our choice of $H_2 (\equiv \int \dd^3 x \,a^4 \mathcal H_2)$ and $H_{\text{int}}$, the tree-level resonance effect is encoded in the interaction picture fields, and we calculate the effects of the non-linear inflaton interaction through the series expansion.
Throughout this work, we assume that $\expval{\delta \phi(\mathbf x,\eta)} = 0$ is satisfied. 
Actually, this condition for one point function is related to the loop correction to the background~\cite{Sloth:2006nu} and we will come back to this point in Sec.~\ref{subsec:tadpole}.

Then, we can express the two-point correlation function as 
\begin{align}
	\expval{\delta \phi_{\mathbf k}(\eta) \delta \phi_{\mathbf k'}(\eta)} = \vev{ \left(T \ee^{-i \int^\eta_{-\infty} \dd \eta' H_{\text{int}}(\eta')}\right)^\dagger \delta \phi_{\mathbf k}(\eta) \delta \phi_{\mathbf k'}(\eta) \left(T \ee^{-i \int^\eta_{-\infty} \dd \eta'' H_{\text{int}}(\eta'')}\right)},
	\label{eq:one_loop_in_in_form}
\end{align}
where $\delta \phi_{\mathbf k}$ is the Fourier mode of $\delta \phi(\mathbf x)$.
Here, we quantize $\delta \phi$ by imposing the following commutation relation:
\begin{align}
		&[\delta \phi(\mathbf x), \Pi_{\delta \phi}(\mathbf y)] = a^2 [\delta \phi(\mathbf x), \delta \phi'(\mathbf y)] = i \delta(\mathbf x - \mathbf y). 
		\label{eq:quantization_cond}
\end{align}
Based on this quantization condition, we can expand $\delta \phi$ with the creation and the annihilation operators ($\hat a(\mathbf k), \hat a^\dagger(\mathbf k)$) as 
\begin{align}
	\delta \phi(\mathbf x,\eta) &= \int \frac{\dd^3 k}{(2\pi)^3} \ee^{i \mathbf k \cdot \mathbf x} \delta \phi_{\mathbf k}(\eta) \nonumber \\
	&= \int \frac{\dd^3 k}{(2\pi)^3} \ee^{i \mathbf k \cdot \mathbf x} \left[ U_k(\eta)\hat a(\mathbf k) +  U_k^{*}(\eta) \hat a^{\dagger}(-\mathbf k) \right],
  \label{eq:d_phi_expand}
\end{align}
where the commutation relations between the creation and annihilation operators satisfy
\begin{align}
		&[\hat a(\mathbf k), \hat a(\mathbf k')] = [\hat a^\dagger(\mathbf k), \hat a^\dagger(\mathbf k')] = 0, \label{eq:aa_comu} \\
		&[\hat a(\mathbf k), \hat a^\dagger(-\mathbf k')] = (2\pi)^3 \delta(\mathbf k + \mathbf k'). \label{eq:aad_comu}
\end{align}
To realize Eq.~(\ref{eq:quantization_cond}), the function $U_k$ must satisfy the Wronskian condition:
\begin{align}
	U_k(\eta){U_k^*}'(\eta)- U_k'(\eta){U_k^{*}}(\eta) = \frac{i}{a^2(\eta)}.
	\label{eq:denom_green}
\end{align}
This Wronskian condition can be rewritten as 
\begin{align}
	|U_k(\eta)|^2 \theta'(k,\eta) =  \frac{1}{2a^2(\eta)},
	\label{eq:u_angle_conserve}
\end{align}
where we have parametrized $U_k(\eta) = |U_k(\eta)|\ee^{-i \theta(k,\eta)}$.
Note that this condition is always satisfied even during and after the resonant amplification of $U_k$, which we will consider in the following sections.\footnote{ 
We can see the analogue between the Wronskian condition and the angular momentum conservation in the two-dimensional dynamics by regarding $U$ and $U^*$ as $x$ and $y$ axes. 
The only difference is the expansion of the universe, which decreases the ``angular momentum''.
Since the inflaton potential must be real to realize real energy, no ``torque'' appears in the equation of motion for $U$, Eq.~(\ref{eq:u_eom}).
For this reason, even if we consider the resonant amplification of the field fluctuations, the ``angular momentum'' always follows Eq.~(\ref{eq:u_angle_conserve}).}

Apart from the Wronskian condition, the function $U_k$ must follow the linear equation of motion, Eq.~(\ref{eq:delta_phi_eom}):
\begin{align}
	U''_k(\eta) + 2 \mathcal H U'_k(\eta) + k^2 U_k(\eta) + a^2 V^{(2)}(\phi) U_k(\eta) = 0,
	\label{eq:u_eom}
\end{align}
where the complex conjugate $U^*_k$ also follows this equation.
As an initial condition of $U_k$, we take the Bunch-Davies vacuum state: 
\begin{align} 
	 U_k(\eta) \simeq \frac{-i}{\sqrt{2k}a(\eta)} \ee^{-i (k \eta + \varphi(k))} \  (\text{in } |k \eta| \gg 1 \text{ and } k^2 \gg a^2 V^{(2)}),
	 \label{eq:u_sol}
\end{align}
where $\varphi(k)$ is an arbitrary constant phase, which does not appear in power spectra.

\begin{figure}[t] 
        \centering \includegraphics[width=0.5\columnwidth]{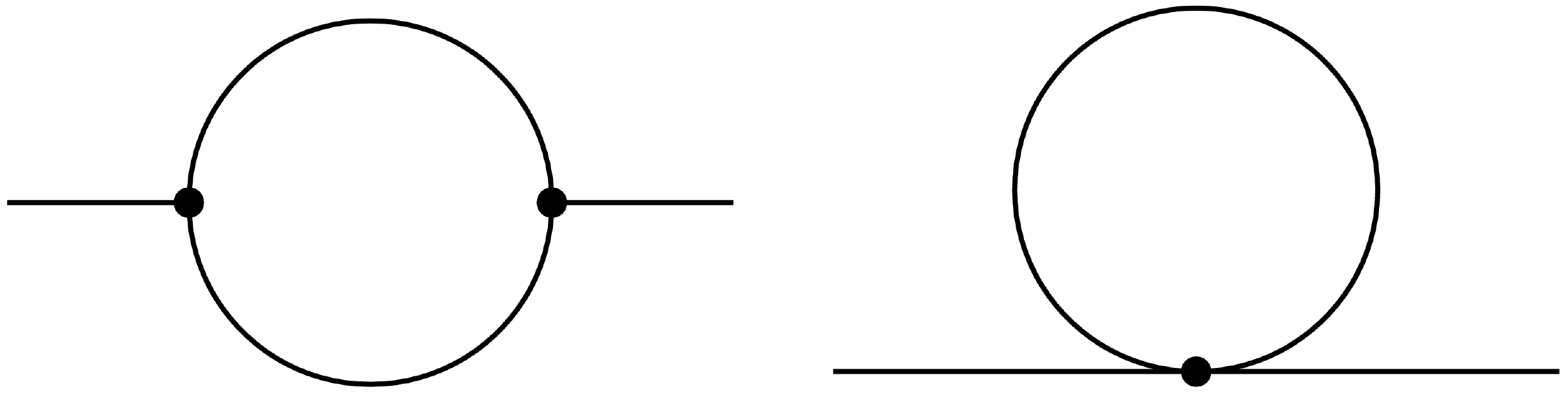}
        \caption{ 
        The Feynman diagrams corresponding to one-loop contributions with two vertices (left) and one vertex (right). 
        }
        \label{fig:loop_diagram}
\end{figure}

For later convenience, we here define the tree-level (no loop) power spectrum as 
\begin{align}
	\vev{\delta \phi_{\mathbf k}(\eta) \delta \phi_{\mathbf k'}(\eta)} &= (2\pi)^3 \delta(\mathbf k + \mathbf k') \frac{2\pi^2}{k^3}\mathcal P_{\delta \phi,\tre}(k,\eta), \\
	\mathcal P_{\delta \phi,\text{tr}}(k,\eta) &\equiv \frac{k^3}{2\pi^2} |U_k(\eta)|^2.
\end{align}
This tree power spectrum encodes complete information on the quadratic Lagrangian $\mathcal L_2$.
The effects of the higher order Lagrangian $\mathcal L_{n(>2)}$ appear as loop corrections to the tree power spectrum in the in-in formalism. 
In the following, we summarize the equations for the one (lowest-order) loop power spectrum, which can be classified into the two-vertex and the one-vertex contributions as shown in Fig.~\ref{fig:loop_diagram}.

%%%%%%%%%%%%%%%%%
\subsection{One loop with two vertices}
%%%%%%%%%%%%%%%%%

From Eq.~(\ref{eq:one_loop_in_in_form}), we can express the one-loop contribution with two vertices as 
\begin{align}
	\expval{\delta \phi_{\mathbf k}(\eta) \delta \phi_{\mathbf k'}(\eta)}_{2\vx} &= \vev{ \left(T \left[-i \int^\eta_{-\infty} \dd \eta' H_{\text{int},3}(\eta') \right]\right)^\dagger \delta \phi_{\mathbf k}(\eta) \delta \phi_{\mathbf k'}(\eta) \left(T \left[-i \int^\eta_{-\infty} \dd \eta'' H_{\text{int},3}(\eta'')\right]\right)} \nonumber \\
	& \quad + \vev{ \delta \phi_{\mathbf k}(\eta) \delta \phi_{\mathbf k'}(\eta) \left(T \left[- \frac{1}{2} \left(\int^\eta_{-\infty} \dd \eta'' H_{\text{int},3}(\eta'') \right)^2\right]\right)}  \nonumber \\
	& \quad + \vev{ \left(T \left[- \frac{1}{2} \left(\int^\eta_{-\infty} \dd \eta'' H_{\text{int},3}(\eta'') \right)^2\right]\right)^\dagger \delta \phi_{\mathbf k}(\eta) \delta \phi_{\mathbf k'}(\eta) }.
	\label{eq:two_vx_one_loop}
\end{align}
For simpler notation, we here express $H_{\text{int},3}(\eta)$ as 
\begin{align}
	H_{\text{int},3}(\eta) &= -\int \dd^3 x \, \frac{\lambda(\eta) }{3} \delta \phi^3(\mathbf x,\eta), \\
	\lambda(\eta) &\equiv -a^4(\eta) \frac{V^{(3)}(\phi)}{2}.
\end{align}
Then, we can rewrite the first term in the right hand side of Eq.~(\ref{eq:two_vx_one_loop}) as 
\begin{align}
	&\vev{ \left(T \left[-i \int^\eta_{-\infty} \dd \eta' H_{\text{int},3}(\eta') \right]\right)^\dagger \delta \phi_{\mathbf k}(\eta) \delta \phi_{\mathbf k'}(\eta) \left(T \left[-i \int^\eta_{-\infty} \dd \eta'' H_{\text{int},3}(\eta'')\right]\right)} \nonumber \\
	&=\vev{ \left(T \left[i \int^\eta_{-\infty} \dd \eta' \frac{\lambda(\eta')}{3} \int \dd^3 y \,\delta \phi^3(\mathbf y,\eta') \right]\right)^\dagger \delta \phi_{\mathbf k}(\eta) \delta \phi_{\mathbf k'}(\eta) \left(T \left[i \int^\eta_{-\infty} \dd \eta'' \frac{\lambda(\eta'')}{3} \int \dd^3 x \,\delta \phi^3(\mathbf x,\eta'') \right]\right)} \nonumber \\
	&=\bra{0} \left(T \left[i \int^\eta_{-\infty} \dd \eta' \frac{\lambda(\eta')}{3} \int \frac{\dd^3 p_1}{(2\pi)^3} \int \frac{\dd^3 p_2}{(2\pi)^3} \delta \phi_{\mathbf p_1}(\eta') \delta \phi_{\mathbf p_2}(\eta') \delta \phi_{-\mathbf p_1-\mathbf p_2}(\eta')\right]\right)^\dagger  \nonumber \\
	& \qquad \times \delta \phi_{\mathbf k}(\eta) \delta \phi_{\mathbf k'}(\eta) \left(T \left[i \int^\eta_{-\infty} \dd \eta'' \frac{\lambda(\eta'')}{3}  \int \frac{\dd^3 p_1}{(2\pi)^3} \int \frac{\dd^3 p_2}{(2\pi)^3}  \delta \phi_{\mathbf p_1}(\eta'') \delta \phi_{\mathbf p_2}(\eta'') \delta \phi_{-\mathbf p_1-\mathbf p_2}(\eta'')\right]\right) \ket{0} \nonumber \\
&=2\int^\eta_{-\infty} \dd \eta' \lambda(\eta')  U^*_k(\eta) U_k(\eta') \int^\eta_{-\infty} \dd \eta'' \lambda(\eta'') U_{k'}(\eta) U^*_{k'}(\eta'')\int \frac{\dd^3 p \, \dd^3 p'}{(2\pi)^6} \vev{\delta \phi_{\mathbf p}(\eta') \delta \phi_{-\mathbf p+\mathbf k}(\eta') \delta \phi_{\mathbf p'}(\eta'')\delta \phi_{-\mathbf p'+\mathbf k'}(\eta'')},	
	\label{eq:one_loop_two_both}
\end{align}
where we have used the following relation:
\begin{equation}
	\int \frac{\dd^3 p}{(2\pi)^3} (\delta \phi_{\mathbf p}(\eta') \delta \phi_{-\mathbf p-\mathbf k}(\eta') )^\dagger = \int \frac{\dd^3 p}{(2\pi)^3} \delta \phi_{\mathbf p+\mathbf k}(\eta')\delta \phi_{-\mathbf p}(\eta')
	= \int \frac{\dd^3 \tilde p}{(2\pi)^3} \delta \phi_{\tilde {\mathbf p}}(\eta')\delta \phi_{-\tilde {\mathbf p} + \mathbf k}(\eta').
\end{equation}
We can express the momentum integrals as
\begin{align}
&\int \frac{\dd^3 p \,\dd^3 p'}{(2\pi)^6} \vev{\delta \phi_{{\mathbf p}}(\eta_1)  \delta \phi_{\mathbf k- {\mathbf p}} (\eta_1) \delta \phi_{{\mathbf p}'}(\eta_2) \delta \phi_{\mathbf k'- {\mathbf p}'} (\eta_2)} \nonumber \\
&= (2\pi)^3 \delta(\mathbf k+ \mathbf k') 2  \int \frac{\dd^3 p}{(2\pi)^3} U_{|\mathbf k- {\mathbf p}|}(\eta_1)U_p(\eta_1) U^*_{|\mathbf k- {\mathbf p}|}(\eta_2)U^*_p(\eta_2) \nonumber \\
&= (2\pi)^3 \delta(\mathbf k+ \mathbf k') \frac{2\pi^2}{k^3} k^6 \int^\infty_0 \dd v \int^{1+v}_{|1-v|} \dd u \frac{uv}{4\pi^4} U_{ku}(\eta_1)U_{kv}(\eta_1) U^*_{k u}(\eta_2)U^*_{k v}(\eta_2),
\end{align}
where $v = p/k$ and $u = |\mathbf k - \mathbf p|/k$. 
Using this expression, we can rewrite the final line of Eq.~(\ref{eq:one_loop_two_both}) as 
\begin{align}
\text{Eq.}~(\ref{eq:one_loop_two_both}) &= (2\pi)^3 \delta(\mathbf k+ \mathbf k') \frac{2\pi^2}{k^3} (2 |U_k(\eta)|^2)  \int^\infty_0 \dd v \int^{1+v}_{|1-v|} \dd u \frac{uv}{4\pi^4} F(k, ku, kv, \eta) F^*(k, ku, kv, \eta),
\label{eq:loop_2vx_first}
\end{align}
where 
\begin{align}
	F(k_1, k_2, k_3, \eta) \equiv k_1^3 \int^\eta_{-\infty} \dd \eta' \lambda(\eta') U_{k_1}(\eta')U_{k_2}(\eta')U_{k_3}(\eta').
\end{align}

Similarly, the second term in the right hand side of Eq.~(\ref{eq:two_vx_one_loop}) becomes 
\begin{align}
&\vev{ \delta \phi_{\mathbf k}(\eta) \delta \phi_{\mathbf k'}(\eta) \left(T \left[- \frac{1}{2} \left(\int^\eta_{-\infty} \dd \eta'' H_{\text{int},3}(\eta'') \right)^2\right]\right)} \nonumber \\
	&= \bra{0}
	\delta \phi_{\mathbf k}(\eta) \delta \phi_{\mathbf k'}(\eta) \left(T \left[- \frac{1}{2} \left(\int^\eta_{-\infty} \dd \eta'' \frac{\lambda(\eta'')}{3}  \int \frac{\dd^3 p_1}{(2\pi)^3} \int \frac{\dd^3 p_2}{(2\pi)^3}  \delta \phi_{\mathbf p_1}(\eta'') \delta \phi_{\mathbf p_2}(\eta'') \delta \phi_{-\mathbf p_1-\mathbf p_2}(\eta'') \right)^2\right]\right) \ket{0} \nonumber \\
	&=-\int^\eta_{-\infty} \dd \eta' \lambda(\eta')  U_k(\eta) U_k^*(\eta') \int^{\eta'}_{-\infty} \dd \eta'' \lambda(\eta'') U_{k'}(\eta) U^*_{k'}(\eta'') \nonumber \\
	& \qquad \times \int \frac{\dd^3 p \, \dd^3 p'}{(2\pi)^6} \vev{\delta \phi_{\mathbf p}(\eta') \delta \phi_{-\mathbf p+\mathbf k}(\eta') \delta \phi_{\mathbf p'}(\eta'')\delta \phi_{-\mathbf p'+\mathbf k'}(\eta'')} + (k \leftrightarrow k')\nonumber \\
	&=-(2\pi)^3 \delta(\mathbf k+ \mathbf k') \frac{2\pi^2}{k^3} (2 U^2_k(\eta)) \int^\infty_0 \dd v \int^{1+v}_{|1-v|} \dd u \frac{uv}{4\pi^4} G(k,ku,kv,\eta),
	\label{eq:one_l_two_v_g}
\end{align}
where we have used the following relation for an arbitrary function $f(\eta)$:
\begin{align}
 T\left[\int^\eta_{-\infty} \dd \eta' f(\eta') \int^\eta_{-\infty} \dd \eta'' f(\eta'') \right] = 2 \int^\eta_{-\infty} \dd \eta' f(\eta') \int^{\eta'}_{-\infty} \dd \eta'' f(\eta'').
\end{align}
The $G$ is defined by
\begin{align}
	G(k_1,k_2,k_3,\eta) \equiv & \ k_1^6 \int^\eta_{-\infty} \dd \eta' \lambda(\eta') U^*_{k_1}(\eta') U_{k_2}(\eta') U_{k_3}(\eta') 
	 \int^{\eta'}_{-\infty} \dd \eta'' \lambda(\eta'') U^*_{k_1}(\eta'')  U^*_{k_2}(\eta'') U^*_{k_3}(\eta'').
\end{align}
Since the third term in the right hand side of Eq.~(\ref{eq:two_vx_one_loop}) is the complex conjugate of the second term, we can easily obtain
\begin{align}
&\vev{ \left(T \left[- \frac{1}{2} \left(\int^\eta_{-\infty} \dd \eta' H_{\text{int},3}(\eta') \right)^2\right]\right)^\dagger \delta \phi_{\mathbf k}(\eta) \delta \phi_{\mathbf k'}(\eta)} \nonumber \\
	&=-(2\pi)^3 \delta(\mathbf k+ \mathbf k') \frac{2\pi^2}{k^3} (2 {U^*_k}^2(\eta)) \int^\infty_0 \dd v \int^{1+v}_{|1-v|} \dd u \frac{uv}{4\pi^4} G^*(k,ku,kv,\eta).
	\label{eq:one_l_two_v_g_comp}
\end{align}

For convenience, we here divide the one-loop contribution with two vertices (Eq.~(\ref{eq:two_vx_one_loop})) into two pieces:
\begin{align}
	\expval{\delta \phi_{\mathbf k}(\eta) \delta \phi_{\mathbf k'}(\eta)}_{2\vx} 
	&\equiv(2\pi)^3 \delta(\mathbf k+ \mathbf k') \frac{2\pi^2}{k^3} \left[\mathcal P^a_{\delta \phi, 2\vx}(k,\eta)+\mathcal P^b_{\delta \phi, 2\vx}(k,\eta)\right].
  \label{eq:p_a_2vx}
\end{align}
The first part is given by 
\begin{align}
	\mathcal P^a_{\delta \phi, 2\vx}(k,\eta) =  \int^\infty_0 \dd v \int^{1+v}_{|1-v|} \dd u \frac{uv}{4\pi^4} I(k, ku, kv, \eta) I^*(k, ku, kv, \eta),
	\label{eq:2vtx_one_loop}
\end{align}
where 
\begin{align}
	I(k,ku,kv,\eta) \equiv k^3 \int^\eta_{-\infty} \dd \eta' \lambda(\eta') 2\, \text{Im} \left[ U_k(\eta) U_k^*(\eta')\right] U_{ku}(\eta') U_{kv}(\eta').
	\label{eq:i_def}
\end{align}
The second part is given by 
\begin{align}
\mathcal P^b_{\delta \phi, 2\vx}(k,\eta) &= 
  8 \int^\infty_0 \dd v \int^{1+v}_{|1-v|} \dd u \frac{uv}{4\pi^4} k^6 \int^\eta_{-\infty} \dd \eta' \int^{\eta'}_{-\infty} \dd \eta'' \lambda(\eta') \lambda(\eta'') \nonumber \\
 & \phantom{\int^\infty_0 } 
 \times \text{Im} \left[U_k(\eta) U^*_k(\eta')\right]\Re\left[U_k(\eta) U^*_k(\eta'')\right] \left(\text{Im}\left[ U_{kv}(\eta')U^*_{kv}(\eta'')\right]\Re\left[U_{ku}(\eta') U^*_{ku}(\eta'')\right] + (u \leftrightarrow v)\right).
	\label{eq:asym_final}
\end{align}
We can easily check that the sum of Eqs.~(\ref{eq:2vtx_one_loop}) and (\ref{eq:asym_final}) is the same as the sum of Eqs.~(\ref{eq:loop_2vx_first}), (\ref{eq:one_l_two_v_g}), and (\ref{eq:one_l_two_v_g_comp}). 
These two parts, Eqs.~(\ref{eq:2vtx_one_loop}) and (\ref{eq:asym_final}), correspond to the two contributions in the equation of motion approach, respectively: 1) the square of the second order perturbations and 2) the product of the first order and the third order perturbations. See Appendix~\ref{app:eom} for details.

%%%%%%%%%%%%%%%%%
\subsection{One loop with one vertex}
%%%%%%%%%%%%%%%%%

From Eq.~(\ref{eq:one_loop_in_in_form}), we can express the one-loop contribution with one vertex as
\begin{align}
	\expval{\delta \phi_{\mathbf k}(\eta) \delta \phi_{\mathbf k'}(\eta)}_{1\vx} &= \vev{ \delta \phi_{\mathbf k}(\eta) \delta \phi_{\mathbf k'}(\eta) \left(T \left[-i \int^\eta_{-\infty} \dd \eta' H_{\text{int},4}(\eta') \right]\right)}  \nonumber \\
	& \quad + \vev{  \left(T \left[-i \int^\eta_{-\infty} \dd \eta' H_{\text{int},4}(\eta') \right]\right)^\dagger \delta \phi_{\mathbf k}(\eta) \delta \phi_{\mathbf k'}(\eta) }.
	\label{eq:one_vx_one_loop}
\end{align}	
For simpler notation, we here reexpress $H_{\text{int},4}(\eta)$ as 
\begin{align}
	H_{\text{int},4}(\eta) &= -\int \dd^3 x \, \frac{\mu(\eta)}{4} \delta \phi^4(\mathbf x,\eta), \\
	\mu(\eta) &\equiv -a^4(\eta) \frac{V^{(4)}(\phi)}{6}.
\end{align}
Then, we can rewrite the first term in Eq.~(\ref{eq:one_vx_one_loop}) as
\begin{align}
	&\vev{ \delta \phi_{\mathbf k}(\eta) \delta \phi_{\mathbf k'}(\eta) \left(T \left[-i \int^\eta_{-\infty} \dd \eta' H_{\text{int},4}(\eta') \right]\right)} \nonumber \\
	&= \vev{ \delta \phi_{\mathbf k}(\eta) \delta \phi_{\mathbf k'}(\eta) \left(i \int^\eta_{-\infty} \dd \eta' \frac{\mu(\eta')}{4} \int \frac{\dd^3 p_1}{(2\pi)^3} \int \frac{\dd^3 p_2}{(2\pi)^3}\int \frac{\dd^3 p_3}{(2\pi)^3} \delta \phi_{\mathbf p_1}(\eta') \delta \phi_{\mathbf p_2}(\eta') \delta \phi_{\mathbf p_3}(\eta') \delta \phi_{-\mathbf p_1 -\mathbf p_2 -\mathbf p_3}(\eta')  \right)} \nonumber \\
	&= (2\pi)^3 \delta(\mathbf k + \mathbf k') i \int^\eta_{-\infty} \dd \eta' \mu(\eta') \int \frac{\dd^3 p_1}{(2\pi)^3} 3 U_{k}(\eta) U_{k}^*(\eta') U_{k}(\eta) U_{k}^*(\eta') U_{p_1}(\eta') U_{p_1}^*(\eta').
	\label{eq:1vx_first_term}
\end{align}
Since the second term in Eq.~(\ref{eq:one_vx_one_loop}) is the complex conjugate of the first term, we obtain
\begin{align}
	&\vev{  \left(T \left[-i \int^\eta_{-\infty} \dd \eta' H_{\text{int},4}(\eta') \right]\right)^\dagger \delta \phi_{\mathbf k}(\eta) \delta \phi_{\mathbf k'}(\eta) } \nonumber \\
	&= (2\pi)^3 \delta(\mathbf k + \mathbf k') (-i) \int^\eta_{-\infty} \dd \eta' \mu(\eta') \int \frac{\dd^3 p_1}{(2\pi)^3} 3 U^*_{k}(\eta) U_{k}(\eta') U^*_{k}(\eta) U_{k}(\eta') U_{p_1}(\eta') U_{p_1}^*(\eta').
	\label{eq:1vx_second_term}
\end{align}
Similar to the two-vertex contribution, we define the power spectrum for the one-vertex contribution as 
\begin{align}
	\expval{\delta \phi_{\mathbf k}(\eta) \delta \phi_{\mathbf k'}(\eta)}_{1\vx} = (2\pi)^3 \delta(\mathbf k + \mathbf k') \frac{2\pi^2}{k^3} \mathcal P_{\delta \phi, 1\vx}(k,\eta).
\end{align}
Combining Eqs.~(\ref{eq:1vx_first_term}) and (\ref{eq:1vx_second_term}), we obtain 
\begin{align}
	\mathcal P_{\delta \phi, 1\vx}(k,\eta)&=
	 -\frac{k^3}{\pi^2} \int^\eta_{-\infty} \dd \eta' \mu(\eta')\, \text{Im}[U_k(\eta) U^*_k(\eta')]  \int \frac{\dd^3 p}{(2\pi)^3} 6\, \text{Re}[U_k(\eta) U^*_k(\eta')] U_p(\eta') U^*_p(\eta').
	\label{eq:1vx_final}
\end{align}
%%

%%%%%%%%%%%%%%%%%%%%%%%%%%%%%%%%
\section{Numerical results in fiducial setup}
\label{sec:num_result}
%%%%%%%%%%%%%%%%%%%%%%%%%%%%%%%%

To make the discussion concrete, we consider a feature model that has been used in the literature to generate large spikes in the power spectrum of curvature perturbations.
Specifically, we consider a resonance model in which oscillatory features appear in the inflaton potential. These oscillatory features induce background oscillations that resonate with the subhorizon quantum modes of inflaton. The resonance mechanism was originally introduced to generate some sizable non-Gaussianities with small corrections to the power spectrum that are compatible with the CMB constraints~\cite{Chen:2008wn,Flauger:2009ab, Flauger:2010ja,Chen:2010bka}. So, the features they generate in the power spectrum are small. On the other hand, on much smaller scales, observational bounds on the power spectrum are much weaker, and phenomenologically large spikes are allowed. Through the introduction of transient oscillatory features in the potential, resonance models are constructed to generate large spikes in the power spectrum only on the small scales (see e.g. Ref.~\cite{Cai:2019bmk,Zhou:2020kkf,Peng:2021zon}). 
Compared to the original resonance models, the background features here have larger oscillation amplitudes that create much broader resonance bands. They are used to boost the amplitude of the power spectrum by orders of magnitude.
These large spikes may seed PBHs or GWs.
To capture the essence of such models, we consider the following hilltop type potential with oscillatory features:
\begin{align}
	V(\phi) = V_0 \left[ 1 - \frac{1-n_s}{2} \frac{\phi^2}{2M_\Pl^2} + 2c\epsilon_0 \, D(\phi, \phi_0, \phi_s, \epsilon_0, \Lambda) \left(-1 + \cos\left( \frac{\phi-\phi_0}{\sqrt{2\epsilon_0} \Lambda M_\Pl} \right) \right)\right] + V_{\text{end}}(\phi),
\label{eq:pot_cos_pbh_planck}	
\end{align}
where $\epsilon_0 \equiv \frac{\phi_0^2}{2}\left( \frac{1-n_s}{2}\right)^2$ and $D$ is the smooth top-hat function for $\phi_0 \lesssim \phi \lesssim \phi_s$, defined as
\begin{align}
	D(\phi, \phi_0,\phi_s,\epsilon_0,\Lambda) &\equiv \left(\frac{1 + \tanh\left(\frac{\phi-\phi_0}{\sqrt{2\epsilon_0} \Lambda M_\Pl}\right)}{2}\right) \left(\frac{1 + \tanh\left(\frac{\phi_s - \phi}{\sqrt{2\epsilon_0} \Lambda M_\Pl}\right)}{2} \right).
\end{align}
The $V_{\text{end}}$ is the modification of the potential around the end of the inflation.
In our analysis, we neglect $V_{\text{end}}$ because it is irrelevant to the resonant amplification. 
The inflaton rolls from $0<\phi < \phi_0$, enters the oscillatory region around $\phi \simeq \phi_0$, and finally exits it around $\phi \simeq \phi_s$.
We will also frequently use the number of e-folds from the beginning of inflation, $N$, as the time variables. E.g.~$N_0$ and $N_s$ denote the time when the inflaton reaches $\phi_0$ and $\phi_s$, respectively.
The parameter $\epsilon_0$ is close to the slow-roll parameter $\epsilon$ at the beginning of the oscillatory feature, $\phi \simeq \phi_0$.
For simplicity, we assume $\epsilon_0$ is small enough that we can regard the Hubble parameter as a constant in the oscillatory region.
The $c$ is the amplitude of the oscillatory feature normalized by $2\epsilon_0 V_0$. 
The $n_s$ controls the scalar tilt on large scales and is set to be consistent with the Planck results~\cite{Planck:2018vyg}.
The $\Lambda$ determines the typical timescale of the oscillations of the background quantities such as the potential and its derivatives.
Assuming the averaged velocity of the inflaton in the oscillatory region is given by $\dot \phi|_{\text{ave}}/H \sim \sqrt{2\epsilon_0}M_\Pl$, we find that $\Lambda^{-1}$ roughly corresponds to the frequency $\omega$ of these background oscillations in unit of $H$, $\Lambda^{-1} \sim \omega/H$.
Also, we can express this with the e-folds for one oscillation as $\Delta N_{\text{osc}} \sim 2\pi \Lambda$.
Throughout this paper, we assume $\Lambda \lesssim \mathcal O(0.1)$ to focus on the resonant amplification of the subhorizon modes.

We can approximate the potential derivatives in the oscillatory region, $\frac{\phi-\phi_0}{\sqrt{2\epsilon_0} \Lambda M_\Pl} \gg 1$ and $\frac{\phi_s-\phi}{\sqrt{2\epsilon_0} \Lambda M_\Pl} \ll 1$, as
\begin{align}
	V^{(1)}(\phi) &\simeq -\frac{\sqrt{2\epsilon_0}V_0}{M_\Pl} \left[ 1 +  \frac{c}{\Lambda} \sin\left( \frac{\phi-\phi_0}{\sqrt{2\epsilon_0} \Lambda M_\Pl} \right) \right],\label{eq:d1_pot} \\
	V^{(2)}(\phi) &\simeq \frac{V_0}{M_\Pl^2}(n_s-1) - \frac{c V_0}{(\Lambda M_\Pl)^2} \cos\left( \frac{\phi-\phi_0}{\sqrt{2\epsilon_0} \Lambda M_\Pl} \right), \\
	V^{(3)}(\phi) &\simeq \frac{c V_0}{\sqrt{2\epsilon_0} (\Lambda M_\Pl)^3} \sin\left( \frac{\phi-\phi_0}{\sqrt{2\epsilon_0} \Lambda M_\Pl} \right), \label{eq:d3_pot}\\	
	V^{(4)}(\phi) &\simeq \frac{c V_0}{2\epsilon_0 (\Lambda M_\Pl)^4} \cos\left( \frac{\phi-\phi_0}{\sqrt{2\epsilon_0} \Lambda M_\Pl} \right), \label{eq:d4_pot}
\end{align}
where, in the expression of $V^{(1)}$, we have used the fact that $|n_s-1| \ll 1$ leads to $\frac{\phi^2}{2}\left( \frac{1-n_s}{2}\right)^2 \simeq \epsilon_0$ in the oscillatory region.
Note that, by multiplying the oscillatory term by $\epsilon_0$ as in Eq.~(\ref{eq:pot_cos_pbh_planck}), we pick a convenient parameterization in which the oscillation amplitude of $V^{(2)}$ is independent of $\epsilon_0$.
The oscillation amplitude of $V^{(2)}$ determines the strength of the parametric resonance, and therefore $c$ and $\Lambda$ are key parameters for the resonant amplification. 
We will discuss the parameter dependence of the amplification magnitude in more detail in Sec.~\ref{sec:necessary_conditions}.

\begin{figure}[h]
  \begin{minipage}[b]{0.49\linewidth}
    \centering
    \includegraphics[keepaspectratio, scale=0.55]{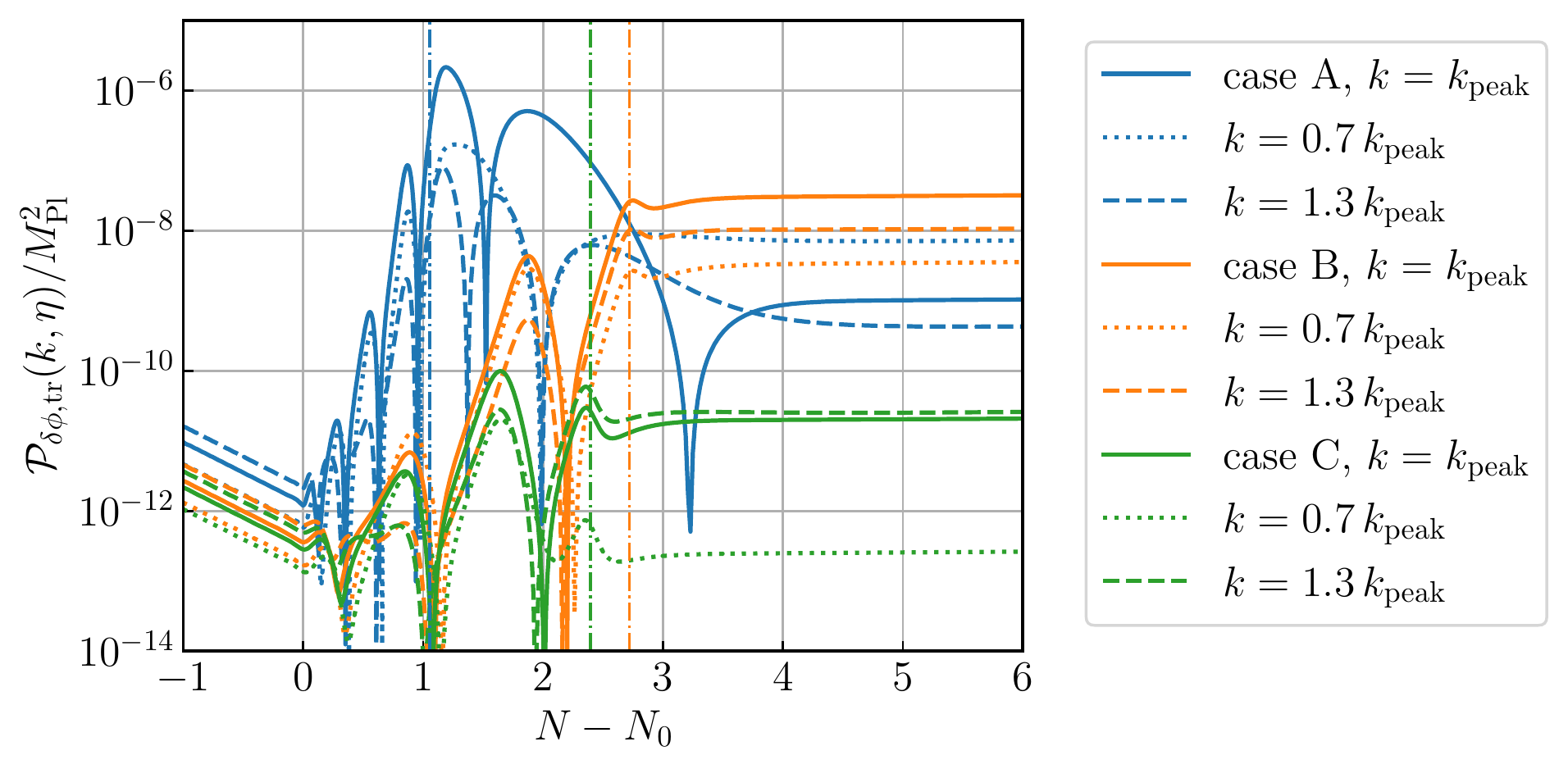}
  \end{minipage}
  \begin{minipage}[b]{0.49\linewidth}
    \centering
    \includegraphics[keepaspectratio, scale=0.55]{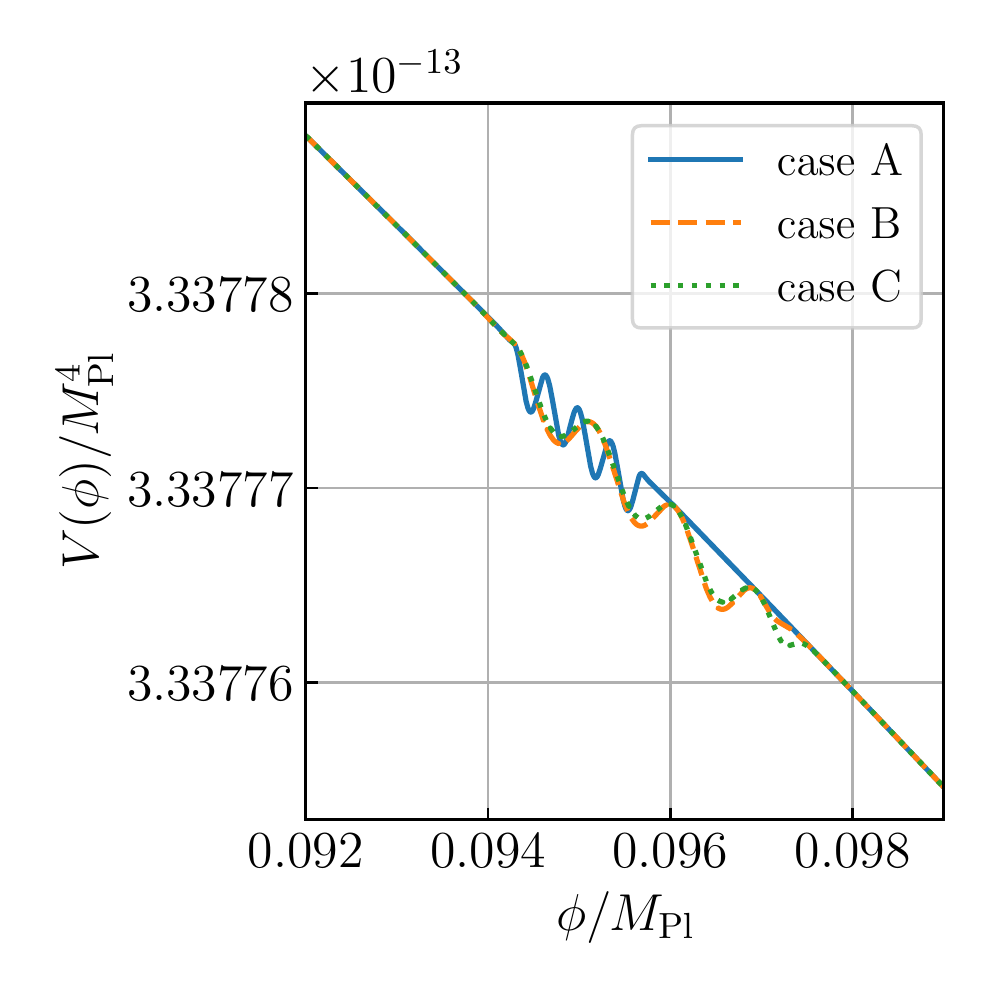}
  \end{minipage}
        \caption{ 
				[Left]: The evolution of the tree-level power spectrum. 
        The vertical dot-dashed lines show $N_s-N_0$ in each case, where the subscripts $0$ and $s$ denote the values at $\phi = \phi_0$ and $\phi = \phi_s$, respectively.
        For all the cases, we have $n_s = 0.97$, $\epsilon_0 = 1 \times 10^{-6}$ ($\phi_0/M_\Pl = 9.428\times 10^{-2}$), and $V_0/M_\Pl^4 = 3.338\times 10^{-13} (= 24\pi^2 \epsilon(\phi_{\text{CMB}}) \mathcal P_\zeta(k_{\text{CMB}})$ with $\phi_{\text{CMB}}/M_\Pl = 7.7\times 10^{-2}$, $k_{\text{CMB}} = 0.05\, \text{Mpc}^{-1}$, and $\mathcal P_\zeta(k_{\text{CMB}}) = 2.1 \times 10^{-9}$).
        Note that $\phi_{\text{CMB}}$ is the field value when $aH = k_{\text{CMB}}$.
        For other parameters, $c = 0.203, \Lambda = 0.04$, and $\phi_s-\phi_0 = \sqrt{2\epsilon_0}M_\Pl$ in case A, $c = 0.22, \Lambda = 0.1$, and $\phi_s-\phi_0 = 2\sqrt{2\epsilon_0}M_\Pl$ in case B, and $c = 0.19, \Lambda = 0.1$, and $\phi_s-\phi_0 = 2.13 \sqrt{2\epsilon_0}M_\Pl$ in case C.
        $k_\pe$ is defined as the scale which realizes the maximum value of the power spectrum during the resonance. % $N-N_s = 5$.
        We find $|k_{\text{peak}}\eta_0| = 21.4$ and $|k_{\text{peak}}\eta_s| = 7.43$ in case A, $|k_{\text{peak}} \eta_0| = 11.4$ and $|k_{\text{peak}}\eta_s| = 0.75$ in case B, and $|k_{\text{peak}} \eta_0| = 10.2$ and $|k_{\text{peak}}\eta_s| = 0.93$ in case C. 
        These parameters give $|\eta_0| = 3.14\times 10^{-5}\,\text{Mpc}$ for case A and $= 3.15\times 10^{-5}\,\text{Mpc}$ for case B anc C.
        [Right]: The potential form given by Eq.~(\ref{eq:pot_cos_pbh_planck}) in case A, B, and C.
        }  
        \label{fig:power_evol_pbh_pl}  
\end{figure}

\begin{figure}%[t]
  \begin{minipage}[b]{0.49\linewidth}
    \centering
    \includegraphics[keepaspectratio, scale=0.53]{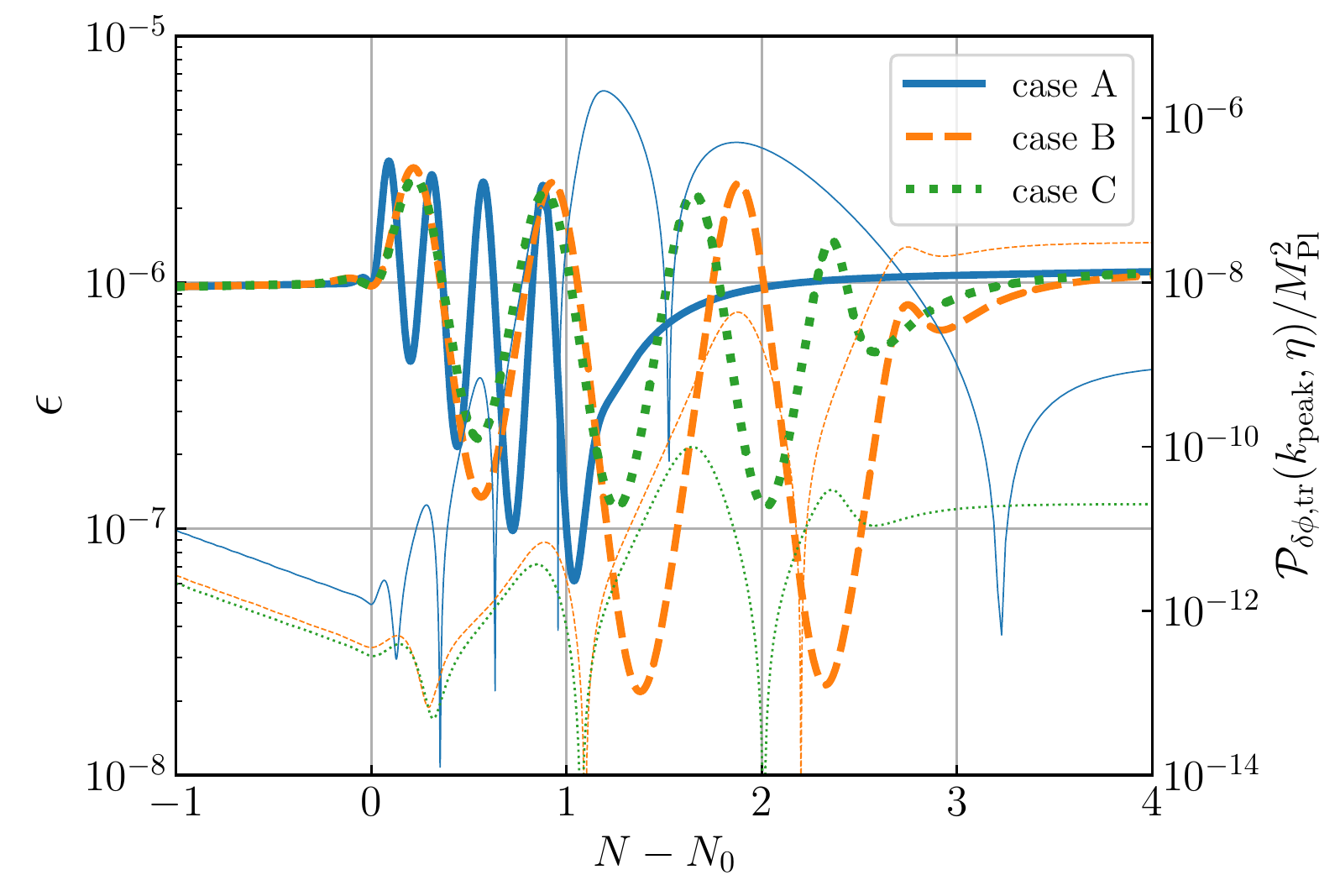}
  \end{minipage}
  \begin{minipage}[b]{0.49\linewidth}
    \centering
    \includegraphics[keepaspectratio, scale=0.53]{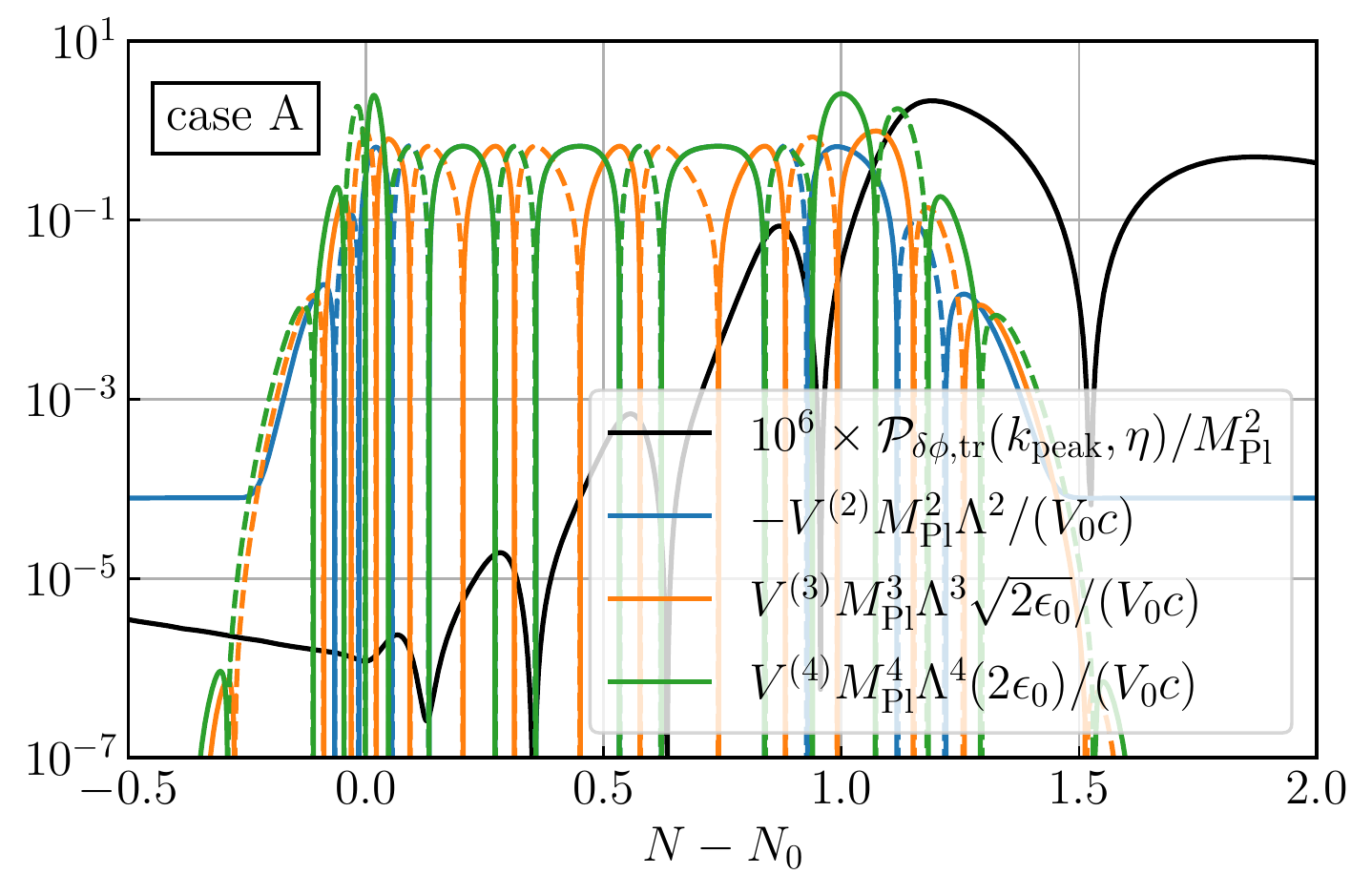}
  \end{minipage}
        \caption{ 
				[Left]: The evolution of $\epsilon$.
        The thin lines are $\mathcal P_{\delta \phi,\tre}(k_\pe,\eta)$ in each case.
        The parameters are the same as in Fig.~\ref{fig:power_evol_pbh_pl}.
        [Right]: 
		       The comparison between the evolution of the tree power spectrum at the peak scale $k_{\rm peak}$ and the potential derivatives. 
        The potential derivatives are normalized so that they are $\simeq \cos((\phi-\phi_0)/(\sqrt{2\epsilon_0}\Lambda M_\Pl))$ or $\sin((\phi-\phi_0)/(\sqrt{2\epsilon_0}\Lambda M_\Pl))$ in $0.1 \lesssim N-N_0 \lesssim 0.9$.
        The parameters are the same as in case A in Fig.~\ref{fig:power_evol_pbh_pl}.
        Note that the blue and green lines almost coincide in $0.1 \lesssim N-N_0 \lesssim 0.9$. 
        For negative values of the quantities, we show their absolute values with dashed lines.
        }  
        \label{fig:eps_evol_mu_lambda}  
\end{figure}

Figure~\ref{fig:power_evol_pbh_pl} shows the evolution of the tree-level power spectrum and the potential form in three fiducial parameter sets, case A, B, and C.
For convenience, we define $k_\pe$ as the scale which realizes the maximum value of the power spectrum during the resonance.
Note that $k_\pe$ can be different from the peak scale at the late time (after the resonance) in general.
The left panel of Fig.~\ref{fig:eps_evol_mu_lambda} shows the evolution of $\epsilon \,(= \dot \phi^2/(2H^2M_\Pl^2))$. 
$\epsilon$ oscillates between $\mathcal O(\epsilon_0)$ and $\mathcal O(0.1 \epsilon_0)$ in case A and C or between $\mathcal O(\epsilon_0)$ and $\mathcal O(0.01 \epsilon_0)$ in case B.
The reason for $\epsilon < \epsilon_0$ is due to the Hubble friction, coming from the expansion of the universe.
From this figure, we can see that the constant $\epsilon = \epsilon_0$ is not a good approximation, but still the oscillation timescale is not far from the rough estimate with $\epsilon = \epsilon_0$ in the oscillatory region. 
In addition, in Fig.~\ref{fig:power_evol_pbh_pl}, we can also see that $N_s - N_0$ is not very far from $(\phi_s - \phi_0)/(\sqrt{2\epsilon_0} \Lambda M_\Pl)$ even in our fiducial parameter sets.
These can be understood as the cancellation of the relatively fast-roll period with $\epsilon > \epsilon_0$ and the relatively slow-roll period with $\epsilon < \epsilon_0$.
We also note that, when the tree power spectrum reaches a local maximum, $\epsilon$ is roughly within $\mathcal O(1)$ deviation from $\epsilon_0$ in our fiducial parameter sets.
The right panel of Fig.~\ref{fig:eps_evol_mu_lambda} compares the evolution of the tree-level power spectrum and the potential derivatives.
From the right panel of Fig.~\ref{fig:eps_evol_mu_lambda}, we can see that, when the tree power spectrum reaches a local maximum within the oscillatory region ($0.1 \lesssim N-N_0 \lesssim 0.9$ in the figure), we find $V^{(2)} > 0$, $V^{(3)} > 0$, and $V^{(4)}<0$.
This is because the sign flip of $V^{(2)}$ from negative to positive (temporally) stops the growth of the tree power spectrum.
This behavior also explains the reason why $\epsilon$ is larger or not much smaller than $\epsilon_0$ at a local maximum time in the left panel of Fig.~\ref{fig:eps_evol_mu_lambda}.
The positive $V^{(2)}$ region is around a local minimum in the oscillatory feature and therefore the inflaton rolls relatively fast compared to that around a local maximum because of the potential energy difference.
We can also see that $V^{(3)}$ and $V^{(4)}$ are exponentially suppressed outside the oscillatory region. 
In particular, the exponential suppression around $N_0$ or $\eta_0$, the conformal time at $N_0$, effectively introduces the lower cutoff of the time integral in the loop calculations.\footnote{
When we numerically perform the time integrals in Eqs.~(\ref{eq:2vtx_one_loop}), (\ref{eq:asym_final}), and (\ref{eq:1vx_final}), we practically introduce the lower cutoff of the time integrals, which is sufficiently earlier than $\eta_0$. Note that the final numerical results do not depend on the lower cutoff of the time integrals if it is sufficiently early because the contribution much before $\eta_0$ is significantly suppressed due to the exponential suppression of $V^{(3)}$ and $V^{(4)}$.}

\begin{figure}%[t] 
        \centering \includegraphics[width=0.48\columnwidth]{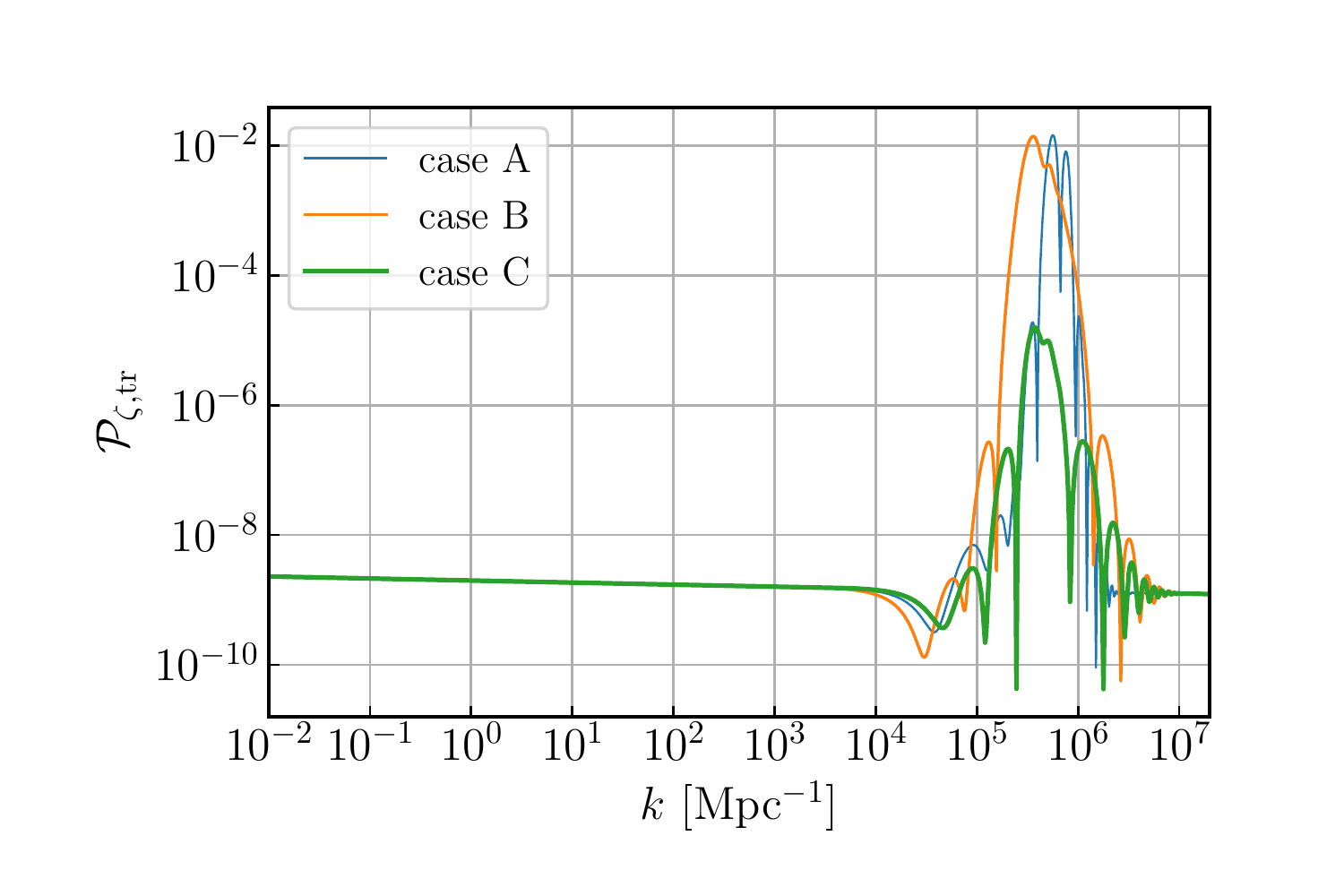}
        \vspace{-15pt}
        \caption{ The tree-level power spectrum of curvature perturbations after the resonant amplification and the horizon exit of the perturbations. 
        The parameters are the same as in Fig.~\ref{fig:power_evol_pbh_pl}.
        }
        \label{fig:pzeta_pbh_pl}
\end{figure}

\begin{figure}%[t] 
        \centering \includegraphics[width=0.75\columnwidth]{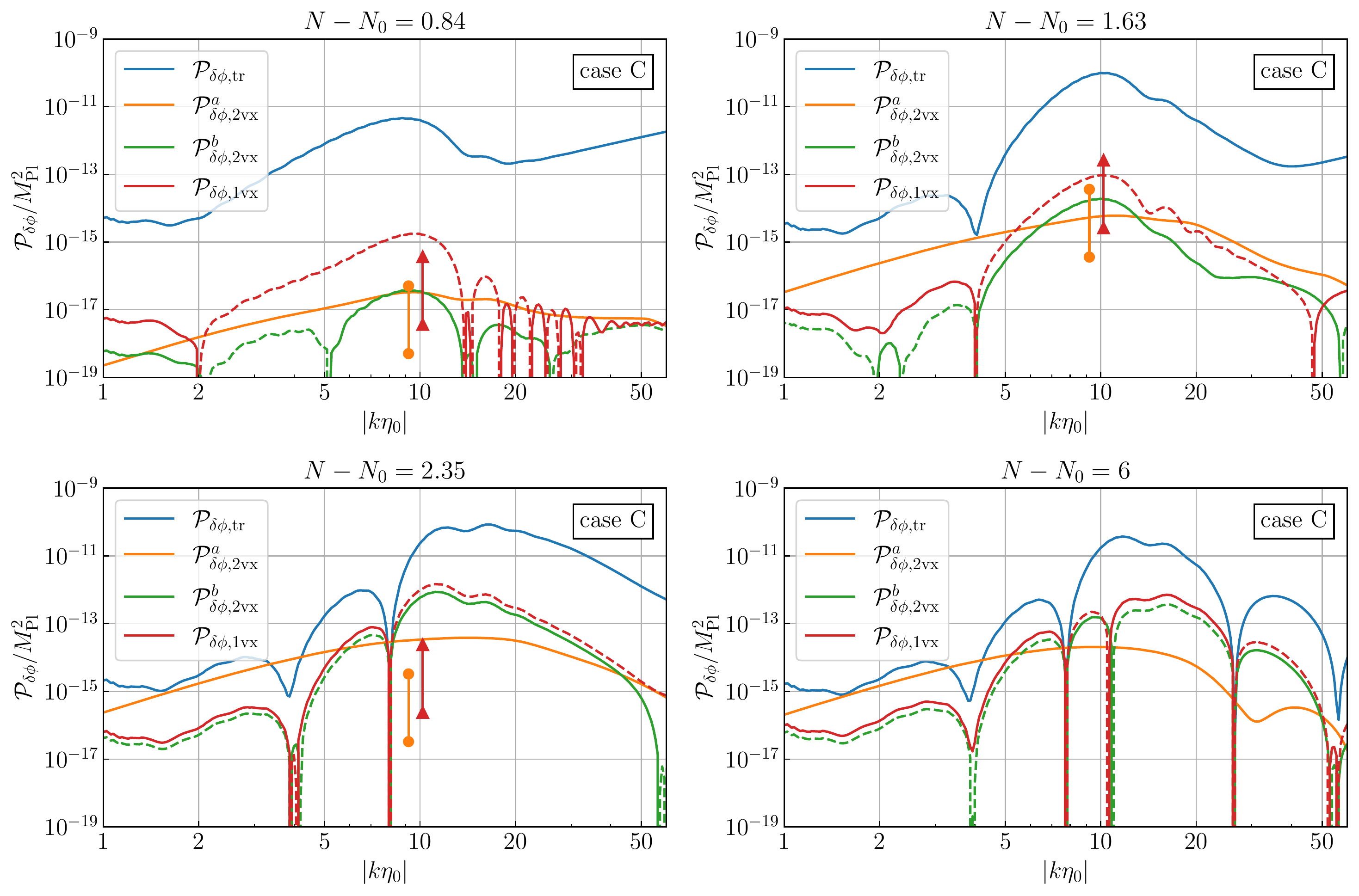}
        \caption{ 
        The power spectra at the times when the tree power spectrum on $k_\pe$ reaches several local maxima during the resonance ($N-N_0 = 0.84 \sim 2.35$) and after the resonance ($N- N_0 = 6$).
        The analytical estimates during the resonance ($\mathcal B$ and $0.01\,\mathcal B$) are also plotted.
        The parameter set is the same as case C in Fig.~\ref{fig:power_evol_pbh_pl}.
        }
        \label{fig:loop_pbh_pl_c}
\end{figure}

Figure~\ref{fig:pzeta_pbh_pl} shows the final tree power spectrum of the curvature perturbations.
Note that we calculate the curvature power spectrum based on the tree power spectrum through $\mathcal P_{\zeta,\tre} = \mathcal P_{\delta \phi,\tre}/(2\epsilon M_\Pl^2)$ during inflation, where $\delta \phi$ is evaluated in the spatially flat gauge.
For case A and B, we tune the parameters to realize $\mathcal P_{\zeta,\tre} \sim \mathcal O(10^{-2})$ on $k \sim \mathcal O(10^5)\, \text{Mpc}^{-1}$, often considered in the LIGO/Virgo PBH scenarios~\cite{Sasaki:2018dmp}.
We here note that the requirement of $\mathcal P_{\zeta,\tre} \sim \mathcal O(10^{-2})$ is based on the assumption of the Gaussian distribution of the perturbations. If we take into account their non-Gaussianities, the requirement can change in general~\cite{Byrnes:2012yx}. 
In this paper, we leave the analysis on the non-Gaussianities for future work and instead just use $\mathcal P_{\zeta,\tre} \sim \mathcal O(10^{-2})$ as a benchmark for simplicity.
On the other hand, for case C, we tune the parameters to keep the one-loop power spectra from dominating over the tree power spectrum, which we will see in Fig.~\ref{fig:loop_pbh_pl_c}.

\begin{figure}%[t] 
        \centering \includegraphics[width=0.75\columnwidth]{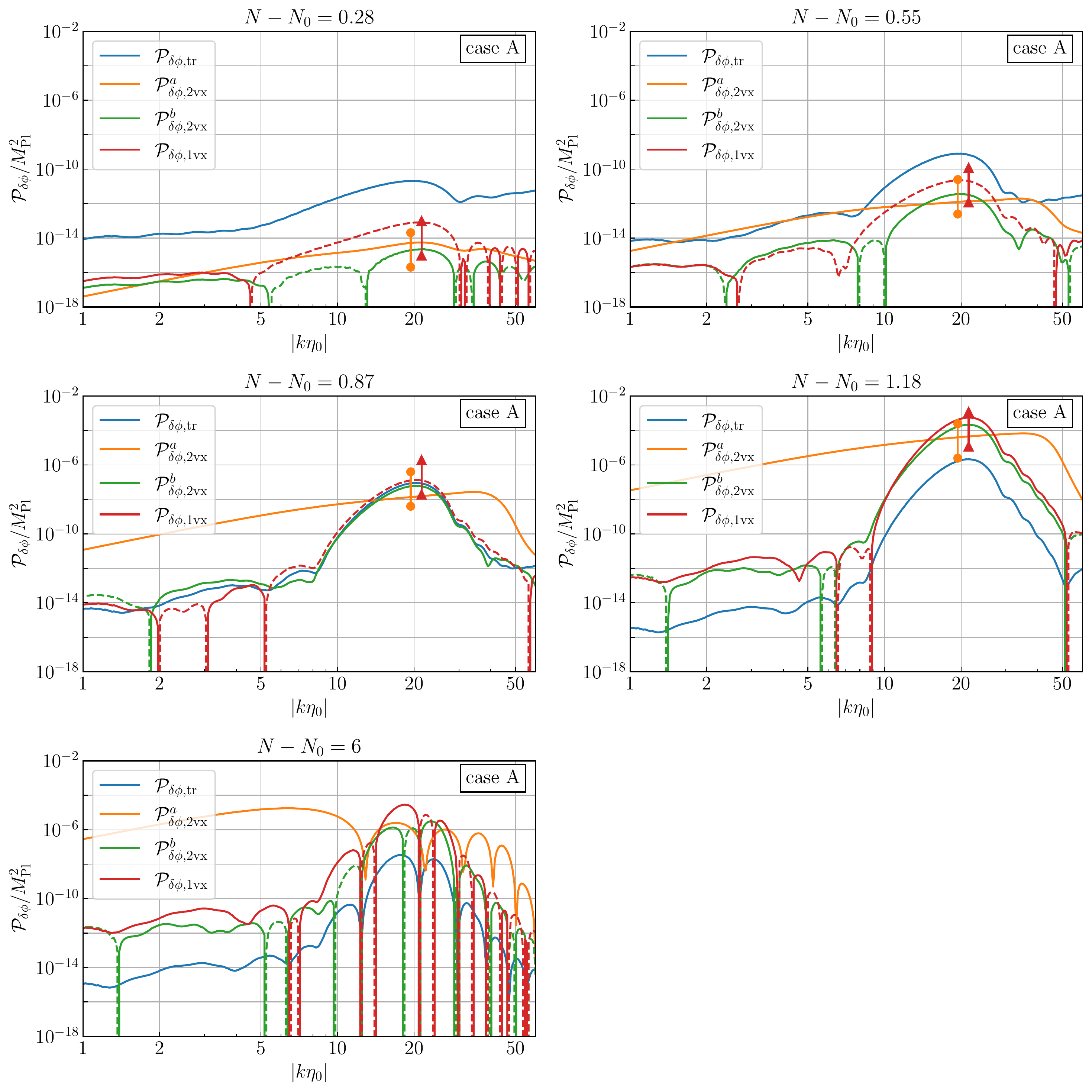}
        \caption{ 
         The power spectra at the times when the tree power spectrum on $k_\pe$ reaches several local maxima during the resonance ($N-N_0 = 0.28 \sim 1.18$) and after the resonance ($N- N_0 = 6$).
        The parameter set is the same as case A in Fig.~\ref{fig:power_evol_pbh_pl}.
        For negative values of the power spectra, we show their absolute values with dashed lines.
        The analytical estimates during the resonance, $\mathcal B$ and $0.01\,\mathcal B$, are plotted with orange circles for $\mathcal P^{a,b}_{\delta \phi, 2\vx}$ and red triangles for $\mathcal P_{\delta \phi, 1\vx}$. See Eqs.~(\ref{eq:b_approx_2vx_s}) and (\ref{eq:b_approx_1vx}) in the next section for the definition of $\mathcal B$.
        We basically put the points of $\mathcal B$ and $0.01\,\mathcal B$ at $k_\pe$, though we slightly shift them to another scale when they are overlapped.
        }
        \label{fig:loop_pbh_pl}
\end{figure}

\begin{figure}%[t] 
        \centering \includegraphics[width=0.75\columnwidth]{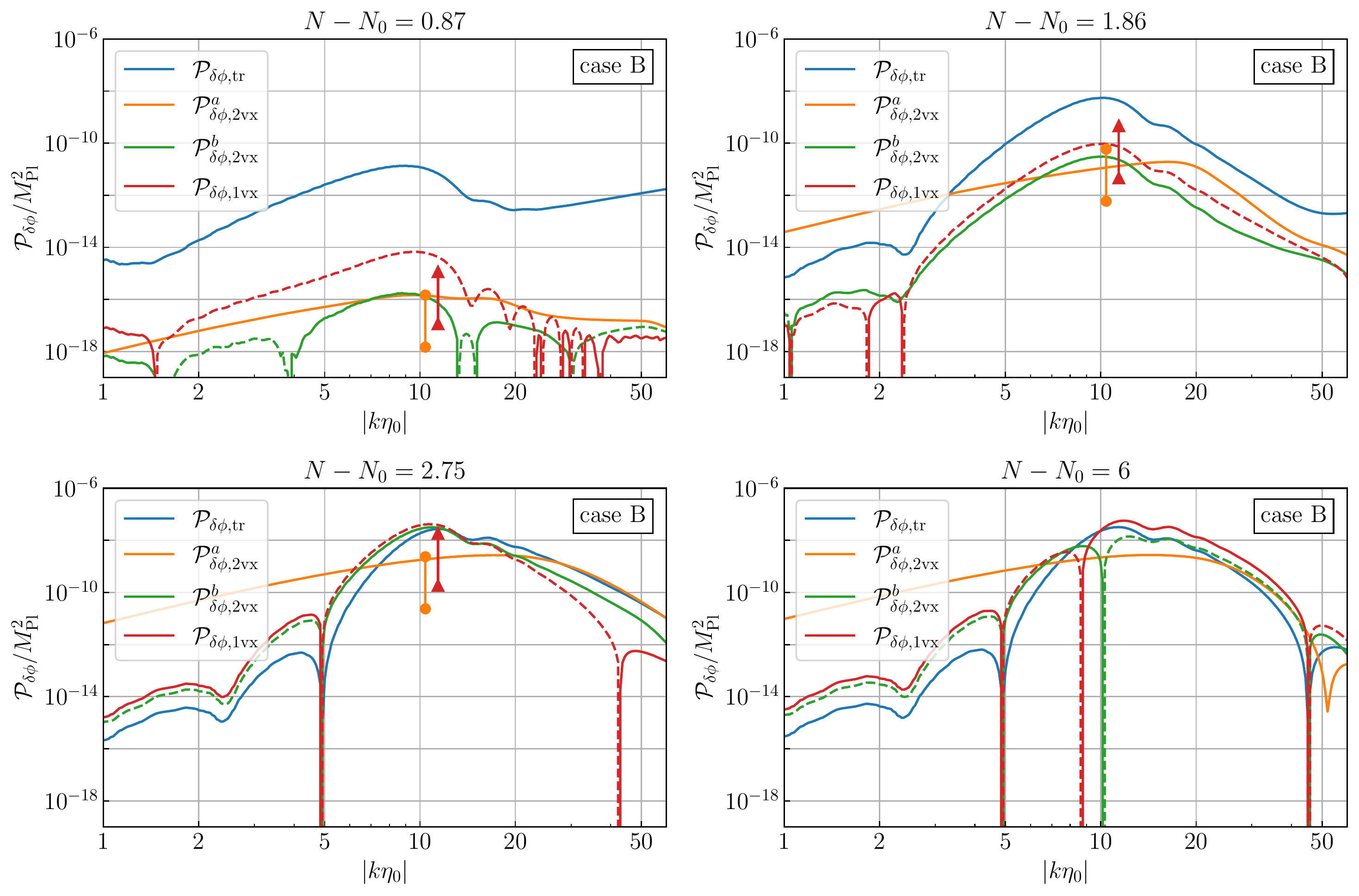}
        \caption{ 
        The power spectra at the times when the tree power spectrum on $k_\pe$ reaches several local maxima during the resonance ($N-N_0 = 0.87 \sim 2.75$) and after the resonance ($N- N_0 = 6$).
        The analytical estimates during the resonance ($\mathcal B$ and $0.01\,\mathcal B$) are also plotted.
        The parameter set is the same as case B in Fig.~\ref{fig:power_evol_pbh_pl}.
        }
        \label{fig:loop_pbh_pl_b}
\end{figure}

Figures~\ref{fig:loop_pbh_pl}-\ref{fig:loop_pbh_pl_c} show the loop contributions at different times. 
Except for the final results in $N-N_0 = 6$, we focus on the times when the tree power spectrum on $k_\pe$ reaches local maxima. 
That is, we focus on the times when $\dd\, \mathcal P_{\delta \phi,\text{tr}}(k_{\text{peak}},\eta)/\dd \eta = 0$ with $\dd^2\, \mathcal P_{\delta \phi,\text{tr}}(k_{\text{peak}},\eta)/\dd \eta^2 < 0$.
We will discuss the results at other times later in Sec.~\ref{subsec:when_tree_small} with analytical estimates.
When we calculate the loop power spectrum, we have neglected the contributions that are not much amplified by the resonance. % by setting the cutoff scales.
Practically, we have set the ultraviolet(UV) and infrared(IR) cutoffs of the wavenumber integrals as $|k_\UV \eta_0| = 61$ and $|k_\IR \eta_0| = 0.1$, while Figures~\ref{fig:loop_pbh_pl}-\ref{fig:loop_pbh_pl_c} show the power spectra in $1 \leq |k\eta_0| \leq 60$.\footnote{A similar procedure is used in Ref.~\cite{Fumagalli:2021mpc}, which discusses GWs induced by amplified perturbations during inflation.}
Since the dominant contribution comes from the resonantly amplified perturbations, the loop power spectrum on the peak scale is insensitive to the cutoff scales unless they cut the peak scale contributions or are very far from the peak scale.\footnote{
	Before the perturbations are amplified significantly, the loop power spectrum obtained with our calculation method is sensitive to the cutoff scales because the peak-scale perturbations are not dominant contributions at that time.
	In this paper, we ignore this issue because we would like to compare the loop and the tree power spectrum after the perturbations are amplified significantly.
}
Strictly speaking, the very small/large scale fluctuations that we have neglected can cause UV/IR divergences. 
However, such divergences originate from the perturbations irrelevant to the amplification and they should be appropriately regularized to be finite values (see e.g. Ref.~\cite{Seery:2010kh} for a review).
Given that the regularization should be done for the contributions from the perturbations that are not amplified, the regularized contributions from the UV/IR modes should be smaller than the contributions from the amplified perturbations, which we are interested in.
In Appendix~\ref{app:cutoff}, we see the limitation of our analysis with the cutoff scales by discussing how large or small the cutoff scales can be without changing our results.

Figures~\ref{fig:loop_pbh_pl}-\ref{fig:loop_pbh_pl_c} show that the power spectra grow until $N-N_0 \simeq 1.2$ in case A, $\simeq 2.8$ in case B, and $\simeq 2.4$ in case C because of the resonant amplification. 
After that, especially in case A, the power spectra decay following the evolution of the perturbations ($U_k(\eta) \propto a^{-1}$ during this period) until they exit the horizon. 
This decay can be interpreted as the redshift of the field fluctuations. 
The oscillation of the final tree power spectrum in $k$ is due to the perturbation evolution from the end of the resonance until the horizon exit of the perturbation. 
Unlike during the resonant amplification, the oscillations of the perturbations after the resonance are determined only by their wavenumbers
 and the perturbations with different wavenumbers experience a different number of oscillations from the end of the resonance until their horizon exits.
Also, the perturbations with smaller wavenumbers exit the horizon earlier, which leads to smaller suppression after the resonance. 
This is why the power spectra on smaller wavenumbers get less suppressed after the resonance.
Finally, the power spectrum exits the horizon and becomes almost constant, which corresponds to the results at $N-N_0 = 6$.
Strictly speaking, the superhorizon $\mathcal P_{\delta \phi}$ continues to change proportionally to $\epsilon$ even after the resonance, which can be seen from $\mathcal P_\zeta = \mathcal P_{\delta \phi}/(2\epsilon M_\Pl^2)$ and the conservation of the curvature perturbations on superhorizon. 
However, the change on superhorizon is very small and slow in our setup with $1-n_s \ll 1$ except around the end of inflation.

In Figs.~\ref{fig:loop_pbh_pl}-\ref{fig:loop_pbh_pl_b}, we can see that the loop power spectra after the resonance are larger than the tree power spectrum in case A and B. 
These are exactly the two cases for the PBH scenarios.
Note that $\mathcal P^b_{\delta \phi, 2\vx}$ and $\mathcal P_{\delta \phi, 1\vx}$ can be negative in general and therefore the total \emph{one-loop} power spectrum can be negative. 
If the total \emph{one-loop} power spectrum is negative, the higher order loop contributions must modify the power spectrum to make it positive. 
Given this, the results in case A and B indicate that not only the one-loop contribution but also the higher order loop contributions cannot be neglected, though we focus only on one-loop power spectrum throughout this work.
In the next section, we obtain analytical estimates of the loop power spectrum to support these numerical results.

\clearpage

%%%%%%%%%%%%%%%%%
\section{Analytical estimates}
\label{sec:analytical_estimates}
%%%%%%%%%%%%%%%%%

In this section, we derive analytical estimates of the loop power spectra on the peak scale and the one-loop correction to the background field in our fiducial model, given by Eq.~(\ref{eq:pot_cos_pbh_planck}).

%%%%%%%%%%%%%%%%%
\subsection{Rough estimates with the equation of motion}
\label{subsec:rough_estimates_eom}
%%%%%%%%%%%%%%%%%

Before proceeding to the concrete calculations in our setup, let us make very rough estimates using the equation of motion. 
Beyond the linear order, the equation of motion for the perturbation is given by
\begin{align}
	\delta \phi'' + 2 \mathcal H \delta \phi' - \nabla^2\delta \phi + a^2 V^{(2)} \delta \phi = -a^2\sum_{n>2} \frac{1}{(n-1)!}V^{(n)} \delta \phi^{n-1}.
	\label{eq:eom_general_n}
\end{align}
To get the rough estimate, let us focus on the modification due to the source terms (the right hand side) over the typical oscillation timescale of the peak-scale mode. Note that, during the resonance, the oscillation period of the peak-scale mode follows that of the background, $1/\omega$, instead of its physical wavenumber, $a(\eta)/k_\pe$; and this behavior can last for a couple e-folds for the broad band resonance. So, in physical time, the period of the oscillation is given by its initial value $\Delta t_{\text{osc}} \sim 1/\omega \sim a_0/k_\pe$ at the onset of the resonance. Converting it to the conformal time, we have $\Delta \eta_{\text{osc}} \sim a_0/(a(\eta) k_\pe)$.
This behavior can also be seen in the left panel of Fig.~\ref{fig:power_evol_pbh_pl}, which shows that the e-fold intervals (that is, the physical time intervals) between the local maxima do not much depend on time.\footnote{For the fiducial parameter sets in Fig.~\ref{fig:power_evol_pbh_pl}, $c$ is not very small and the resonance band is not so narrow, which allows the resonant amplification to continue even if $\Delta \eta_{\text{osc}}$ changes in time.
}

Then, we can roughly approximate the left hand side in Eq.~(\ref{eq:eom_general_n}) as $\sim (a(\eta)/a_0)^2 k_\pe^2 \delta \phi$.
For convenience, we express $\delta \phi$ as $\delta \phi = \delta \phi^{(1)} + \delta \phi^{(2)} + \delta \phi^{(3)} + \cdots$ with the superscript indicating the perturbation order. 
For the second order perturbation, we find
\begin{align}
   \delta \phi^{(2)} \sim  \frac{a_0^2 V^{(3)}}{k_\pe^2} \left(\delta {\phi^{(1)}}\right)^2,
\end{align}
Then, we can approximate the one-loop correction from the second order perturbations as 
\begin{align}
  \left(\delta \phi^{(2)}\right)^2 \sim \left(\frac{a^2_0 V^{(3)}}{k_\pe^2} \right)^2 \left(\delta \phi^{(1)} \right)^4.
  \label{eq:delta_22_eom}
\end{align}
This corresponds to the order of the $\mathcal P^a_{\delta \phi, 2\vx}$, which we will see in the next subsection (see also Appendix~\ref{app:eom}). 
On the other hand, for the third order perturbation, we find 
\begin{align}
   \delta \phi^{(3)} \sim  \frac{a_0^2 V^{(3)}}{k_\pe^2} \delta {\phi^{(1)}} \delta {\phi^{(2)}} + \frac{a_0^2 V^{(4)}}{k_\pe^2} \left(\delta {\phi^{(1)}}\right)^3,
\end{align}
where we have ignored $\mathcal O(1)$ coefficients in this expression.
Then, we can approximate the one-loop correction from this third order perturbations as 
\begin{align}
  \delta \phi^{(3)} \delta \phi^{(1)} \sim  \left(\frac{a^2_0 V^{(3)}}{k_\pe^2} \right)^2 \left(\delta \phi^{(1)} \right)^4 + \frac{a^2_0 V^{(4)}}{k_\pe^2}\left(\delta \phi^{(1)} \right)^4,
  \label{eq:delta_13_eom}
\end{align}
where we have ignored $\mathcal O(1)$  coefficients again. 
The first term in this equation corresponds to $\mathcal P^b_{\delta \phi, 2\vx}$ (see Appendix~\ref{app:eom}) and the second term corresponds to $\mathcal P_{\delta \phi, 1\vx}$.
From the above estimates, we can see that large potential derivatives from oscillatory features enable the loop power spectrum to be larger than the tree power spectrum even though the tree power spectrum normalized by $M_\Pl^2$ is much smaller than unity.

Apart from the perturbations, the background evolution is also modified by the perturbations through 
\begin{align}
	\phi'' + 2 \mathcal H \phi' + a^2 V^{(1)} = -a^2\sum_{n>1} \frac{1}{(n-1)!}V^{(n)} \expval{\delta \phi^{n-1}}.
\end{align}
Similar to the above perturbation cases, let us focus on the modification of the field evolution over the typical oscillation timescale, $\Delta \eta_{\text{osc}} \sim a_0/(a(\eta) k_\pe)$.
Then, we can approximate the modification of the background field value from $V^{(3)}$ as
\begin{align}
	\Delta \phi	&\sim \frac{a_0^2 V^{(3)}}{k_\pe^2} \left( \delta \phi^{(1)}\right)^2.
	\label{eq:delta_1_eom}	
\end{align}
We will see that the one-loop correction to the background in our fiducial setup is of the same order as this estimate.

%%%%%%%%%%%%%%%%%
\subsection{The two-vertex contributions}
\label{subsec:symm_approx}
%%%%%%%%%%%%%%%%%

We now obtain analytical estimates in our concrete example.
First, we discuss the two-vertex contributions.
In particular, we only discuss the order of $\mathcal P^a_{\delta \phi, 2\vx}$ explicitly in this subsection, though we can easily see that the order of $\mathcal P^b_{\delta \phi, 2\vx}$ on the peak scale is also the same as that of $\mathcal P^a_{\delta \phi, 2\vx}$ in a similar manner.
From Eq.~(\ref{eq:2vtx_one_loop}), we can express $\mathcal P^a_{\delta \phi,2\vx}$ as 
\begin{align}
 \mathcal P^a_{\delta \phi, 2\vx}(k,\eta) = \int^{v_\UV(k)}_{v_\IR(k)} \dd v \int^{\min[1+v,v_\UV(k)]}_{\max[|1-v|,v_\IR(k)]} \dd u \frac{uv}{4\pi^4} I(k, ku, kv, \eta) I^*(k, ku, kv, \eta).
 \label{eq:mathcal_p_delta_sym}
\end{align}
Note that we have explicitly introduced the UV/IR cutoff with $v_\UV(k)$ and $v_\IR(k)$.
We assume that the integrations over $u$ and $v$ include the peak scale contributions.
We also note that the drop of $\mathcal P^a_{\delta \phi,2\vx}$ around $k \simeq 2 k_\pe$ in Figs.~\ref{fig:loop_pbh_pl}-\ref{fig:loop_pbh_pl_c} is due to the momentum conservation, which prohibits two peak-scale modes in the loop from contributing to the power spectrum on $k \gtrsim 2k_\pe$.

\begin{figure}[t]
  \begin{minipage}[b]{0.49\linewidth}
    \centering
    \includegraphics[keepaspectratio, scale=0.55]{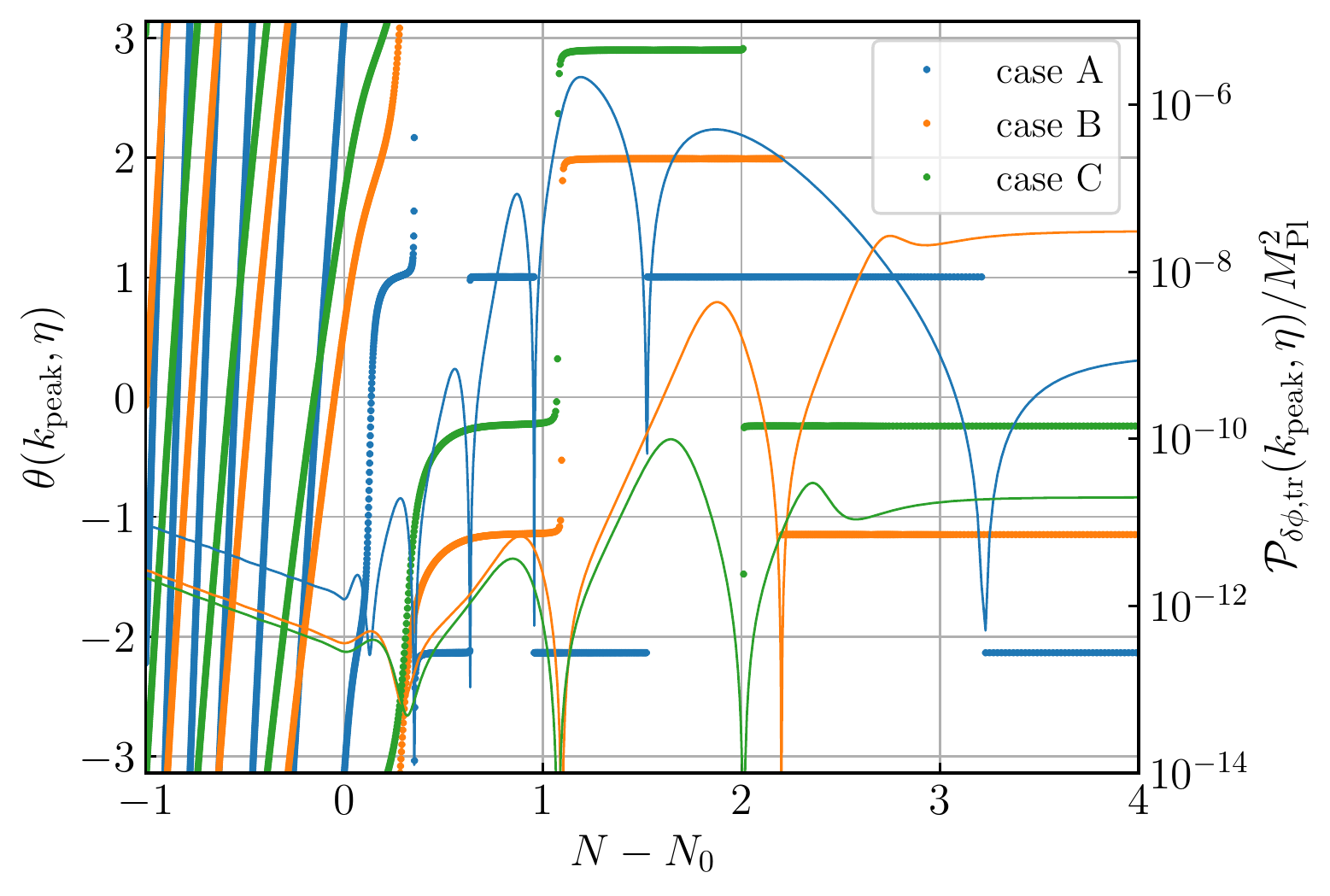}
  \end{minipage}
  \begin{minipage}[b]{0.49\linewidth}
    \centering
    \includegraphics[keepaspectratio, scale=0.55]{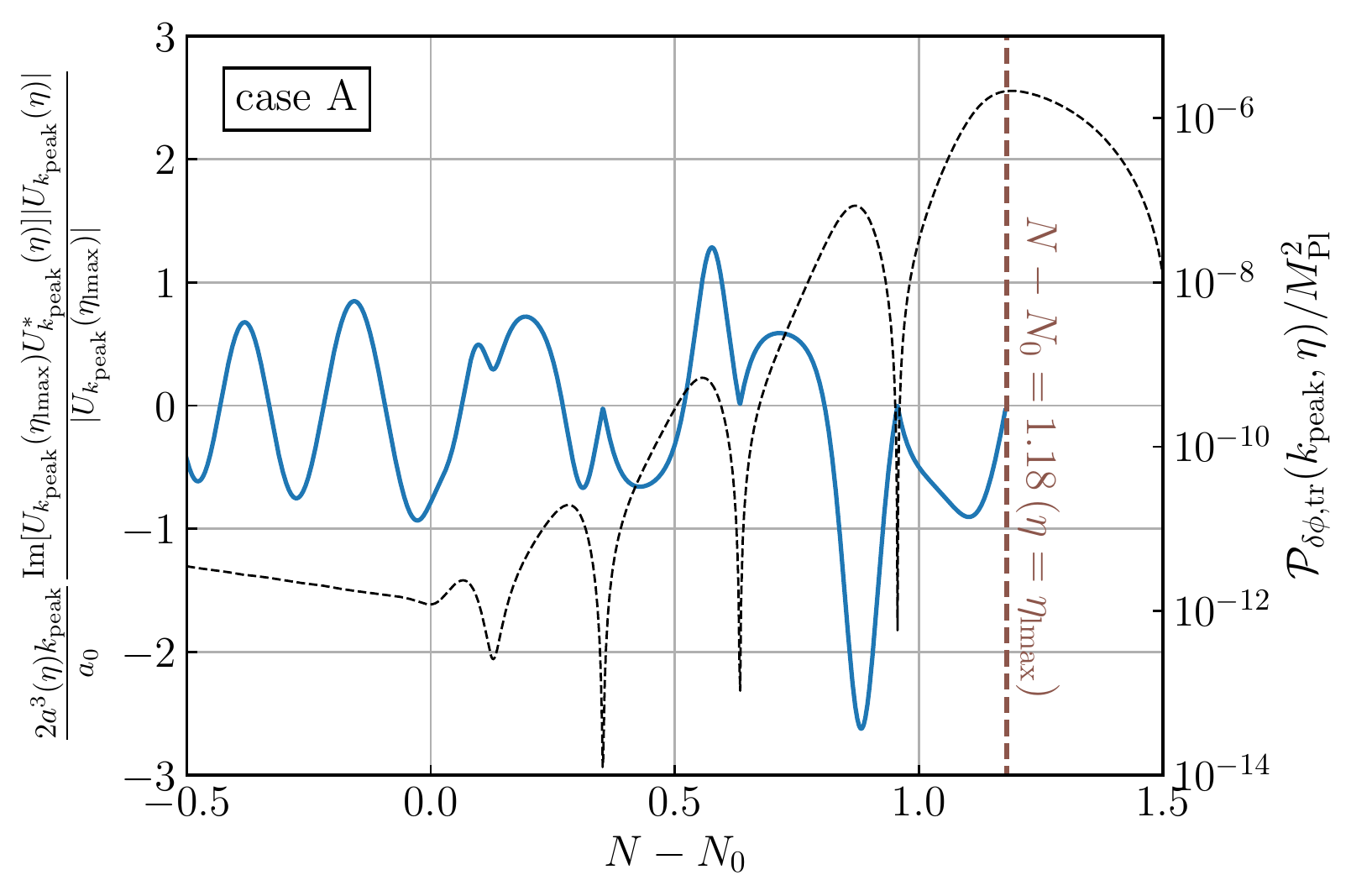}
  \end{minipage}
        \caption{ 
        [Left]: The evolution of $\theta$, defined through $U_k(\eta) = |U_k(\eta)|\ee^{-i\theta(k,\eta)}$ with $-\pi < \theta \leq \pi$.
        The parameters are the same as in Fig.~\ref{fig:power_evol_pbh_pl}.
        In addition, as an initial condition, we take $\theta(k_\pe,\eta) = -\pi/2$ at $|k_\pe \eta_0| = 100$.
        For comparison, we also plot the evolution of $\mathcal P_{\delta \phi,\tre}(k_\pe,\eta)$ in Fig.~\ref{fig:power_evol_pbh_pl} with thin solid lines.
        [Right]: The evolution of the left hand side of Eq.~(\ref{eq:im_factor_o1}) (blue solid) and $\mathcal P_{\delta \phi,\tre}(k_\pe,\eta)$ (black dashed). 
        The parameters are the same as case A in Fig.~\ref{fig:power_evol_pbh_pl} and $\eta_\lmax$ is taken to be at $N-N_0 = 1.18$.
        }  
        \label{fig:theta_evol}        
\end{figure}

Given that the contribution much before $\eta_0$ is negligible because of the exponential suppression of $V^{(3)}$ (see the right panel of Fig.~\ref{fig:eps_evol_mu_lambda}), we can approximate the function $I$ (Eq.~(\ref{eq:i_def})) as
\begin{align}
	I(k,ku,kv,\eta) \simeq k^3 \int^\eta_{\eta_0} \dd \eta' \lambda(\eta') 2\, \text{Im} \left[ U_k(\eta) U^*_k(\eta')\right] U_{ku}(\eta') U_{kv}(\eta').
	\label{eq:i_def_again}
\end{align}
At first glance, one might expect $\mathcal P^a_{\delta \phi, 2\vx} \stackrel{?}{\propto} |U|^8 \propto \mathcal P_{\delta \phi,\tre}^4$, but it is incorrect. 
The point lies in the $\text{Im}[\cdots]$, which is related to the fact that $\text{Im}\left[ U_k(\eta) U^*_k(\eta') \right]$ is proportional to the Green function, see Eq.~\eqref{eq:green_exp}. 
Hereafter, for simplicity, we focus on the loop power spectrum around the peak scale, $k \sim k_\pe$.
From the Wronskian condition Eq.~(\ref{eq:u_angle_conserve}), we can express the imaginary part as 
\begin{align}
	\text{Im}\left[ U_k(\eta) U^*_k(\eta') \right] &= -|U_k(\eta)||U_k(\eta')| \sin\left[ \theta(k,\eta) - \theta(k,\eta') \right]\nonumber \\
	&= -|U_k(\eta)||U_k(\eta')| \sin\left[ \int^\eta_{\eta'} \dd \eta'' \left(\frac{1}{2a^2(\eta'')|U_k(\eta'')|^2} \right) \right].
	\label{eq:im_sin_factor}
\end{align}
The left panel of Fig.~\ref{fig:theta_evol} shows the evolution of $\theta$. 
From this figure, we can see that $\theta$ becomes almost constant when $|U_k|$ is large.
On the other hand, $\theta$ changes by $\sim \pi$ when $|U_k|$ passes through a near-zero value. 
Note that the almost constant value of $\theta$ at the time of the large $|U_k|$ depends on the initial condition of $\theta$, which is arbitrary. 
However, the final power spectra do not depend on the initial $\theta$ because the power spectra only depend on the difference of $\theta$ at different times, as we will see below. 

Here, we focus on the loop power spectrum at the time the power spectrum reaches a local maximum, which corresponds to Figs.~\ref{fig:loop_pbh_pl}-\ref{fig:loop_pbh_pl_c} except for the $N-N_0 = 6$ case, though the following order estimate of the power spectrum is valid only if $\eta$ is close to a local maximum time. 
We will discuss the case where $\eta$ is far from the local maximum later in Sec.~\ref{subsec:when_tree_small}.
For convenience, we define $\eta_\lmax$ as a time the mode with $k_\pe$ reaches a local maximum and consider $\eta = \eta_\lmax$ in the following.
Note that there are several $\eta_\lmax$'s as the mode with $k_\pe$ oscillates.
Since the integrand of Eq.~(\ref{eq:i_def_again}) becomes large when $|U_{k_\pe}(\eta')|$ is large, let us focus on the contribution with $\eta'$ being around one of the local maximum times of the tree power spectrum.
The order of the typical displacement of $\eta'$ from the local maximum time is $\sim \Delta \eta_{\text{osc}}(\eta')$, estimated below Eq.~\eqref{eq:eom_general_n}.
If the local maximum time for $\eta'$ is different from $\eta_\lmax$, 
the sine factor in Eq.~(\ref{eq:im_sin_factor}) becomes zero around the local maximum time for $\eta'$ because the resonant amplification results from the match of one oscillation of $U_{k_\pe}$ and two oscillations of the background, which keeps the difference of $\theta$'s at two local maximums for $|U_{k_\pe}|$ close to $\sim q \pi$ with $q$ being an integer.
Then, we can estimate the sine factor contributed by the local peak near $\eta'$ to be $\sim \Delta \eta_{\text{osc}}(\eta')/(2a^2(\eta')|U_k^2(\eta')|)$.
Note that this approximation for the sine factor is valid even if the local maximum for $\eta'$ is the same as that at $\eta_\lmax$.
Then, we can approximate the typical order of the imaginary part on $k \sim k_\pe$ as
\begin{align}
	\text{Im}\left[ U_{k}(\eta_\lmax) U^*_{k}(\eta') \right] &\sim \pm|U_{k}(\eta_\lmax)| |U_{k}(\eta')| \left(\frac{\Delta \eta_{\text{osc}}(\eta')}{2a^2(\eta')|U_{k}(\eta')|^2} \right) \nonumber \\
	&\sim \pm\frac{\Delta \eta_{\text{osc}}(\eta')}{2a^2(\eta')} \frac{|U_{k}(\eta_\lmax)|}{|U_{k}(\eta')|},
	\label{eq:im_uu_approx}
\end{align}
where, if the local maximum time for $\eta'$ is the same as $\eta_\lmax$, the overall sign is negative.
Since the typical timescale is given by $\Delta \eta_{\text{osc}}(\eta') \sim a_0/(a(\eta')k_\pe)$, we can expect 
\begin{align}
	\frac{2a^3(\eta') k_\pe}{a_0} \frac{\text{Im}\left[ U_{k_\pe}(\eta_\lmax) U^*_{k_\pe}(\eta') \right] |U_{k_\pe}(\eta')|}{|U_{k_\pe}(\eta_\lmax)|} \sim \pm \mathcal O(1).
	\label{eq:im_factor_o1}
\end{align}
The right panel of Fig.~\ref{fig:theta_evol} shows the evolution of the left hand side of Eq.~\eqref{eq:im_factor_o1} with one of the parameter sets and indicates that this rough estimate works well.
From Eqs.~(\ref{eq:i_def_again}) and (\ref{eq:im_uu_approx}), we can see that the dominant contribution comes from $\eta' \sim \eta_\lmax$ and the power spectrum obeys $\mathcal P^a_{\delta \phi, 2\vx} \propto |U|^4 \propto \mathcal P_{\delta \phi,\tre}^2$.
Substituting Eq.~(\ref{eq:im_uu_approx}) into Eq.~(\ref{eq:i_def_again}), we obtain 
\begin{align}
	I(k,ku,kv,\eta_\lmax) &\sim -k^3 \Delta \eta_{\text{osc}}^2(\eta_\lmax) \frac{\lambda(\eta_\lmax)}{a^2(\eta_\lmax)} U_{ku}(\eta_\lmax) U_{kv}(\eta_\lmax) \nonumber \\
	&\sim -k^3 \left(\frac{a_0}{a(\eta_\lmax) k_\pe}\right)^2 \frac{\lambda(\eta_\lmax)}{a^2(\eta_\lmax)} U_{ku}(\eta_\lmax) U_{kv}(\eta_\lmax) ~,
	\label{eq:i_approx}
\end{align}
where we have used the fact that all the oscillating quantities ($\lambda$ and $U_k$) have the same oscillation timescale, $\Delta \eta_{\text{osc}}$.

For the loop power spectrum on $k \sim k_\pe$, the dominant contribution comes from $u \sim v \sim 1$ with $\mathcal O(1)$ width in the normalized wavenumber integrals of Eq.~(\ref{eq:mathcal_p_delta_sym}) and therefore we can approximate it as 
\begin{align}
 \mathcal P^a_{\delta \phi, 2\vx}(k,\eta_\lmax) &\sim \int^{v_\UV(k)}_{v_\IR(k)} \dd v \int^{\min[1+v,v_\UV(k)]}_{\max[|1-v|,v_\IR(k)]} \dd u  \frac{uv}{4\pi^4} k^6 \left( \frac{a_0}{a(\eta_\lmax)k_\pe}\right)^4 \frac{\lambda^2(\eta_\lmax)}{a^4(\eta_\lmax)} |U_{k u}(\eta_\lmax) U_{k v}(\eta_\lmax)|^2  
 \nonumber \\
 &\sim \frac{1}{4\pi^4} k_\pe^2 \left( \frac{a_0}{a(\eta_\lmax)}\right)^4 \frac{\lambda^2(\eta_\lmax)}{a^4(\eta_\lmax)} |U_{k_\pe}(\eta_\lmax)|^4 \nonumber \\
 &\sim \left(\frac{\lambda(\eta_\lmax)}{k_\pe^2 a^2(\eta_\lmax)}\right)^2 \left( \frac{a_0}{a(\eta_\lmax)}\right)^4 \mathcal P_{\delta\phi, \tre}^2(k_\pe,\eta_\lmax) \nonumber \\
 &\sim \left(\frac{a_0^2 V^{(3)}(\phi(\eta_\lmax))}{2 k_\pe^2 }\right)^2 \mathcal P_{\delta\phi, \tre}^2(k_\pe,\eta_\lmax).
 \label{eq:p_2vx_sym_est}
\end{align}
Note that the final line is of the same order of magnitude as the rough estimate with the equation of motion, Eq.~(\ref{eq:delta_22_eom}).
From Eq.~(\ref{eq:d3_pot}), we can get 
\begin{align}
	\left(\frac{a_0^2 V^{(3)}(\phi(\eta_\lmax))}{2 k_\pe^2 }\right)^2	&\sim \mathcal O\left(\frac{a^4_0}{4 k_\pe^4} \frac{c^2 V_0^2}{2 \epsilon_0 \Lambda^6 M_\Pl^6} \right) \nonumber \\ 
	&\sim \mathcal O\left(\frac{9 c^2}{8 \epsilon_0 (k_\pe \eta_0)^4 \Lambda^6 M_\Pl^2 }\right), 
	\label{eq:lambda_squ}
\end{align}
where we have focused on the oscillatory region of the potential and approximated the trigonometric factors as $\mathcal O(1)$.
To compare Eq.~(\ref{eq:p_2vx_sym_est}) with the numerical results, we here define the estimated value as 
\begin{align}
		\mathcal B_{2\vx}(\eta) \equiv \frac{9 c^2}{8 \epsilon_0 (k_\pe \eta_0)^4 \Lambda^6 M_\Pl^2} \mathcal P_{\delta\phi, \tre}^2(k_\pe,\eta).
		\label{eq:b_approx_2vx_s}
\end{align}
Note again this estimated value applies not only to $\mathcal P^a_{\delta \phi,2\vx}$ but also to $\mathcal P^b_{\delta \phi,2\vx}$.
Since we have ignored $\mathcal O(1)$ factors and, in particular, approximated all trigonometric factors as $\mathcal O(1)$, we roughly expect that the numerical results are between $\mathcal O(\mathcal B_{2\vx})$ and  $\mathcal O(0.01\, \mathcal B_{2\vx})$, which are mostly consistent with Figs.~\ref{fig:loop_pbh_pl}-~\ref{fig:loop_pbh_pl_c}. 
Note that, although we plot the estimated value even in the case of $N-N_0=2.35$ in Figs.~\ref{fig:loop_pbh_pl_c} (case C), that case does not correspond to the case of $\eta_\lmax$ (a local maximum time during the resonance of \emph{the $k_\pe$ mode}) because the perturbation on $k_\pe$ at that time is smaller than that at the previous local maximum ($N-N_0 = 1.63$), which means that the resonant amplification for the perturbation on $k_\pe$ ends before $N-N_0 = 2.35$.
In addition, the peak of the tree power spectrum extends over $10 \lesssim |k\eta_0| \lesssim 20$ at $N-N_0 = 2.35$.
Our analytical estimates are based on the assumption that the resonance peak scale is one scale, but that assumption is not valid in that case.
For these reasons, the numerical result in that case is not between $\mathcal O(\mathcal B_{2\vx})$ and  $\mathcal O(0.01\, \mathcal B_{2\vx})$.

%%%%%%%%%%%%%%%%%
\subsection{The one-vertex contribution}
%%%%%%%%%%%%%%%%%

Next, we discuss the one-vertex contribution.
From Eq.~(\ref{eq:1vx_final}), the one-vertex power spectrum is given by 
\begin{align}
	\mathcal P_{\delta \phi, 1\vx}(k,\eta) &\simeq - \frac{k^3}{\pi^2} \int^\eta_{\eta_0} \dd \eta' \mu(\eta') \text{Im}\left[U_k(\eta) U^*_k(\eta')\right] \int^{p_\UV}_{p_\IR} \frac{\dd^3 p}{(2\pi)^3} 6\, \text{Re} \left[U_k(\eta) U^*_k(\eta') \right] U_p(\eta') U^*_p(\eta'), 
	\label{eq:1vx_ana}
\end{align}
where we have set $\eta_0$ as the lower bound of the time integral due to the exponential suppression of $V^{(4)}$ before $\eta_0$ and introduced the cutoff scales that do not cut the peak scale contribution.
Similar to $\mathcal P^a_{\delta \phi, 2\vx}$, when $\eta = \eta_\lmax$, the dominant contribution comes from $\eta' \sim \eta_\lmax$.
Then, in the case of $k \sim k_\pe$, we can approximate the above as 
\begin{align}
	\left|\mathcal P_{\delta \phi, 1\vx}(k,\eta_\lmax)\right| &\sim \left|\Delta \eta_{\text{osc}}^2(\eta_\lmax) \frac{\mu(\eta_\lmax)}{a^2(\eta_\lmax)} 6\, \frac{k^3}{2\pi^2}|U_k(\eta_\lmax)|^2 \int^{p_\UV}_{p_\IR} \frac{\dd p}{p} \mathcal P_{\delta \phi,\tre}(p,\eta_\lmax)\right| \nonumber \\
	&\sim \left|\frac{6 \mu(\eta_\lmax)}{k_\pe^2 a^2(\eta_\lmax)} \right|\left(\frac{a_0}{a(\eta_\lmax)}\right)^2 \mathcal P_{\delta \phi, \tre}(k,\eta_\lmax) \mathcal P_{\delta \phi,\tre}(k_\pe,\eta_\lmax) \nonumber \\
	&\sim \left|\frac{a_0^2 V^{(4)}(\phi(\eta_\lmax))}{k_\pe^2 }\right| \mathcal P_{\delta \phi, \tre}(k,\eta_\lmax) \mathcal P_{\delta \phi,\tre}(k_\pe,\eta_\lmax),
	\label{eq:1vx_ana2}
\end{align}
where we have used $\Delta \eta_{\text{osc}} \sim a_0/(a(\eta_\lmax)k_\pe)$ and $\int^{p_\UV}_{p_\IR} \frac{\dd p}{p} \mathcal P_{\delta \phi,\tre}(p,\eta_\lmax) \sim \mathcal P_{\delta \phi,\tre}(k_\pe,\eta_\lmax)$.
Note that the final line is of the same order of magnitude as the rough estimate with the equation of motion, Eq.~(\ref{eq:delta_13_eom}).
From this equation, we can see that $\mathcal P_{\delta \phi, 1\vx}(k) \propto \mathcal P_{\delta \phi, \tre}(k) \mathcal P_{\delta \phi, \tre}(k_\pe)$ at least on $k \sim k_\pe$, which are consistent with the numerical results in Figs.~\ref{fig:loop_pbh_pl}--\ref{fig:loop_pbh_pl_c}, showing that $|\mathcal P_{\delta \phi, 1\vx}(k)|$ has a similar $k$-dependence as $\mathcal P_{\delta \phi, \tre}(k)$.
Using Eq.~(\ref{eq:d4_pot}), we obtain 
\begin{align}
	 \left|\frac{a_0^2 V^{(4)}(\eta_\lmax)}{k_\pe^2 } \right|
	&\sim \left|\frac{a_0^2 c V_0 }{2\epsilon_0 k_\pe^2 \Lambda^4 M_\Pl^4 }\right| \nonumber \\ 
	&\sim \left|\frac{3 c }{2\epsilon_0(k_\pe \eta_0)^2 \Lambda^4 M_\Pl^2  }\right|, 
	\label{eq:mu_approx}
\end{align}
where we have approximated the trigonometric factor as $\mathcal O(1)$ again.
Similar to the two-vertex contribution, we here define the estimated value in Eq.~(\ref{eq:1vx_ana2}) as 
\begin{align}
		\mathcal B_{1\vx}(\eta) \equiv \frac{3 c }{2\epsilon_0(k_\pe \eta_0)^2 \Lambda^4 M_\Pl^2 } \mathcal P_{\delta\phi, \tre}^2(k_\pe,\eta).
		\label{eq:b_approx_1vx}
\end{align}
In Figs.~\ref{fig:loop_pbh_pl}-\ref{fig:loop_pbh_pl_c}, we can see that the numerical results are mostly between $\mathcal O(\mathcal B_{1\vx})$ and $\mathcal O(0.01\, \mathcal B_{1\vx})$, except for the result for the case of $N-N_0=2.35$ in Figs.~\ref{fig:loop_pbh_pl_c} (case C). See the end of the previous subsection for the reason for the inconsistency.

%%%%%%%%%%%%%%%%%
\subsection{When the tree power spectrum becomes small}
\label{subsec:when_tree_small}
%%%%%%%%%%%%%%%%%

\begin{figure}[t] 
        \centering \includegraphics[width=0.8\columnwidth]{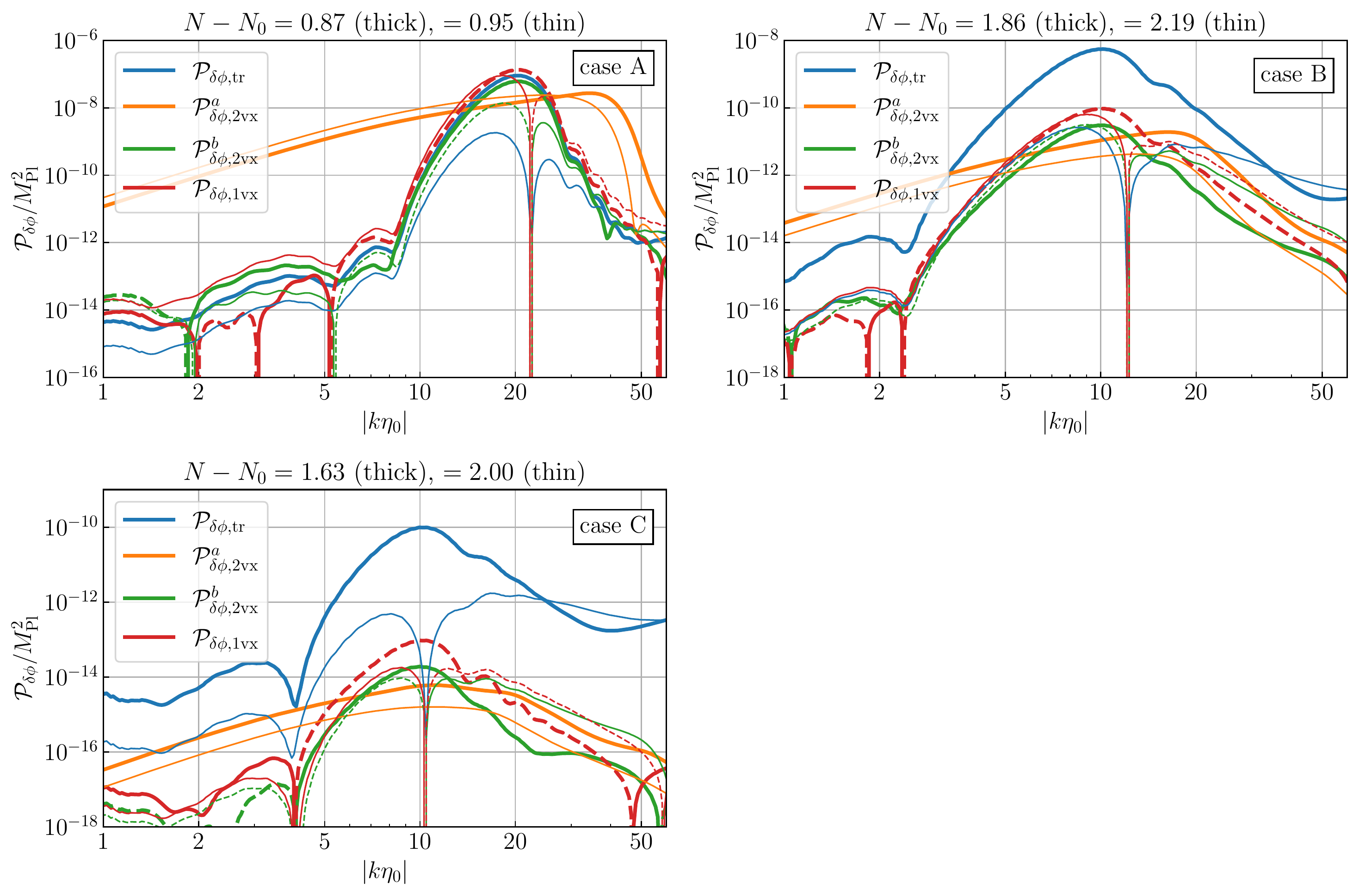}
        \caption{ 
        Comparison between the power spectra at the times when $\mathcal P_{\delta \phi,\tre}(k_\pe,\eta)$ reaches a local maximum (thick) and becomes much smaller than the local maximum (thin).
				The parameter sets are the same as in Fig.~\ref{fig:power_evol_pbh_pl}.
        }
        \label{fig:loop_lmax_lmin}
\end{figure}

Although we have so far focused on the loop power spectra at the time when the tree power spectrum on the peak scale reaches a local maximum, $\eta$ can be different from a local maximum time. 
Here, we discuss the loop power spectra at the time when the tree power spectrum becomes much smaller than a local maximum.
Figure~\ref{fig:loop_lmax_lmin} compares the power spectra at the times when $\mathcal P_{\delta \phi,\tre}(k_\pe,\eta)$ reaches a local maximum ($N-N_0 = 0.87$ (case A), $=1.86$ (case B), $=1.63$ (case C)) and becomes much smaller than the local maximum ($N-N_0 = 0.95$ (case A), $=2.19$ (case B), $=2.00$ (case C)).
Note that the latter times are between the two adjacent local maximum times: for example, $N-N_0 = 0.95$ in case A is between the two local maximum times, $N-N_0 = 0.87$ and $N-N_0 = 1.18$ in case A.
From this figure, we can see that the orders of the loop power spectra do not change much.
In the following, we discuss the reason for this behavior.

Hereafter, we take $\eta$ to be the time when the tree power spectrum becomes much smaller just after some $\eta_\lmax (< \eta)$.
In that case, the dominant contributions for all the loop power spectra come from $\eta' \sim \eta_\lmax$ because of the large tree power spectrum around $\eta_\lmax$. 
At $\eta' = \eta_\lmax$, the imaginary part that appears in all the loop power spectra can be approximated as 
\begin{align}
	\left|\text{Im}\left[ U_k(\eta) U^*_k(\eta_\lmax) \right]\right| 
	&= \left|-|U_k(\eta)||U_k(\eta_\lmax)| \sin\left[ \int^\eta_{\eta_\lmax} \dd \eta'' \left(\frac{1}{2a^2(\eta'')|U_k(\eta'')|^2} \right) \right] \right|\nonumber \\
  &\sim \left|-|U_k(\eta)||U_k(\eta_\lmax)| \sin\left[\int^{U_{k}(\eta)}_{U_{k}(\eta_\lmax)} \dd U_{k} \left(\frac{1}{2a^2(\eta_\lmax)|U_{k}|^2 |U_{k}|'} \right)\right] \right| \nonumber \\
	&\sim |U_k(\eta)||U_k(\eta_\lmax)| \left| \sin\left[\frac{\Delta \eta_{\text{osc}}(\eta_\lmax)}{2a^2(\eta_\lmax)|U_{k}(\eta)| |U_{k}(\eta_\lmax)|} \right] \right|\nonumber \\
	&\sim \frac{\Delta \eta_{\text{osc}}(\eta_\lmax)}{2a^2(\eta_\lmax)},
\end{align}
where we have used $a(\eta) \sim a(\eta_\lmax)$, $\left|U_k(\eta)'\right| \sim |U_k(\eta_\lmax)|/\Delta \eta_{\text{osc}}(\eta_\lmax)$, and $|U_k(\eta)| \ll |U_k(\eta_\lmax)|$.
This relation shows that, even if $\eta$ corresponds to the time when the tree power spectrum becomes small, the order of the imaginary part does not change from that in the case of $\eta = \eta_\lmax$ and therefore the order of $\mathcal P^a_{\delta\phi, 2\vx}$ does not change.
Although  $\mathcal P^b_{\delta\phi, 2\vx}$ and $\mathcal P_{\delta\phi, 1\vx}$ still have one more $U_k(\eta)$ in $\text{Re}[\cdots]$ in Eqs.~(\ref{eq:asym_final}) and (\ref{eq:1vx_final}), $|U_k(\eta_\lmax)/U_k(\eta)|$ in Fig.~\ref{fig:loop_lmax_lmin} is not far from $\mathcal O(1)$ except around the scales on which $U_k(\eta)$ crosses a near-zero value.
Since the oscillations of the coefficients $\lambda$ and $\mu$ can make $\mathcal O(1)$ changes, $\mathcal P^b_{\delta\phi, 2\vx}$ and $\mathcal P_{\delta\phi, 1\vx}$ do not necessarily decrease even if we consider the time for the smaller tree power spectrum, except on the scales for the near-zero crossing of $U_k(\eta)$.

%%%%%%%%%%%%%%%%%
\subsection{Tadpole contribution and energy conservation bound}
\label{subsec:tadpole}
%%%%%%%%%%%%%%%%%

\begin{figure}[t] 
        \centering \includegraphics[width=0.25\columnwidth]{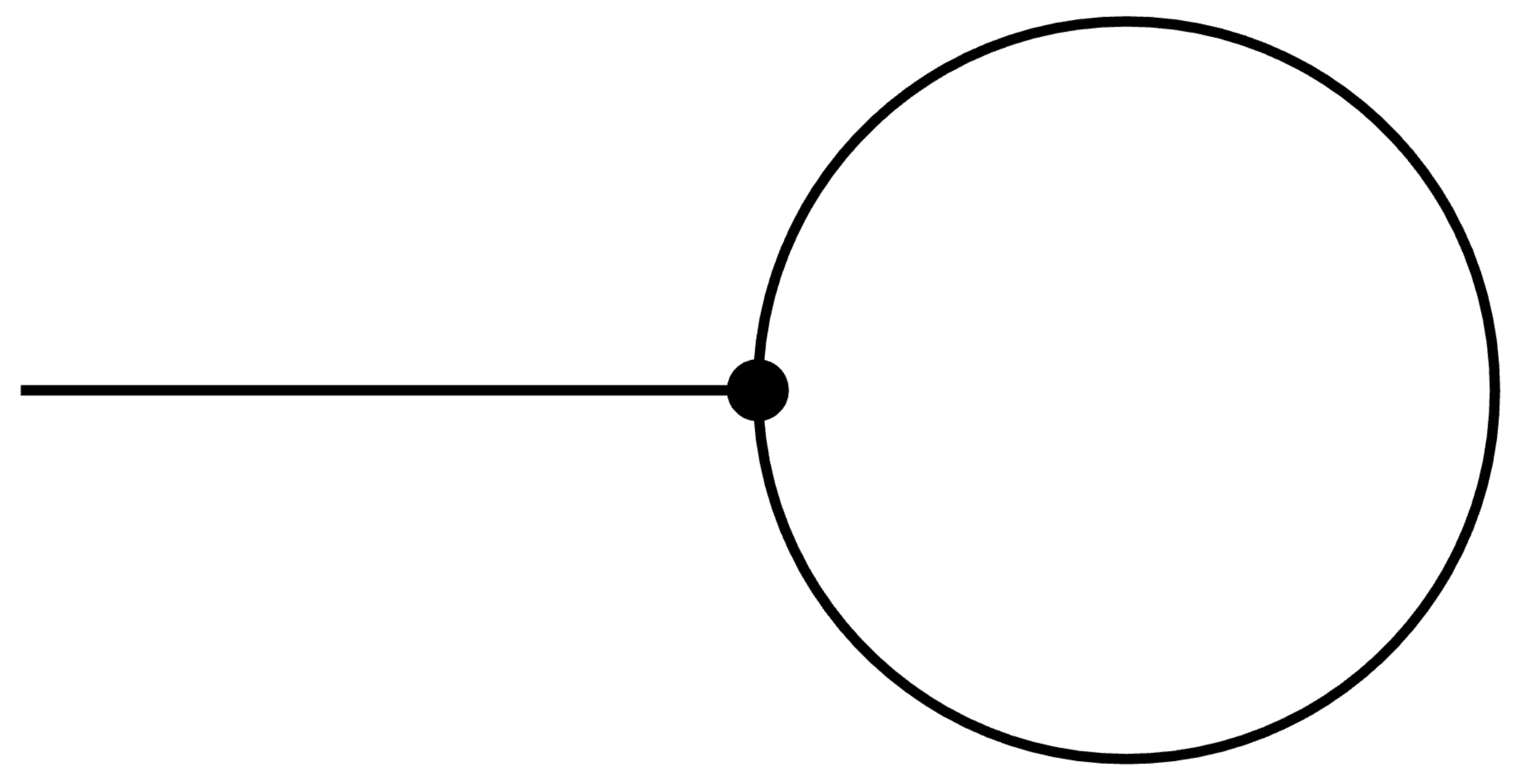}
        \caption{ 
        The Feynman diagram corresponding to the tadpole contribution.
        }
        \label{fig:tadpole}
\end{figure}

As mentioned in Sec.~\ref{sec:in_in}, we have so far assumed $\expval{\delta \phi(\bf x,\eta)} = 0$.
However, to keep $\expval{\delta \phi(\bf x,\eta)} = 0$, we need to continue to modify the background value of $\phi$ according to its perturbations, though we do not take into account the modification of the background throughout this work.
This modification can be interpreted as the loop correction to the background~\cite{Sloth:2006nu}. 
At one loop level, this modification can be associated with the tadpole diagram, shown in Fig.~\ref{fig:tadpole}.
In particular, in Ref.~\cite{Sloth:2006nu}, the authors discussed how the tadpole contribution affects the background evolution of the inflaton in the case where no perturbation amplification occurs.
Also, the tadpole contribution in the existence of spectator fields was discussed in Ref.~\cite{Senatore:2009cf}.
In this subsection, we estimate the order of the tadpole contribution and derive the condition for the negligible modification of the background. 
Then, we also see the relation between the tadpole condition and the energy conservation bound on the perturbations~\cite{Adshead:2014sga,Mirbabayi:2014jqa,Inomata:2021zel}.

From Eq.~(\ref{eq:in_in_form}), the tadpole contribution is given by 
\begin{align}
	\expval{\delta \phi(\bf x,\eta)}_{1 \text{loop}} = &\vev{ \int \frac{\dd^3 q}{(2\pi)^3} \ee^{i \bf q \cdot \bf x} \delta \phi_{\mathbf q}(\eta) \left(T \left[-i \int^\eta_{-\infty} \dd \eta' H_{\text{int},3}(\eta') \right]\right)}  \nonumber \\
	& \quad + \vev{  \left(T \left[-i \int^\eta_{-\infty} \dd \eta' H_{\text{int},3}(\eta') \right]\right)^\dagger  \int \frac{\dd^3 q}{(2\pi)^3} \ee^{i \bf q \cdot \bf x} \delta \phi_{\mathbf q}(\eta) }.
	\label{eq:one_point}
\end{align}
The first term can be expressed as 
\begin{align}
	&\vev{ \int \frac{\dd^3 q}{(2\pi)^3} \ee^{i \bf q \cdot \bf x} \delta \phi_{\mathbf q}(\eta) \left(T \left[-i \int^\eta_{-\infty} \dd \eta' H_{\text{int},3}(\eta') \right]\right)} \nonumber \\
	&= \vev{ \int \frac{\dd^3 q}{(2\pi)^3} \ee^{i \bf q \cdot \bf x} \delta \phi_{\mathbf q}(\eta) \left(T \left[i \int^\eta_{-\infty} \dd \eta' \frac{\lambda(\eta')}{3}  \int \frac{\dd^3 p_1}{(2\pi)^3} \int \frac{\dd^3 p_2}{(2\pi)^3}  \delta \phi_{\mathbf p_1}(\eta') \delta \phi_{\mathbf p_2}(\eta') \delta \phi_{-\mathbf p_1-\mathbf p_2}(\eta')\right]\right)} \nonumber \\
	&= \int \dd^3 q\, \delta(\mathbf q) \ee^{i \mathbf q \cdot \mathbf x} U_{q}(\eta) \left( i \int^\eta_{-\infty} \dd \eta' \lambda(\eta') \int \frac{\dd^3 p_1}{(2\pi)^3} |U_{p_1}(\eta')|^2  U_q^*(\eta')\right).
\end{align}
Given that the second term in Eq.~(\ref{eq:one_point}) is the complex conjugate of the first term, we can rewrite Eq.~(\ref{eq:one_point}) as
\begin{align}
	\expval{\delta \phi({\bf x},\eta)}_{1 \text{loop}} &= -\int \dd^3 q\, \delta(\mathbf q) \ee^{i \mathbf q \cdot \mathbf x} \left( \int^\eta_{-\infty} \dd \eta' \lambda(\eta') \int \frac{\dd^3 p}{(2\pi)^3} |U_{p}(\eta')|^2  2 \, \text{Im} [ U_q(\eta) U_q^*(\eta')] \right).
	\label{eq:one_point_2}
\end{align}
Since the delta function $\delta(\mathbf q)$ makes us focus on the superhorizon $U_q$, we here approximate it as 
\begin{align}
	U_q(\eta) \simeq \frac{H}{\sqrt{2q^3}}(1+i q \eta) \ee^{-i q \eta} \simeq \frac{H}{\sqrt{2 q^3}} \left( 1 + \frac{(q\eta)^2}{2} - \frac{i (q \eta)^3}{3} + \mathcal O((q\eta)^4)\right),
	\label{eq:u_q_approx}
\end{align}
where note that we do not use this approximation for $U_p$ in Eq.~(\ref{eq:one_point_2}).
Strictly speaking, $U_{q}$ oscillates during the resonance proportionally to the oscillation of $\sqrt{\epsilon}$.
However, $\epsilon$ does not significantly deviate from $\epsilon_0$ in $c \ll 1$ and $\Lambda \ll 1$ and, even in our fiducial parameter sets with $c \sim \mathcal O(0.1)$ and $\Lambda \sim \mathcal O(0.1)$, the deviation of $\epsilon$ from $\epsilon_0$ is $\mathcal O(1)$ at a local maximum time (see the left panel of Fig.~\ref{fig:eps_evol_mu_lambda}). 
In the following, we focus on the local maximum time and use Eq.~(\ref{eq:u_q_approx}) as a simple approximation of $U_q$.
Substituting Eq.~(\ref{eq:u_q_approx}) into Eq.~(\ref{eq:one_point_2}), we obtain
\begin{align}
	\expval{\delta \phi(\bf x,\eta_\lmax)}_{1 \text{loop}} &\simeq -\int \dd^3 q\, \delta(\mathbf q) \ee^{i \mathbf q \cdot \mathbf x} \left( \int^{\eta_\lmax}_{\eta_0} \dd \eta' \lambda(\eta') \int \frac{\dd^3 p}{(2\pi)^3} |U_{p}(\eta')|^2  \frac{H^2 ({\eta'}^3 - \eta_\lmax^3)}{3} \right),
	\label{eq:one_point_3}
\end{align}
where we have introduced the lower bound of the time integral due to the exponential suppression of $V^{(3)}$ before $\eta_0$.
Assuming that the resonant amplification is strong enough that the contribution within one oscillation from the local maximum at $\eta_\lmax$ is dominant, which is the case in our fiducial parameter sets in Fig.~\ref{fig:power_evol_pbh_pl},
we can approximate Eq.~(\ref{eq:one_point_3}) as 
\begin{align}
	|\expval{\delta \phi(\bf x,\eta_\lmax)}_{1 \text{loop}}| &\sim \left| H^2 \eta^2_\lmax \Delta \eta_{\text{osc}}^2(\eta_\lmax) \lambda(\eta_\lmax) \int^{p_\UV}_{p_\IR} \frac{\dd p}{p} \mathcal P_{\delta \phi,\tre}(p, \eta_\lmax)\right| \nonumber \\
	&\sim \left|\frac{a_0^2 V^{(3)}}{2k_\pe^2}\right| \mathcal P_{\delta \phi,\tre}(k_\pe, \eta_\lmax) \nonumber \\
	&\sim \frac{3c}{2\sqrt{2 \epsilon_0}\Lambda M_\Pl } \mathcal P_{\delta \phi,\tre}(k_\pe, \eta_\lmax),
	\label{eq:one_point_4}
\end{align}
where we have used $|k_{\text{peak}}\eta_0| \sim \Lambda^{-1}$ and approximated the trigonometric factor as $\mathcal O(1)$ again.
Note that this is of the same order of magnitude as the rough estimate with the equation of motion, Eq.~(\ref{eq:delta_1_eom}).
To satisfy $\expval{\delta \phi(\mathbf x, \eta)} = 0$ exactly, the background must absorb the one-loop contribution given by Eq.~(\ref{eq:one_point_4}). 
This modification of the background becomes important when it is comparable to or larger than the field width for one oscillation of the oscillatory feature.
Conversely, we can roughly expect that the background modification is negligible if 
\begin{align} 
	&\left|\expval{\delta \phi(\bf x,\eta_\lmax)}_{1 \text{loop}}\right| < \mathcal O(\sqrt{2 \epsilon_0} \Lambda M_\Pl) \nonumber \\
	\Rightarrow \quad 
	&\frac{3c}{4 \epsilon_0 \Lambda^2 M_\Pl^2 } \mathcal P_{\delta \phi,\tre}(k_\pe, \eta_\lmax) < \mathcal O(1).
	\label{eq:tadpole_cond}
\end{align}
The left hand side for the final local maximum during the resonance becomes $2.0\times 10^2$ for case A in $N-N_0 = 1.18$, $0.45$ for case B in $N-N_0 = 2.75$, and $4.2\times 10^{-4}$ for case C in $N-N_0 = 2.35$.
This indicates that our calculation in case A (and possibly case B) could be modified if we take into account the loop correction to the background.

Actually, this condition from the tadpole contribution can be related to the upper bound on the perturbations from the energy conservation~\cite{Adshead:2014sga,Mirbabayi:2014jqa,Inomata:2021zel}.
In the rest of this subsection, let us see the relation.
In single-field inflation models, the energy density is given by~\cite{Inomata:2021zel}
	\begin{align}
	\rho = \frac{1}{2} \partial^\mu \phi \partial_\mu \phi + V(\phi) - \partial^0 \phi \partial_0 \phi.
	\end{align}
Using this, we can obtain the following expression for the energy density of the amplified field fluctuations:
	\begin{align}
	\rho_f \simeq \frac{1}{2a^2} \left[ (\delta \phi')^2 + (\partial_i \delta \phi)^2 \right] + V^{(2)}(\phi) \delta \phi^2,
	\end{align}
where we have neglected higher order perturbations.
Here, let us focus on the energy density at $\eta_\lmax$, which leads to $\delta \phi' = 0$ on the peak scale. 
Then, we can approximate the amplified field fluctuation energy as
	\begin{align}
	\rho_f(\eta_\lmax) \sim \frac{k_\pe^2}{2a^2(\eta_\lmax)} \mathcal P_{\delta \phi}(k_\pe,\eta_\lmax) + V^{(2)}(\phi) \mathcal P_{\delta \phi}(k_\pe,\eta_\lmax).
	\end{align}
Note that $V^{(2)}$ is positive at $\eta_\lmax$ (see Fig.~\ref{fig:eps_evol_mu_lambda}). 
The order of the second derivative of the oscillatory potential can be approximated as 
\begin{align}
	V^{(2)}(\phi) &\simeq -\frac{3cH^2 }{\Lambda^2} \cos\left( \frac{\phi-\phi_0}{\sqrt{2\epsilon_0}\Lambda M_\Pl}\right) \nonumber \\
	&\simeq \mathcal O\left(\frac{3 c k_\pe^2 }{a_0^2} \right),
\end{align}
where we have used $\Lambda \sim |k_\pe \eta_0|^{-1}$ and approximated the trigonometric factor as $\mathcal O(1)$.
Then, we obtain 
\begin{align}
	\rho_f(\eta_\lmax) \sim \max \left [\frac{k_\pe^2}{2a^2(\eta_\lmax)},  \frac{3c k_\pe^2}{a_0^2} \right] \mathcal P_{\delta \phi}(k_\pe,\eta_\lmax).
\end{align}
In the case of a very small $c$, the spatial derivative contribution (the first coefficient) is dominant. 
On the other hand, the potential term (the second coefficient) is dominant in the case where $c$ is not very small, such as our fiducial parameter sets in Fig.~\ref{fig:power_evol_pbh_pl}, because the spatial derivative contribution is suppressed by the growing scale factor.
The fluctuation amplification on subhorizon scales can be regarded as the excitation of the inflaton field and the excitation energy should come from the background energy, which is the sum of the potential and kinetic energy of the inflaton.
Since the difference between the potential energies before and after the resonance is $<\mathcal O(\epsilon_0 H^2 M_\Pl^2)$ in the case where the resonance occurs within $< \mathcal O(1)$ e-folds, the energy conservation law requires
	\begin{align} 
	\rho_f < \mathcal O(\epsilon_0 H^2 M_\Pl^2),
	\label{eq:rho_f_bound}
		\end{align}
where we have used the fact that the inflaton kinetic energy is given by $\rho_{\text{kin}} = \dot \phi^2/2 = \epsilon H^2 M_\Pl^2$.
Note that this bound only applies to the amplified fluctuation energy. If the fluctuation amplification does not occur, this bound just corresponds to the eternal inflation bound~\cite{Guth:2007ng}.
We also remark that, in Eq.~(\ref{eq:rho_f_bound}), we do not take into account the expansion of the universe for simplicity, which makes $\epsilon < \epsilon_0$ due to the Hubble friction during some time periods.
In this sense, Eq.~(\ref{eq:rho_f_bound}) should be considered as a rough bound in the case where the resonance continues for more than $\mathcal O(1)$ e-folds.
Since the main focus of this paper is to discuss the one-loop power spectrum, we do not discuss the effect of the Hubble friction further in this paper and just use Eq.~(\ref{eq:rho_f_bound}) for simplicity in the following.

The violation of the energy conservation bound means that we cannot neglect the backreaction from the perturbations to the background evolution.
In other words, the linear perturbation theory, which neglects the backreaction, does not give reliable results once this bound is violated.
Using Eq.~(\ref{eq:rho_f_bound}), we can obtain
\begin{align}
	&\max \left [\frac{k_\pe^2}{2a^2(\eta_\lmax)},  \frac{3c k_\pe^2}{a_0^2} \right] \mathcal P_{\delta \phi}(k_\pe,\eta_\lmax) < \mathcal O(\epsilon_0 H^2 M_\Pl^2) \nonumber \\
	\Rightarrow \quad &
	\max \left [\frac{a_0^2}{2a^2(\eta_\lmax) },  3c \right] \frac{1}{\epsilon_0 \Lambda^2 M_\Pl^2} \mathcal P_{\delta \phi}(k_\pe,\eta_\lmax) < \mathcal O(1), % \nonumber \\
\end{align}
where we have used $|k_\pe \eta_0| \sim \Lambda^{-1}$. 
This energy conservation bound is related to the tadpole contribution, which describes the one-loop backreaction from the amplified fluctuations.
We find that, in $3c > a_0^2/(2 a^2(\eta_\lmax))$, this condition becomes the same as the tadpole condition Eq.~(\ref{eq:tadpole_cond}) except for $\mathcal O(1)$ coefficients, which can be easily changed.
On the other hand, if $c$ is so small that $3c < a_0^2/(2 a^2(\eta_\lmax))$, the energy conservation bound is different from the tadpole bound. 
However, we here note that we derived Eq.~(\ref{eq:one_point_4}) by focusing on the contribution within one oscillation from the local maximum at $\eta_\lmax$.
In the case of a very small $c$, the tadpole contribution could be larger than Eq.~(\ref{eq:one_point_4}) because the perturbation amplification can occur through many oscillations with small amplifications of each oscillation and the one-loop backreaction from each oscillation could be accumulated through the many oscillations.
For this reason, the bound Eq.~(\ref{eq:tadpole_cond}) could be too conservative in the case of a very small $c$. 
This means that, even if we consider a very small $c$, there still remains the possibility that the energy conservation bound becomes consistent with the tadpole condition. We leave the analysis on that case for future work.
In our fiducial parameter sets, this energy conservation bound is satisfied in case C, while marginally in case B and not in case A, though the bound is a rough one since we neglect the Hubble friction.
This indicates that, in case A (and possibly case B), the linear perturbation theory is unreliable not only due to the dominant loop corrections but also due to the violation of the energy conservation (or the backreaction from the amplified perturbations).
The latter effect should be related to the effects of loop diagrams on the background, such as the tadpole diagram, in the in-in formalism.

%%%%%%%%%%%%%%%%%%%%%%%%%%%%%%%%
\section{Necessary conditions for subdominant loop power spectrum}
\label{sec:necessary_conditions}
%%%%%%%%%%%%%%%%%%%%%%%%%%%%%%%%

In this section, we discuss the necessary conditions for the loop power spectrum to be subdominant in the oscillatory feature model. 
One condition for the subdominant loop is that the loop power spectrum should be smaller than the tree power spectrum at the late time (e.g. $N-N_0 = 6$ in Figs.~\ref{fig:loop_pbh_pl}-{\ref{fig:loop_pbh_pl_c}}). 
However, since the one-loop power spectrum can both increase and decrease in time, as seen in Fig.~\ref{fig:loop_lmax_lmin}, the comparison between loop and tree power spectra at the late time could give too weak conditions.
When the one-loop spectrum dominates over the tree spectrum at some point before the late time, the higher order loop power spectrum could be more important and possibly break the perturbation theory.
Once the perturbation theory breaks down, we must follow the evolution of the quantities in a non-perturbative way, such as lattice simulations~\cite{Caravano:2021pgc}, even if the perturbation theory predicts the tree power spectrum finally dominates over the one-loop power spectrum at the late time. 
Given this, in the following, we discuss the conditions for the one-loop power spectra to become smaller than the tree power spectrum at the time when the tree power spectrum on $k_\pe$ reaches its global maximum.
In other words, we derive the necessary conditions for the subdominant loop power spectrum throughout the evolution.
Note that our conditions can still be too weak because the tree power spectrum decreases during some time period and can in principle become smaller than the loop power spectrum at some point even if the above condition is satisfied. 
In addition, as we can see in the case of $N-N_0 = 2.35$ in Fig.~\ref{fig:loop_pbh_pl_c}, our analytical estimates that we use in the following can underestimate true results when the peak of the power spectrum is broad.
For these reasons, we stress that the conditions that we discuss in this section are \emph{not sufficient} conditions.

We denote by $\eta_{e}$ the global maximum time for the tree power spectrum on $k_\pe$, which corresponds to the end of the resonance of the peak-scale perturbations.
Then, the power spectrum of the superhorizon curvature perturbations can be approximated as $\mathcal P_{\zeta,\tre}(k_\pe) \sim \frac{\mathcal P_{\delta \phi, \tre}(k_\pe,\eta_e)}{2\epsilon_0 (k_\pe \eta_e)^2 M_\Pl^2}$, where the additional $(k_\pe \eta_e)^{-2}$ is due to the redshift of the amplified perturbations from the end of the resonance until their horizon exits.
Then, we can rewrite the analytical estimates, Eqs.~(\ref{eq:b_approx_2vx_s}) and (\ref{eq:b_approx_1vx}) at $\eta_e$ as 
\begin{align}
	\label{eq:p_2vx_s_ana_est}
	 \mathcal P_{\delta \phi, 2\vx}(k_{\text{peak}},\eta_e)
	 &\sim \frac{9 c^2 R}{8 \epsilon_0 (k_\pe \eta_0)^4 \Lambda^6 M_\Pl^2 } \mathcal P_{\delta\phi, \tre}^2(k_\pe,\eta_e) 
	 \sim \frac{9 c^2 R}{4 (k_\pe \eta_0)^4 \Lambda^6}   (k_\pe \eta_e)^2  \mathcal P_{\delta\phi, \tre}(k_\pe,\eta_e) \mathcal P_{\zeta,\tre}(k_\pe), \\
	\label{eq:p_1vx_ana_est}
	 |\mathcal P_{\delta \phi, 1\vx}(k_{\text{peak}},\eta_e)|
	 &\sim  \frac{3 c R}{2\epsilon_0 (k_\pe \eta_0)^2 \Lambda^4 M_\Pl^2 } \mathcal P_{\delta\phi, \tre}^2(k_\pe,\eta_e)
	 \sim \frac{3 c R}{(k_\pe \eta_0)^2 \Lambda^4 } (k_\pe \eta_e)^2 \mathcal P_{\delta\phi, \tre}(k_\pe,\eta_e) \mathcal P_{\zeta,\tre}(k_\pe), 
\end{align}
where the  $R (\simeq \mathcal O(1-0.01))$ denotes the uncertainty of the analytical estimates. 
We have also omitted the superscript of $\mathcal P_{\delta \phi, 2\vx}$ because the  amplitudes of $\mathcal P^{a,b}_{\delta \phi, 2\vx}$ at the peak scale are almost the same.
Here, we note that the above analytical estimates are for the case where $U_{k_\pe}$ grows by $\gg \mathcal O(1)$ in one oscillation during the resonance, as already mentioned in the context of the tadpole contribution in Sec.~\ref{subsec:tadpole}.
In other words, we have derived the above estimates by focusing on the contributions from one or two of the oscillations. 
However, even if the growth of $U_{k_\pe}$ in one oscillation is small, the large amplification of $U_{k_\pe}$ can be realized through many oscillations, which can be the case when $c$ and $\Lambda$ are very small.
In that case, the right hand side of Eqs.~(\ref{eq:p_2vx_s_ana_est}) and (\ref{eq:p_1vx_ana_est}) could possibly become larger because the contribution of each oscillation could be accumulated through the many oscillations.
In this paper, we do not discuss this possibility in detail and instead just use the above analytical estimates for simplicity, which means that the following necessary conditions could possibly be too weak.
However, even if we ignore that possibility, we will still see that a smaller $c$ leads to stronger conditions on the tree power spectrum to realize the subdominant loop power spectrum.

From Eqs.~(\ref{eq:p_2vx_s_ana_est}) and (\ref{eq:p_1vx_ana_est}), we can express the necessary conditions for the subdominant loop power spectrum making use of the power spectrum of the field fluctuations or the curvature perturbations: 
\begin{align}
	 \frac{\mathcal P_{\delta \phi, 2\vx}(k_{\text{peak}},\eta_e)}{\mathcal P_{\delta\phi, \tre}(k_\pe,\eta_e)}
	 &\sim  \frac{9 c^2 R}{8\epsilon_0 \Lambda^2 M_\Pl^2  } \mathcal P_{\delta \phi,\tre}(k_\pe,\eta_e) \sim \frac{9 c^2  R}{4\Lambda^4} \left(\frac{\eta_e}{\eta_0}\right)^2 \mathcal P_{\zeta,\tre}(k_\pe)< 1, \label{eq:cond_2vx_s} \\
	 \left|\frac{\mathcal P_{\delta \phi, 1\vx}(k_{\text{peak}},\eta_e)}{\mathcal P_{\delta\phi, \tre}(k_\pe,\eta_e)}\right|
	 &\sim \frac{3 c  R}{2\epsilon_0 \Lambda^2M_\Pl^2 }  \mathcal P_{\delta \phi,\tre}(k_\pe,\eta_e) \sim \frac{3 c  R}{\Lambda^4 } \left(\frac{\eta_e}{\eta_0}\right)^2 \mathcal P_{\zeta,\tre}(k_\pe)< 1, \label{eq:cond_1vx} 
\end{align}
where we have used the relation $|k_{\text{peak}}\eta_0| \sim \mathcal O(\Lambda^{-1})$. 

Here, to see the model parameter dependence more clearly, let us rewrite these necessary conditions with the amplification magnitude. 
To do so, we express the tree power spectrum as 
\begin{equation}
\label{eq:P_end_res}    \mathcal P (k,\,\eta_e)=
    \mathcal X^2 (k)\mathcal P (k,\,\eta_0),
\end{equation}
where $\mathcal X$ is the amplification factor of the perturbations which is normalized by the value at the beginning of the oscillatory feature.
Eq.~(\ref{eq:P_end_res}) holds regardless of whether we consider the power spectrum of $\delta\phi$ or $\zeta$ except for $\mathcal O(1)$ factor, which we neglect for simplicity. They are related by $\mathcal P_{\zeta,\tre} \sim \mathcal P_{\delta\phi,\tre}/(2\epsilon_0 M_\Pl^2)$ at $\eta_0$ and a local maximum, where $\epsilon$ does not significantly deviate from $\epsilon_0$.
Since the power spectrum on subhorizon scales redshifts as $a^{-2}$, the power spectrum at the horizon crossing is related to the value at the reference time 
$\lvert\eta_0\rvert>\lvert\eta_{\hc}\rvert$ by 
\begin{equation}
    \overline{\mathcal{P}}(k_{\rm peak},\,\eta_{\hc}) \simeq \left(\frac{\eta_\hc}{\eta_0}\right)^2 \overline{\mathcal{P}}(k_{\rm peak},\,\eta_{0})=e^{2(N_0-N_{\hc})} \overline{\mathcal{P}}(k_{\rm peak},\,\eta_{0}),
\end{equation}
where the subscript ``h.c.'' represents the value at the horizon crossing and the overline indicates the power spectrum without the oscillatory feature.
Using this, we can rewrite the superhorizon power spectrum $\mathcal P_{\zeta,\tre}(k_\pe)$ as
\begin{align}
    \mathcal P_{\zeta,\tre}(k_\pe)&\simeq \mathcal P_{\zeta,\tre}(k_\pe,\,\eta_{\hc}) \simeq \ee^{-2(N_{\hc}-N_e)}\mathcal P_{\zeta,\tre}(k_\pe,\,\eta_e)=e^{-2(N_{\hc}-N_e)}\mathcal X^2(k_\pe)\mathcal P_{\zeta,\tre}(k_\pe,\,\eta_0)\notag \\
    & \simeq \ee^{-2(N_{\hc}-N_e)}e^{2(N_{\hc}-N_0)}\mathcal X^2(k_\pe)\overline{\mathcal P}_{\zeta,\tre}(k_\pe,\,\eta_{\hc}) \simeq \ee^{2\Delta N_{\text{res}}}\mathcal X^2(k_\pe)\, A_s\left(\frac{k_\pe}{k_{\rm CMB}}\right)^{n_s-1}, \label{eq:p_zeta_expression}
\end{align}
where $N_e$ is the e-folds at $\eta_e$ and we have used $\overline{\mathcal P}_{\zeta,\tre}(k_\pe,\,\eta_{\hc}) \simeq A_s (k_\pe/k_\text{CMB})^{n_s-1}$ with $A_s = 2.1\times 10^{-9}$~\cite{Planck:2018vyg}.
In the first equality of this equation, we have used that, also in our model, the spectrum on superhorizon scales is essentially equal to the one at the horizon crossing, since the amplification mechanism works on subhorizon scales. In the fourth equality, we have used the fact that the spectra at $\eta_0$ in the featureless and feature model are the same.
$\Delta N_{\text{res}}$ is the duration of the resonance, which is $\Delta N_{\text{res}} \simeq N_{e}-N_0$ in our fiducial setup.
From this expression, we can see that $\ee^{2 \Delta N_{\text{res}}} \mathcal X^2(k_\pe)$ corresponds to the amplification magnitude normalized by the power spectrum without feature and $\ee^{2 \Delta N_{\text{res}}} \mathcal X^2(k_\pe) \sim \mathcal O(10^7)$ is often considered in the context of PBHs.
For convenience, we here define $\mathcal Y(k) \equiv \ee^{\Delta N_{\text{res}}} \mathcal X(k)$.
Using this result and writing the scale factor as $a(N)=(-1/(H\eta_0))\exp(N-N_0)$, we can rewrite the conditions, Eqs.~(\ref{eq:cond_2vx_s}) and (\ref{eq:cond_1vx}), as 
\begin{align}
    \label{eq:2vx_sym_n_cond}
    \frac{9c^2 R}{4\Lambda^4} e^{-2\Delta N_{\text{res}}} \mathcal{Y}^2(k_\pe) \,A_s\left(\frac{k_\pe}{k_{\rm CMB}}\right)^{n_s-1}< 1, \\
    \label{eq:1vx_sym_n_cond}    
    \frac{3c R }{\Lambda^4} e^{-2\Delta N_{\text{res}}} \mathcal{Y}^2(k_\pe)\,A_s\left(\frac{k_\pe}{k_{\rm CMB}}\right)^{n_s-1} < 1, 
\end{align}
where we have used $|k_\pe\eta_0|\sim \mathcal O(\Lambda^{-1})$ again.

\begin{figure}[t]
  \begin{minipage}[b]{0.4\linewidth}
    \centering
    \includegraphics[keepaspectratio, scale=0.55]{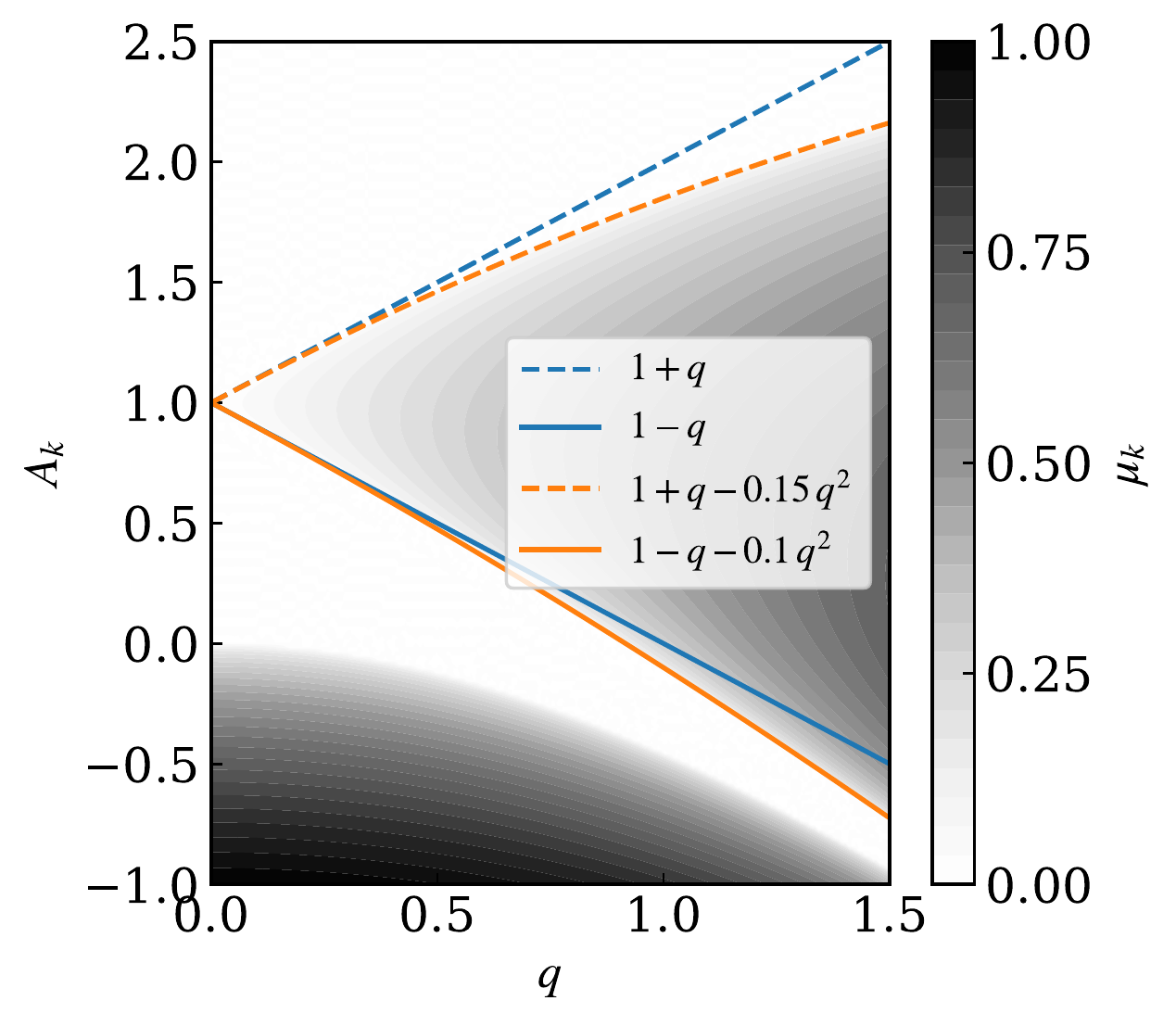}
  \end{minipage}
  \begin{minipage}[b]{0.49\linewidth}
    \centering
    \includegraphics[keepaspectratio, scale=0.55]{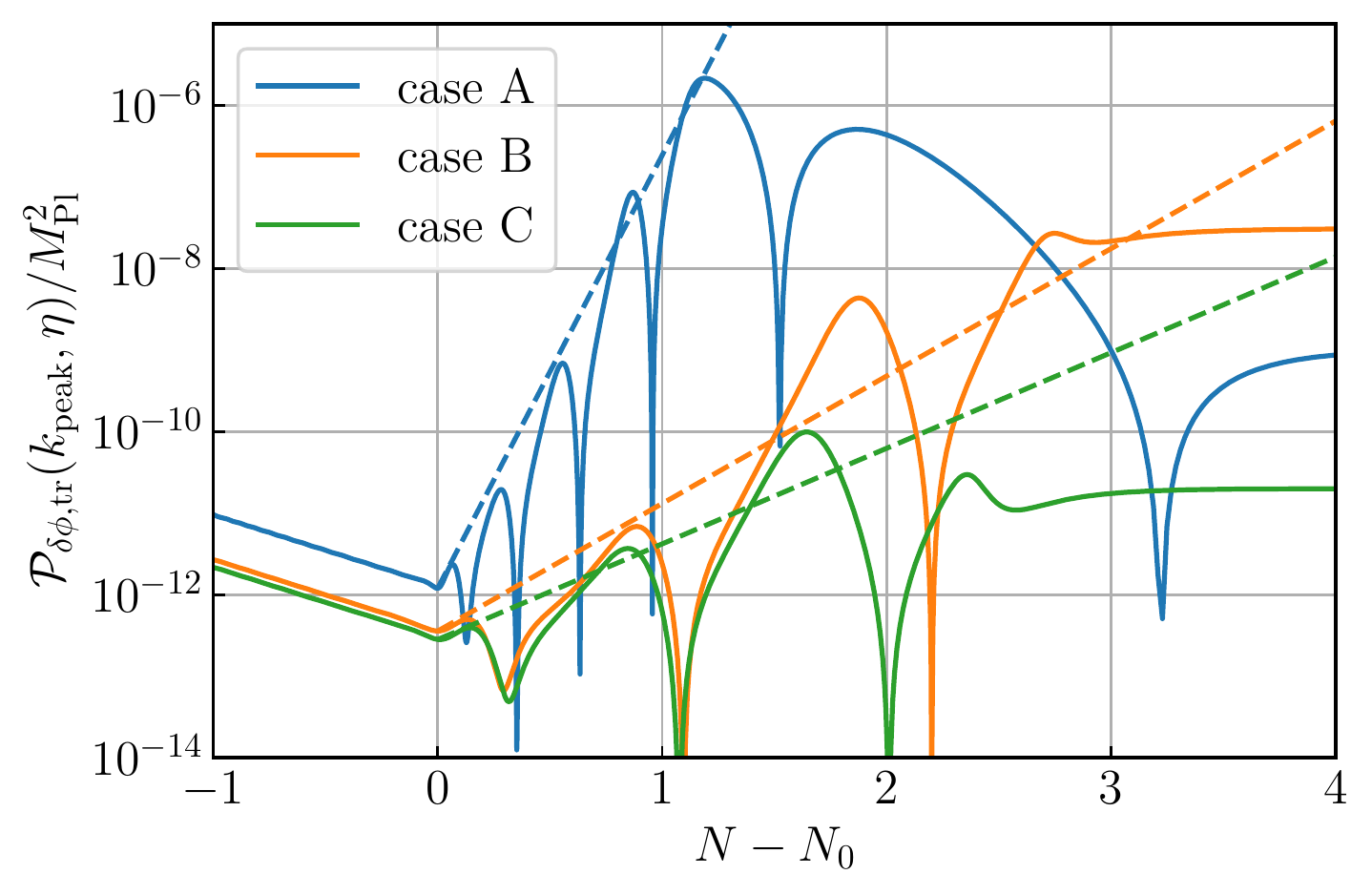}
  \end{minipage}
        \caption{ [Left]: The instability chart zoomed in the region $q\lesssim \mathcal{O}(1)$. We plot the lines $1+q$ and $1-q$ in blue and the lines $1+q-0.15 q^2$ and $1-q-0.1q^2$ in orange. 
        [Right]: Comparison between the numerical results of $\mathcal P_{\delta \phi,\tre}(k_\pe,\eta)$ (solid lines) and the approximate formula with Eq.~(\ref{eq:amplification}), $\mathcal P_{\delta \phi,\tre}(k_\pe,\eta_0) \, \ee^{3\left(\frac{c}{\Lambda} -1 \right) (N-N_0)}$ (dashed lines). 
        }  
        \label{fig:mathieu_and_analytical}
\end{figure}

We here roughly relate the amplification factor $\mathcal Y (k_\pe)$ to the model parameters.
To this end, we start from the linear equation of motion Eq.~(\ref{eq:delta_phi_eom}) with the physical time:
\begin{equation}
\label{eq:EOM}
	\ddot{\delta \phi}_{\mathbf k} + 3 H \dot{\delta \phi}_{\mathbf k} + \frac{k^2}{a^2} \delta \phi_{\mathbf k} + V^{(2)}(\phi) \delta \phi_{\mathbf k} = 0.
\end{equation}
We here approximate the field evolution as $\phi-\phi_0\simeq\sqrt{2\epsilon_0} M_\Pl H(t-t_0)$. Although this is a good approximation in $c \ll 1$ and $\Lambda \ll 1$, it becomes a very crude approximation in our fiducial parameter sets with $c \sim \mathcal O(0.1)$ and $\Lambda \sim \mathcal O(0.1)$, where $\epsilon$ oscillates between $\mathcal O(\epsilon_0)$ and $\mathcal O(0.1\epsilon_0 \-- 0.01\epsilon_0)$. However, as we have seen in Sec.~\ref{sec:num_result}, the approximation of $\epsilon = \epsilon_0$ can still give the typical oscillation timescale even in that case and therefore we expect that this approximation could work similarly. Given this, we use this approximation in the following. 
Then, defining $\delta\phi\equiv a^{-3/2}\Phi$ and neglecting terms proportional to $\dot{H}$, we can rewrite Eq.~\eqref{eq:EOM} as
\begin{equation}
    \label{eq:mathieu}
    \frac{\dd^2{\Phi}_{\mathbf k}(z)}{\dd z^2}+\left[A_k(z)-2 q \cos(2z) \right]\Phi_{\mathbf k}(z)=0,
\end{equation}
where
\begin{align}
    &2 z\equiv
\frac{H}{\Lambda}(t-t_0), \ A_k(z)=4\Lambda^2\left[\left(k \eta_0\right)^2 \ee^{-4 \Lambda z}-\frac{9}{4}   \right], \ q= 6 c. %\notag.
\end{align}
This equation is a Mathieu-like equation with the time-dependent $A_k$.
The $\Phi(z)$ grows proportionally to $\ee^{\mu_k z}$ and the instability chart on $\mu_k$ is shown in the left panel of Fig.~\ref{fig:mathieu_and_analytical}.  
Using that the Floquet index $\mu_k$ is roughly equal to $q/2$ at the peak scale~\cite{Kofman:1997yn}, we can estimate the amplification factor as
\begin{align}
\label{eq:amplification}    
    \mathcal X(k_\pe) &\simeq a^{-\frac{3}{2}} \ee^{\frac{q \Delta z_{\text{res}}}{2}} \simeq  \ee^{\frac{3}{2}\left(\frac{c}{\Lambda}-1\right) \Delta N_{\text{res}}}, \\
\label{eq:amplification_y} 
    \mathcal Y(k_\pe) &= \ee^{\Delta N_{\text{res}}} \mathcal X(k_\pe) \simeq \ee^{\left(\frac{3c}{2\Lambda}- \frac{1}{2}\right) \Delta N_{\text{res}}},
\end{align}
where $\Delta z_{\text{res}}$ is the duration of the resonance in $z$ and we have multiplied by $a^{-3/2}$ because we are interested in the amplification of $\delta\phi$, not $\Phi$. 
The right panel of Fig.~\ref{fig:mathieu_and_analytical} compares the numerical results and the approximate formula with Eq.~(\ref{eq:amplification}).
We can see that, while Eq.~(\ref{eq:amplification}) underestimates the resonant growth of the numerical results in case B and C, it works well for case A.
The reason for the discrepancy in case B and C is that a longer oscillatory period (a larger $\Lambda$) leads to a larger difference between the period lengths of $V^{(2)}> 0$ and $V^{(2)}<0$ in the oscillatory region, which makes our approximation worse.

One might think that a smaller $c$ could realize a subdominant loop power spectrum with $\mathcal Y(k_\pe)$ fixed because the loop power spectra are proportional to $c$ or $c^2$. However, it does not work.
The point is that, due to the time dependence of $A_k$, the modes can spend only a short time in the resonance band in the case of a very small $c$. 
From the left panel of Fig.~\ref{fig:mathieu_and_analytical}, we can see that each mode experiences the resonance when the following inequalities are satisfied:\footnote{Note that a more refined condition is actually  $1-q - 0.1q^2\lesssim A_k(z)\lesssim 1+q - 0.15 q^2$. However, since we just want to make order estimates, we use the simpler condition in Eq.~\eqref{eq:condition_resonance}.}
\begin{equation}
\label{eq:condition_resonance}
1-q\lesssim A_k(z)\lesssim 1+q.    
\end{equation}
Then, we can approximate the duration of the resonance as
\begin{equation}
    \Delta N_{\text{res}} \simeq {\rm min}\,[N_{\text{out}},\,N_s] - {\rm max}\,[N_{\text{in}},\,N_0],
    \label{eq:n_s_approx}
    \end{equation}
 where $N_s$ is the time when the feature is turned off, i.e. $\phi(N_s)=\phi_s$, and $N_{\text{in}/\text{out}}$ are the times when the peak scale perturbation enters/exits the resonance band (Eq.~(\ref{eq:condition_resonance})) in the case without the cutoff of the oscillatory features. 
We can obtain $N_{\text{in}/\text{out}}$ by imposing $A_{k_\pe}=1+q$ and $A_{k_\pe}=1-q$, respectively.
Then, we obtain
\begin{align}
N_{\text{in}}&\simeq N_0+\log (2l) -\frac{1}{2}  \log \left(1+6 c+ 9 \Lambda^2\right),\\
N_{\text{out}}&\simeq N_0+\log (2l) -\frac{1}{2}  \log \left(1-6 c+ 9 \Lambda^2\right) \label{eq:n_e_approx},
\end{align}
where we have substituted $k=k_\pe = -l (\Lambda\eta_0)^{-1}$ into Eq.~(\ref{eq:mathieu}) with $l$ being an $\mathcal O(1)$ constant.
Then, we can get the following upper bound on $\Delta N_{\text{res}}$ in $c \ll 1$:
\begin{align}
	\Delta N_{\text{res}} \lesssim 6c \quad (\text{in}\  c \ll 1).
	\label{eq:delta_n_bound}
\end{align}
Note that, in our fiducial cases in Fig.~\ref{fig:power_evol_pbh_pl}, $c$ is still large and Eq.~(\ref{eq:n_e_approx}) is meaningless and therefore this bound on $\Delta N_{\text{res}}$ is not applicable. Also, this bound on $\Delta N_{\text{res}}$ is a rough bound for the growth of $\Phi$, not $\delta \phi$. Because of this, the actual upper bound could be smaller than this upper bound.

From Eqs.~(\ref{eq:amplification_y}) and (\ref{eq:delta_n_bound}), we find 
\begin{align}
    \mathcal Y(k_\pe) \lesssim  \ee^{\left(\frac{9c}{\Lambda}-3\right) c} \sim \ee^{\frac{9c^2}{\Lambda}} \quad (\text{in}\  c \ll 1).
\end{align}
From this expression, we can see that, if the amplification magnitude $\mathcal Y(k_\pe)$ is fixed in $c \ll 1$, the lower bound of $c/\Lambda^{1/2}$ is also fixed.
This means that a smaller $c$ leads to a smaller $\Lambda$ if $\mathcal Y(k_\pe)$ is fixed.
Since the necessary conditions, Eqs.~(\ref{eq:2vx_sym_n_cond}) and (\ref{eq:1vx_sym_n_cond}), are proportional to $(c/\Lambda^{1/2})^{n} \Lambda^{-m}$ with $n>0$ and $m>0$, a smaller $c$ results in larger left hand sides of Eqs.~(\ref{eq:2vx_sym_n_cond}) and (\ref{eq:1vx_sym_n_cond}) with $\mathcal Y(k_\pe)$ fixed.
The physical interpretation is as follows.
A small $c$ corresponds to a small amplitude of the oscillatory features and therefore, having the same amplification requires oscillatory features with larger frequency.
The larger frequency enhances the potential derivatives and therefore the loop power spectrum. 
From these results, we can conclude that, in order to realize a smaller loop power spectrum with $\mathcal Y(k_\pe)$ fixed, we need to consider larger $c$ and $\Lambda$. 
Actually, we can see this tendency by comparing the numerical results in Figs.~\ref{fig:loop_pbh_pl} (case A) and Figs.~\ref{fig:loop_pbh_pl_b} (case B), where larger $c$ and $\Lambda$ lead to a smaller ratio of the loop power spectrum to the tree power spectrum.
Since we have seen that the loop power spectrum dominates over the tree power spectrum with the $\mathcal O(10^7)$ amplification even in the case of $c=0.22$ and $\Lambda = 0.1$ (case B)\footnote{One might think a larger $\phi_s$ (a longer oscillatory region) could realize a smaller $c$ and therefore a smaller loop spectrum with $\Lambda$ and $\mathcal Y(k_\pe)$ fixed in case B. 
However, the resonance for each perturbation stops once the perturbation exits the horizon even if we take a larger $\phi_s$ with $c=0.22$ and $\Lambda = 0.1$.
This means that a larger $\phi_s$ does not help to decrease the loop power spectrum with $\Lambda$ and $\mathcal Y(k_\pe)$ fixed in case B, where the parameters are already tuned so that the horizon exit of the peak scale is close to the end of the oscillatory region ($|k_\pe \eta_s| = 0.75$ in case B). 
}, the remaining parameter region for the subdominant loop is $c \geq \mathcal O(0.1)$ and $\Lambda \geq \mathcal O(0.1)$. 
However,
for $c \geq \mathcal O(0.1)$ and $\Lambda \geq \mathcal O(0.1)$, the oscillation timescale becomes close to or longer than the Hubble timescale and the perturbation amplification does not occur on subhorizon.
We leave the analysis in that case for future work.
Given the result in case B, we can find that the enhancement of the power spectrum with subdominant one-loop corrections requires $\mathcal Y^2(k_\pe) < \mathcal O(10^7)$ for the typical oscillatory feature models with $\Lambda < 0.1$.

%%%%%%%%%%%%%%%%%%%%%%%%%%%%%%%%
\section{Conclusion}
\label{sec:conclusion}
%%%%%%%%%%%%%%%%%%%%%%%%%%%%%%%%

The amplification of the small-scale perturbations during inflation has recently attracted much interest in the context of PBHs and SGWB.
In the literature, the PBH abundance and the amount of GWs are often calculated with the tree-level power spectrum, based on the linear equation of motion.
However, the perturbation evolution is generally affected by higher order perturbations, whose effects appear as loop corrections to the tree power spectrum in the in-in formalism.
In particular, when the perturbations are amplified during inflation, it is not obvious whether the loop corrections are negligible.

Motivated by this, in this work, we have focused on the loop power spectrum in the models that predict large amplification of the perturbations during inflation.
We stress that such a check on the loop power spectrum with the amplified perturbations in the in-in formalism, which is crucial to the reliability of the model prediction, has never be performed before. 
In particular, we have considered an inflaton potential with oscillatory features as a fiducial model. 
We have numerically calculated the loop power spectrum by taking into account the amplified perturbations in the in-in formalism and also obtained their analytical estimates on the peak scale.
In the analytical estimates, we have seen that the Wronskian condition of the perturbations plays an important role in obtaining the correct order of the estimates.
Apart from the power spectrum, we have also analytically discussed the one-loop contribution to the background, called the tadpole contribution. 
We have derived the condition for the tadpole contribution not to change our results and shown that the condition is consistent with the energy conservation bound on the perturbations at least in the case where the power spectrum is amplified by $\gg \mathcal O(1)$ through one oscillation during the resonance.

\begin{figure}[t] 
        \centering \includegraphics[width=0.5\columnwidth]{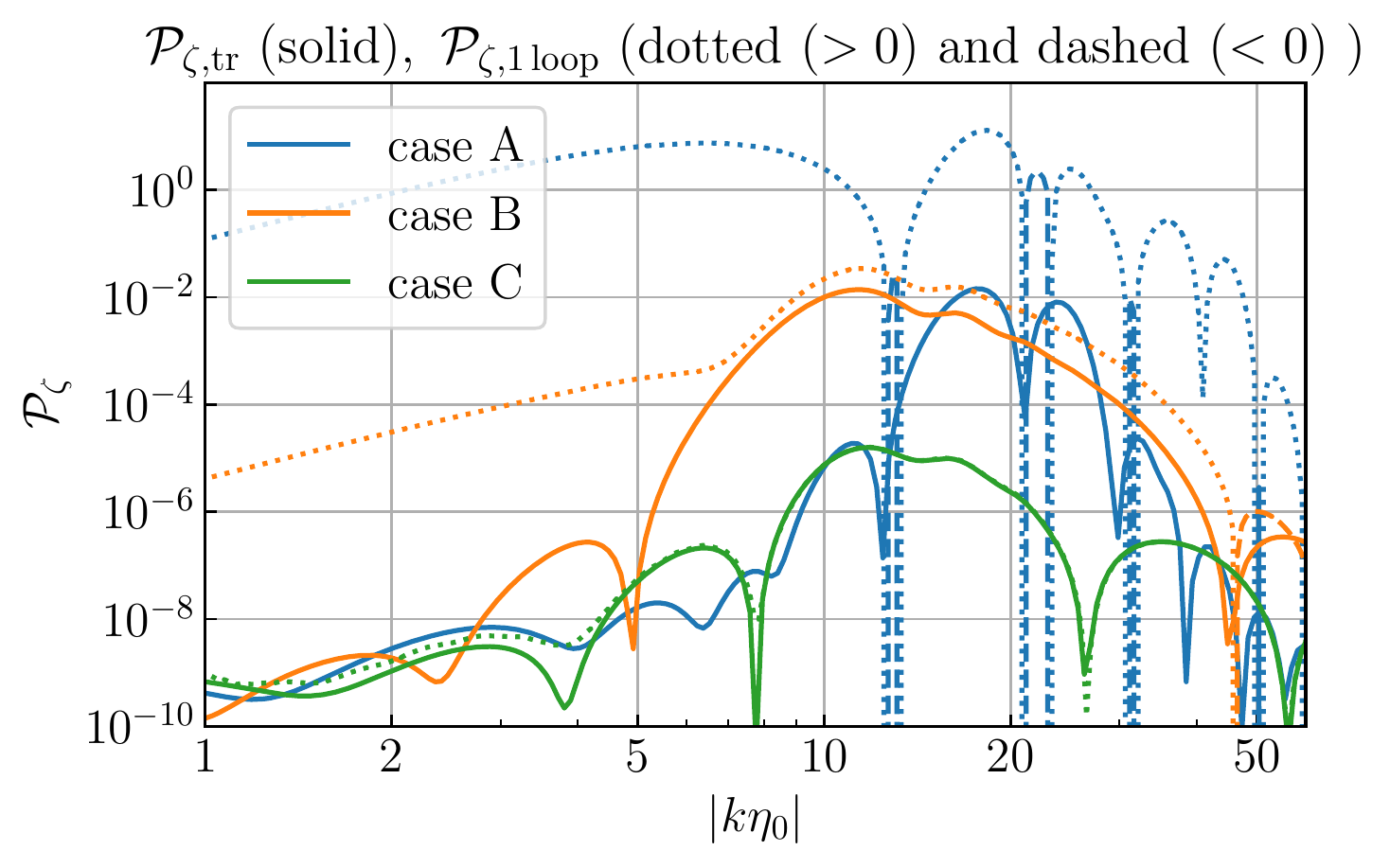}
        \caption{ 
        The tree power spectrum and the total one-loop power spectrum on superhorizon scales well after the resonance. 
				We plot the total one-loop power spectrum of curvature perturbations using $\mathcal P_{\zeta, 1\,\text{loop}}(k) = \frac{1}{2\epsilon(\eta_c) M_\Pl^2} \mathcal P_{\delta \phi, 1\,\text{loop}}(k,\eta_c)$, where $\mathcal P_{\delta \phi, 1\,\text{loop}} \equiv \mathcal P_{\delta \phi,\tre} + \mathcal P^a_{\delta \phi, 2\vx} + \mathcal P^b_{\delta \phi, 2\vx} + \mathcal P_{\delta \phi, 1\vx}$ and $\eta_c$ is well after the resonance and the horizon exit of the perturbations.
				For the one-loop power spectrum, we show their absolute values with dotted/dashed lines denoting positive/negative values of the power spectra.
        The parameters are the same as in Fig.~\ref{fig:power_evol_pbh_pl}.
        }
        \label{fig:final_summary}
\end{figure}

In addition, with the analytical estimates, we have obtained the necessary conditions for the loop power spectrum to be subdominant. 
Then, we have shown that, for a fixed amplification magnitude, a smaller amplitude of the oscillatory features leads to a larger loop power spectrum.
If the necessary conditions are not satisfied in the oscillatory feature model, the calculations of the amount of PBHs and SGWB with the tree power spectrum are unreliable.
Using the necessary conditions and the numerical results, we have also shown that the one-loop power spectrum \emph{typically} becomes larger than the tree power spectrum in the case where the tree power spectrum is amplified by $\mathcal O(10^7)$, often considered in the PBH scenarios.
The ``\emph{typically}'' here means that, in our analysis, we have focused only on the typical case where the oscillatory features cause the resonant amplification on subhorizon scales with the oscillation timescale much shorter than the Hubble timescale.
This situationroughly corresponds to $\Lambda \leq 0.1$ in our potential, Eq.~(\ref{eq:pot_cos_pbh_planck}).
We summarize the results in Fig.~\ref{fig:final_summary}, which compares the tree power spectrum and the power spectrum at one loop level in our fiducial setups.
Case A and B correspond to the $\mathcal O(10^7)$ amplification of the tree power spectrum $\mathcal P_{\zeta,\tre}$ and case C corresponds to the $\mathcal O(10^4)$ amplification. 
Only in case C, the one-loop power spectrum does not dominate over the tree power spectrum throughout the evolution.
In particular, the total one-loop power spectrum can be negative in case A and B. This indicates that the higher-order loop contributions must modify the power spectrum to make it positive.
Note that the large potential derivatives from the oscillatory features enable the loop power spectrum to be dominant even though the tree power spectrum is much smaller than unity.
In the marginal case of the typical oscillatory feature with $\Lambda = 0.1$, the requirement of the subdominant loop power spectrum leads to an upper bound of the enhancement of the power spectrum as $< \mathcal O(10^7)$.
Also, although we have focused on the single-field inflation model throughout this work, we can easily generalize the necessary conditions to the case where fluctuations of a spectator field are amplified by the oscillatory features in its potential, instead of the inflaton fluctuations.
In that case, we need to use the expressions without $\mathcal P_{\zeta,\tre}$ in Eqs.~(\ref{eq:cond_2vx_s}) and (\ref{eq:cond_1vx}) and regard the parameters as those in the spectator field potential.

The main message of this work is that careful consideration of the loop power spectrum is required when we discuss the models that predict large amplification of field fluctuations during inflation. 
Although we have focused only on the oscillatory feature model in this paper, other amplification models could similarly predict the loop power spectrum dominating over the tree power spectrum, whose analyses are left for future work. 
We also note that, although we have shown the problems, our results do not give the solution to the computation of the power spectrum dominated by the loop contributions as we do not discuss the higher order loop contributions.
When the one-loop power spectrum is larger than the tree power spectrum, it is natural to expect that the higher order loop power spectrum becomes larger than the one-loop power spectrum.
In fact, in case A and B of our fiducial setups, the higher order loop contributions must modify the one-loop power spectrum to realize the positive power spectrum.
In addition, it is not obvious whether we can use the perturbation theory when the higher order contributions are dominant. 
To fully address this issue, non-perturbative methods, such as lattice simulations~\cite{Caravano:2021pgc}, might be required.

%%%%%%%%%%%%%%%%%%%%%%%%%%%%%%%%
\acknowledgments
%%%%%%%%%%%%%%%%%%%%%%%%%%%%%%%%
We thank Wayne Hu, Atsuhisa Ota, Lucas Pinol, Akhil Premkumar, Zekai Wang, and Zhong-Zhi Xianyu for useful comments and discussions.
K.I. is supported by the Kavli Institute for Cosmological Physics at the University of Chicago through an endowment from the Kavli Foundation and its founder Fred Kavli. K.I thanks the 
Fonds Alexandre Friedmann - Fondation de l’\'Ecole polytechnique for supporting his visit to IAP.
S.RP is supported by the European Research Council under the European Union's Horizon 2020 research and innovation programme (grant agreement No 758792, Starting Grant project GEODESI).

%%%%%%%%%%%%%%%%%%%%%%%%%%%%%%%%
%%%%%%%%%%%%%%%%%%%%%%%%%%%%%%%%
%%%%%%%%%%%%%%%%%%%%%%%%%%%%%%%%
\appendix

%%%%%%%%%%%%%%%%%%%%%%%%%%%%%%%%
\section{Metric perturbations}
\label{app:met_pertb}
%%%%%%%%%%%%%%%%%%%%%%%%%%%%%%%%

In this appendix, we show that the metric perturbations in the spatially flat gauge are negligible in our fiducial setup.
Note that, in this appendix, we set $M_\Pl = 1$, ignore the overall sign in the order approximations, and assume the potential derivatives are given by Eqs.~(\ref{eq:d1_pot})-(\ref{eq:d4_pot}) with $c < 1$ and $\Lambda < 1$. 

We take the ADM formalism, where the metric in the spatially flat gauge is expressed as
\begin{align}
	\dd s^2 = g_{\mu \nu} \dd x^\mu \dd x^\nu = -N^2 \dd t^2 + h_{ij} (\dd x^i + N^i \dd t) (\dd x^j + N^j \dd t),
\end{align}
with
\begin{align}
	h_{ij} &= a^2 \delta_{ij},\\
	N &= 1 + \alpha_1 + \alpha_2, \\
	N_i &= h_{ij} N^j = \partial_i (\vartheta_1 + \vartheta_2) + \beta_{2i}, 
\end{align}
where we have neglected vector and tensor perturbations.
Similar to the main text, we use a notation where the same subscript spatial indexes are contracted, e.g. $\partial_i \delta \phi \partial_i \delta \phi = \delta^{ij} \partial_i \delta \phi \partial_j \delta \phi$.
The subscript numbers denote the order in perturbations and $\beta_{2j}$ is the transverse component, satisfying $\partial_i \beta_{2i} = 0$.
We do not show $\beta_{1i}$ because it is a vector perturbation, which we neglect here.
Note that $N$ does not denote the e-folds in this appendix.

First, let us see the cubic interaction term.
The cubic action in the spatially flat gauge is given by~\cite{Maldacena:2002vr}\footnote{Note that our $\vartheta$ corresponds to $a^2 \chi$ in Ref.~\cite{Maldacena:2002vr}.}
\begin{align}
	S_3 =& \int \dd t\, \dd^3 x\, a^3 \left( - \frac{\dot \phi}{4 H} \delta \phi \delta \dot \phi^2 -  \frac{\dot \phi}{4 a^2 H} \delta \phi (\partial_i \delta \phi)^2 - \frac{1}{a^{2}}\delta \dot \phi \partial_i \vartheta_1 \partial_i \delta \phi \right. \nonumber \\
	&
	+ \frac{3 \dot \phi^3}{8 H} \delta \phi^3 - \frac{\dot \phi^5}{16H^3} \delta \phi^3 - \frac{\dot \phi V^{(2)}}{4 H} \delta \phi^3 - \frac{V^{(3)}}{6} \delta \phi^3 + \frac{\dot \phi^3}{4H^2} \delta \phi^2 \delta \dot \phi + \frac{\dot \phi^2}{4 a^2 H} \delta \phi^2 \partial^2 \vartheta_1 \nonumber \\
	& \left. 
	+ \frac{\dot \phi}{4 a^4 H} (- \delta \phi \partial_i \partial_j \vartheta_1 \partial_i \partial_j \vartheta_1 + \delta \phi \partial^2 \vartheta_1 \partial^2 \vartheta_1) \right),
	\label{eq:cubic_action}
\end{align}
where we have taken the shorthand notation $\partial^2 = \partial_i^2$ and used $\alpha_1 = \frac{\dot \phi}{2H} \delta \phi$. % (= \sqrt{\frac{\epsilon}{2}} \delta \phi)$.
The $\vartheta_1$ is given by 
\begin{align}
	\frac{\partial^2 \vartheta_1}{a^2} &= \frac{\dot \phi^2}{2 H^2} \frac{\dd}{\dd t} \left(- \frac{H}{\dot \phi} \delta \phi\right) \nonumber \\
	&= \epsilon \frac{\dd}{\dd t} \left( - \frac{\delta \phi}{\sqrt{2\epsilon}} \right) \nonumber \\
	&= - \sqrt{\frac{\epsilon}{2}} \delta \dot \phi + \frac{\sqrt{\epsilon} s H}{2\sqrt{2}} \delta \phi,
\end{align}
where we have used $\dot \phi>0$ and $s \equiv \dot \epsilon/(\epsilon H) \sim \mathcal O(c \Lambda^{-1})$.
In the oscillatory region, we can approximate the third potential derivative term as (see Eq.~(\ref{eq:d3_pot})) 
\begin{align}
	\frac{V^{(3)}}{6} \delta \phi^3 &\simeq V_0 \frac{c}{\sqrt{2\epsilon_0} \Lambda^3} \sin\left( \frac{\phi-\phi_0}{\sqrt{2 \epsilon_0} \Lambda}\right) \delta \phi^3 \nonumber \\
	&\simeq \mathcal O\left( \epsilon^{-1/2}_0 c \Lambda^{-3} H^2 \right) \delta \phi^3.
	\label{eq:third_deriv_pot}
\end{align}
Here, we focus on the peak scale contributions and approximate all the time and spatial derivatives as $\delta \dot \phi \sim \partial_i \delta \phi/a \sim \mathcal O(\Lambda^{-1} H \delta \phi)$.
Strictly speaking, this overestimates the contribution of $\partial_i \delta \phi/a$ because we have $\delta \dot \phi > \partial_i \delta \phi/a$ in $a(\eta)>a_0$ during the resonance. However, this overestimation just leads to a conservative constraint on our model and does not change our conclusion.
Then, neglecting the terms in $\mathcal O(\epsilon^{3/2}\delta \phi^3)$ or higher order in $\epsilon$, we can rewrite the cubic action Eq.~(\ref{eq:cubic_action}) as 
\begin{align}
	S_3 \simeq & \int \dd t \dd^3 x\, a^3 \left( - \frac{\dot \phi}{4 H} \delta \phi \delta \dot \phi^2 -  \frac{\dot \phi}{4 a^2 H} \delta \phi (\partial_i \delta \phi)^2 - \frac{1}{a^{2}}\delta \dot \phi \partial_i \vartheta_1 \partial_i \delta \phi 
	 - \frac{\dot \phi V^{(2)}}{4 H} \delta \phi^3 - \frac{V^{(3)}}{6} \delta \phi^3 \right).
\end{align}
With the approximation of $\delta \dot \phi \sim \partial_i \delta \phi/a \sim \mathcal O(\Lambda^{-1} H \delta \phi)$, we find the order of the terms except for the last two terms as 
\begin{align}
	 - \frac{\dot \phi}{4 H} \delta \phi \delta \dot \phi^2 - \frac{\dot \phi}{4 a^2 H} \delta \phi (\partial_i \delta \phi)^2 - \frac{1}{a^{2}}\delta \dot \phi \partial_i \vartheta_1 \partial_i \delta \phi
	 \simeq \mathcal O( \sqrt{\epsilon} \Lambda^{-2} H^2 \delta \phi^3).
	\label{eq:the_other_terms}	 
\end{align}
The order of the second derivative potential term is 
\begin{align}
	- \frac{\dot \phi V^{(2)}}{4 H} \delta \phi^3 \simeq \mathcal O(\sqrt{\epsilon} c \Lambda^{-2}H^2 \delta \phi^3).
	\label{eq:second_deriv_pot}	
\end{align}
Comparing Eqs.~(\ref{eq:third_deriv_pot}), (\ref{eq:the_other_terms}), and (\ref{eq:second_deriv_pot}), we find that the condition for the third potential derivative term to be dominant is 
\begin{align}
	c \gg \epsilon_0\Lambda,
	\label{eq:cond_for_cubic}
\end{align}
where we have used $c < 1$ and $\epsilon \lesssim \mathcal O(\epsilon_0)$. % and $\epsilon \simeq \epsilon_0$.

Next, let us see the quartic interaction. 
From Ref.~\cite{Seery:2006vu}, the quartic action in the spatially flat gauge can be expressed as 
\begin{align}
S_4 =& \int \dd t\, \dd^3 x\, a^3 \left( - \frac{1}{24} V^{(4)} \delta \phi^4 + \frac{1}{2a^4} \partial_{(i} \beta_{2j)} \partial_i \beta_{2j} + \frac{1}{2a^4} \partial_j \vartheta_1 \partial_j \delta \phi \partial_m \vartheta_1 \partial_m \delta \phi - \frac{1}{a^2} \delta \dot \phi (\partial_j \vartheta_2 + \beta_{2j} ) \partial_j \delta \phi \right.\nonumber \\
& + \left( \alpha_1^2 \alpha_2 - \frac{1}{2} \alpha_2^2 \right) (-6 H^2 + \dot \phi^2) + \frac{\alpha_1}{2} \left[ - \frac{1}{3} V^{(3)} \delta \phi^3 - 2 \alpha_1^2 V^{(1)} \delta \phi + \alpha_1 \left( - \frac{1}{a^2} (\partial_i \delta \phi)^2 - V^{(2)} \delta \phi^2 \right) \right. \nonumber \\
& \left. \left. - \frac{2}{a^4} \partial_i \partial_j \vartheta_2 \partial_i \partial_j \vartheta_1 + \frac{2}{a^4} \partial^2 \vartheta_2 \partial^2 \vartheta_1 - \frac{2}{a^4} \partial_i \beta_{2j} \partial_i \partial_j \vartheta_1 + \frac{2}{a^2} \dot \phi (\partial_j \vartheta_2 + \beta_{2j}) \partial_j \delta \phi + \frac{2}{a^2} \delta \dot \phi \partial_j \vartheta_1 \partial_j \delta \phi \right] \right),
\label{eq:quartic_action}
\end{align}
where 
\begin{align}
	\alpha_2 =&\, \frac{\alpha_1^2}{2} + \frac{1}{2H} \partial^{-2} \left( \partial_j \delta \dot \phi \partial_j \delta \phi + \delta \dot \phi \partial^2 \delta \phi + \frac{1}{a^2} \partial^2 \alpha_1 \partial^2 \vartheta_1 - \frac{1}{a^2} \partial_i \partial_j \alpha_1 \partial_i \partial_j \vartheta_1 \right) \nonumber \\
	\simeq &\,  \mathcal O(\Lambda^{-1} \delta \phi^2), \\
	\frac{4H}{a^2} \partial^2\vartheta_2 =& -\frac{1}{a^2} (\partial_i \delta \phi)^2 - V^{(2)} \delta \phi^2 + \frac{1}{a^4} (\partial^2 \vartheta_1)^2 - \frac{1}{a^4} \partial_i \partial_j \vartheta_1\partial_i \partial_j \vartheta_1 - \delta \dot \phi^2 + 2 \frac{\dot \phi}{a^2} \partial_i \vartheta \partial_i \delta \phi \nonumber \\
	& + 2 \alpha_1 \left( \frac{4H}{a^2}\partial^2 \vartheta_1 + 2 \dot \phi \delta \dot \phi \right) - (3 \alpha_1^2 - 2 \alpha_2)(-6H^2 + \dot \phi^2) \nonumber \\
	\simeq &\, \mathcal O(\Lambda^{-2} H^2 \delta \phi^2), \nonumber \\
	\frac{1}{2a^2}\beta_{2j} =&\, \partial^{-4} \left( \frac{1}{a^2} \partial^2 \alpha_1 \partial_j \partial^2 \vartheta_1 - \frac{1}{a^2} \partial_m \partial_j \alpha_1 \partial_m \partial^2 \vartheta_1 + \frac{1}{a^2} \partial_m \alpha_1 \partial_m \partial_j \partial^2 \vartheta_1 - \frac{1}{a^2} \partial_j \alpha_1 \partial^4 \vartheta_1 \right. \nonumber \\
	&\qquad
	 - \frac{1}{a^2} \partial_m \partial_j \partial_i \alpha_1 \partial_i \partial_m \vartheta_1 + \frac{1}{a^2} \partial^2 \partial_i \alpha_1 \partial_i \partial_j \vartheta_1 - \frac{1}{a^2} \partial_j \partial_i \alpha_1 \partial_i \partial^2 \vartheta_1 + \frac{1}{a^2} \partial_m \partial_i \alpha_1 \partial_m \partial_i \partial_j \vartheta_1 \nonumber \\
	&\quad
	\left. \phantom{\frac{1}{2}} + \partial_m \partial_j \delta \dot \phi \partial_m \delta \phi - \partial^2 \delta \dot \phi \partial_j \delta \phi + \partial_j \delta \dot \phi \partial^2 \delta \phi - \partial_m \delta \dot \phi \partial_m \partial_j \delta \phi \right) \nonumber \\
	\simeq &\, \mathcal O(a^{-1} \delta \phi^2),
\end{align}
where we have neglected the higher order $\epsilon$ terms in the order estimation.

From Eq.~(\ref{eq:d4_pot}), we can approximate the fourth potential derivative term in the oscillatory region as
\begin{align}
	- \frac{1}{24} V^{(4)} \delta \phi^4 &\simeq -\frac{cV_0}{48 \epsilon_0 \Lambda^4} \cos\left(\frac{\phi-\phi_0}{\sqrt{2\epsilon_0}\Lambda} \right) \delta \phi^4 \nonumber \\
	&\simeq \mathcal O(\epsilon^{-1}_0 c \Lambda^{-4} H^2 \delta \phi^4).
	\label{eq:v4_approx}
\end{align}
Neglecting the terms in $\mathcal O(\epsilon \delta \phi^3)$ or higher order in $\epsilon$, we can rewrite the quartic action Eq.~(\ref{eq:quartic_action}) as 
\begin{align}
S_4 \simeq& \int \dd t\, \dd^3 x a^3 \left( - \frac{1}{24} V^{(4)} \delta \phi^4 + \frac{1}{2a^4} \partial_{(i} \beta_{2j)} \partial_i \beta_{2j}  - \frac{1}{a^2} \delta \dot \phi (\partial_j \vartheta_2 + \beta_{2j} ) \partial_j \delta \phi - \frac{1}{2} \alpha_2^2 (-6 H^2) + \frac{\alpha_1}{2} \left[ - \frac{1}{3} V^{(3)} \delta \phi^3 \right] \right).
\label{eq:quartic_action2}
\end{align}
The order of the second term is 
\begin{align}
	\frac{1}{2a^4} \partial_{(i} \beta_{2j)} \partial_i \beta_{2j} \simeq \mathcal O(\Lambda^{-2}H^2 \delta \phi^4).
	\label{eq:s4_2}
\end{align}
The order of the third term is 
\begin{align}
	 - \frac{1}{a^2} \delta \dot \phi (\partial_j \vartheta_2 + \beta_{2j} ) \partial_j \delta \phi \simeq \mathcal O(\Lambda^{-3} H^2 \delta \phi^4). 
	\label{eq:s4_3}	 
\end{align}
The order of the fourth term is 
\begin{align}
	- \frac{1}{2} \alpha_2^2 (-6 H^2) \simeq \mathcal O(\Lambda^{-2} H^2 \delta \phi^4). 
	\label{eq:s4_4}	 	
\end{align}
The order of the fifth term is 
\begin{align}
	\frac{\alpha_1}{2} \left[ - \frac{1}{3} V^{(3)} \delta \phi^3 \right] \simeq \mathcal O(c \Lambda^{-3}H^2 \delta \phi^4).
	\label{eq:s4_5}	 	
\end{align}
Comparing Eqs.~(\ref{eq:v4_approx}) and (\ref{eq:s4_2})-(\ref{eq:s4_5}), we find that the condition for the fourth potential derivative term to be dominant in the quartic action is 
\begin{align}
	c \gg \epsilon_0 \Lambda,
\end{align}
where we have used $c < 1$. 
Note that this is the same condition for the third potential derivative term to be dominant in the cubic action, Eq.~(\ref{eq:cond_for_cubic}).
Since our fiducial parameter sets in Fig.~\ref{fig:power_evol_pbh_pl} satisfy this condition, we can safely neglect the metric perturbations in our analysis.

%%%%%%%%%%%%%%%%%%%%%%%%%%%%%%%%
\section{Equation of motion approach}
\label{app:eom}
%%%%%%%%%%%%%%%%%%%%%%%%%%%%%%%%

Although we have obtained the one-loop power spectra by calculating the in-in formalism in the main text, we can also obtain the loop power spectra by solving the equation of motion~\cite{Musso:2006pt}.
In this appendix, we show that the one-loop power spectra from the equation of motion indeed match those from the in-in formalism. 

In the following, we solve the following equation of motion of the scalar field fluctuations order by order:
\begin{align}
  \delta \phi'' + 2 \mathcal H \delta \phi' - \nabla^2 \delta \phi + a^2 \frac{\partial^2 V}{\partial \phi^2} \delta \phi = -a^2 \sum_{n=3} \frac{1}{(n-1)!} V^{(n)}(\delta \phi)^{n-1},
  \label{eq:eom_delta_phi}
\end{align}
where the right-hand side represents the higher-order corrections. 
Note again we express the field fluctuations as $\delta \phi = \delta \phi^\fo + \delta \phi^\so + \delta \phi^\tho$ with the superscript indicating the perturbation order.

%%%%%%%%%%%%%%%%%
\subsection{Second order perturbations}
%%%%%%%%%%%%%%%%%

First, let us discuss the second order contributions to the one-loop power spectrum. 
For the second order perturbations, we can reexpress Eq.~(\ref{eq:eom_delta_phi}) as 
\begin{align}
  {\delta \phi^\so}'' + 2 \mathcal H {\delta \phi^\so}' - \nabla^2 {\delta \phi^\so} + a^2 \frac{\partial^2 V}{\partial \phi^2} \delta \phi^\so = -a^2 \frac{1}{2} V^{(3)}(\delta \phi^\fo)^{2}.
\end{align}
In Fourier space, this equation becomes 
\begin{align}
  {\delta \phi^\so_{\mathbf k}}'' + 2 \mathcal H {\delta \phi^\so_{\mathbf k}}' + k^2 \delta \phi^\so_{\mathbf k} + a^2 \frac{\partial^2 V}{\partial \phi^2} \delta \phi^\so_{\mathbf k} = -a^2 \frac{1}{2} V^{(3)} \int \frac{\dd^3 q}{(2\pi)^3} \delta \phi^\fo_{\mathbf k - \mathbf q} \delta \phi^\fo_{\mathbf q},
\label{eq:so_eom}  
\end{align}
where $\delta \phi^\fo$ is expanded with the creation and the annihilation operators in the same way as Eq.~(\ref{eq:d_phi_expand}):
\begin{align}
  \delta \phi^\fo(\mathbf x,\eta) &= \int \frac{\dd^3 k}{(2\pi)^3} \ee^{i \mathbf k \cdot \mathbf x} \delta \phi^\fo_{\mathbf k}(\eta) \nonumber \\
  &= \int \frac{\dd^3 k}{(2\pi)^3} \ee^{i \mathbf k \cdot \mathbf x} \left[ U_k(\eta)\hat a(\mathbf k) +  U^*_k(\eta) {\hat a}^{\dagger}(-\mathbf k) \right].
  \label{eq:delta_phi_expand}
\end{align}

Solving Eq.~(\ref{eq:so_eom}) with Green function method, we can express the solution as 
\begin{align}
  \delta \phi^\so_{\mathbf k} = \int^\eta_{-\infty} \dd \eta' g_k(\eta;\eta') \mathcal S_{\mathbf k}(\eta'),
  \label{eq:so_delta_phi}
\end{align}
where 
\begin{align}
  &\mathcal S_{\mathbf k}(\eta) \equiv \frac{\lambda(\eta)}{a^2(\eta)} \int \frac{\dd^3 q}{(2\pi)^3} \delta \phi^\fo_{\mathbf k - \mathbf q}(\eta) \delta \phi^\fo_{\mathbf q}(\eta), 
  \end{align}
and the Green function satisfies 
\begin{align}
  g_k'' + 2\mathcal H g_k' + k^2 g_k + a^2 \frac{\partial^2 V}{\partial \phi^2} g_k = \delta(\eta-\eta').
  \label{eq:green}
\end{align}
The concrete expression of this Green function is given by
\begin{align}
  g_k(\eta;\eta') &= \frac{U_k(\eta) U_k^*(\eta') - U_k^*(\eta) U_k(\eta')}{U_k'(\eta')U_k^*(\eta')- U_k(\eta'){U_k^{*}}'(\eta')} \nonumber \\
  &= -2 a^2(\eta')\text{Im}[U_k(\eta)U^*(\eta')],
  \label{eq:green_exp}
\end{align}
where we have used Eq.~(\ref{eq:denom_green}) in the second line. 
After some calculation, we obtain the following expression:
\begin{align}
  \vev{\delta \phi^\so_{\mathbf k}(\eta) \delta \phi^\so_{\mathbf k'} (\eta)} &= 
\int^\eta_{-\infty} \dd \eta_1 \int^\eta_{-\infty} \dd \eta_2 \, g_{k}(\eta; \eta_1) g_{k'}(\eta; \eta_2)\vev{\mathcal S_{\mathbf k} (\eta_1) \mathcal S_{\mathbf k'}(\eta_2)} \nonumber \\
&= (2\pi)^3 \delta(\mathbf k+ \mathbf k') \frac{2\pi^2}{k^3} \mathcal P^a_{\delta \phi, 2\vx}(k,\eta),
  \label{eq:delta_phi_so}
\end{align}
where $\mathcal P^a_{\delta \phi, 2\vx}(k,\eta)$ is defined in Eq.~(\ref{eq:p_a_2vx}) and we have used Eq.~(\ref{eq:green_exp}) in the second equality.

%%%%%%%%%%%%%%%%%
\subsection{Third order perturbations}
%%%%%%%%%%%%%%%%%

Next, let us see the third order perturbations. 
The contributions from the third order perturbations can be divided into two, 1): the two-vertex contribution proportional to $(V^{(3)})^2$, and 2): the one-vertex contribution proportional to $V^{(4)}$.

\subsubsection{Two-vertex contribution}

First, let us focus on the third-order equation that includes $(V^{(3)})^2$:
\begin{align}
  \delta {\phi^\tho}'' + 2 \mathcal H \delta {\phi^\tho}' - \nabla^2 \delta \phi^\tho + a^2 \frac{\partial^2 V}{\partial \phi^2} \delta \phi^\tho = -a^2 \frac{V^{(3)}}{2} \left( \delta \phi^{(2)} \delta \phi^{(1)} + \delta \phi^{(1)} \delta \phi^{(2)} \right),
\end{align}
In Fourier space, this equation becomes 
\begin{align}
  \delta {\phi_{\mathbf k}^\tho}'' + 2 \mathcal H \delta {\phi_{\mathbf k}^\tho}' + k^2 \delta \phi^\tho_{\mathbf k} + a^2 \frac{\partial^2 V}{\partial \phi^2} \delta \phi^\tho_{\mathbf k} = -a^2 \frac{V^{(3)}}{2} \int \frac{\dd^3 q}{(2\pi)^3} \left(\delta \phi^\so_{\mathbf q} \delta \phi^\fo_{\mathbf k - \mathbf q} + \delta \phi^\fo_{\mathbf k - \mathbf q} \delta \phi^\so_{\mathbf q} \right). 
\end{align}
Then, similar to Eq.~(\ref{eq:so_delta_phi}), we can obtain
\begin{align}
  \delta \phi^\tho_{\mathbf k}(\eta) &= \int^\eta_{-\infty} \dd \eta' g_k(\eta,\eta') \frac{\lambda(\eta')}{a^2(\eta')} \int \frac{\dd^3 q}{(2\pi)^3} \left(\delta \phi^\so_{\mathbf q}(\eta') \delta \phi^\fo_{\mathbf k - \mathbf q}(\eta') + \delta \phi^\fo_{\mathbf k - \mathbf q}(\eta') \delta \phi^\so_{\mathbf q}(\eta') \right) \nonumber \\
  &= \int \frac{\dd^3 q}{(2\pi)^3} \int \frac{\dd^3 q'}{(2\pi)^3}  \int^\eta_{-\infty} \dd \eta' g_k(\eta;\eta') \frac{\lambda(\eta')}{a^2(\eta')} \int^{\eta'}_{-\infty} \dd \eta'' g_q(\eta';\eta'') \frac{\lambda(\eta'')}{a^2(\eta'')} \nonumber \\
  & \qquad \times   \left[\delta \phi^\fo_{\mathbf q - \mathbf q'}(\eta'') \delta \phi^\fo_{\mathbf q'}(\eta'') \delta \phi^\fo_{\mathbf k - \mathbf q}(\eta') + \delta \phi^\fo_{\mathbf k - \mathbf q}(\eta') \delta \phi^\fo_{\mathbf q - \mathbf q'}(\eta'') \delta \phi^\fo_{\mathbf q'}(\eta'')  \right].
\end{align}
Using this expression, we obtain the following one-loop power spectrum from the third order perturbations:
\begin{align}
  \vev{\delta \phi^\fo_{\mathbf k'}(\eta) \delta \phi^\tho_{\mathbf k} (\eta)} + \vev{\delta \phi^\tho_{\mathbf k}(\eta) \delta \phi^\fo_{\mathbf k'} (\eta)} &= (2\pi)^3 \delta(\mathbf k+ \mathbf k') \frac{2\pi^2}{k^3} \mathcal P^b_{\delta \phi, 2\vx}(k,\eta),
  \label{eq:first_third_eom}
\end{align}
where $\mathcal P^b_{\delta \phi,2\vx}$ is defined in Eq.~(\ref{eq:asym_final}).

\subsubsection{One-vertex contribution}

Finally, we discuss the one-vertex contribution that includes $V^{(4)}$.
To this end, we solve the following third-order equation:
\begin{align}
  \delta {\phi^\tho}'' + 2 \mathcal H \delta {\phi^\tho}' - \nabla^2 \delta \phi^\tho + a^2 \frac{\partial^2 V}{\partial \phi^2} \delta \phi^\tho = -a^2 \frac{V^{(4)}}{6} (\delta \phi^\fo)^{3}.
\end{align}
In Fourier space, this equation becomes 
\begin{align}
  \delta {\phi_{\mathbf k}^\tho}'' + 2 \mathcal H \delta {\phi_{\mathbf k}^\tho}' + k^2 \delta \phi^\tho_{\mathbf k} + a^2 \frac{\partial^2 V}{\partial \phi^2} \delta \phi^\tho_{\mathbf k} = -a^2 \frac{V^{(4)}}{6} \int \frac{\dd^3 q}{(2\pi)^3} \int \frac{\dd^3 p}{(2\pi)^3} \delta \phi^\fo_{\mathbf p} \delta \phi^\fo_{\mathbf q} \delta \phi^\fo_{\mathbf k - \mathbf p - \mathbf q}.
\end{align}
Solving this equation, we obtain 
\begin{align}
  \delta {\phi_{\mathbf k}^\tho} = \int^\eta_{-\infty} \dd \eta'\, \frac{\mu(\eta')}{a^2(\eta')} g_{k}(\eta; \eta') \int \frac{\dd^3 q}{(2\pi)^3} \int \frac{\dd^3 p}{(2\pi)^3} \delta \phi^\fo_{\mathbf p}(\eta') \delta \phi^\fo_{\mathbf q}(\eta') \delta \phi^\fo_{\mathbf k - \mathbf p - \mathbf q}(\eta').
\end{align}
Using this, we finally obtain 
\begin{align}
    \vev{\delta \phi^\fo_{\mathbf k'}(\eta) \delta \phi^\tho_{\mathbf k} (\eta)} + \vev{\delta \phi^\tho_{\mathbf k}(\eta) \delta \phi^\fo_{\mathbf k'} (\eta)} &= (2\pi)^3 \delta(\mathbf k+ \mathbf k') \frac{2\pi^2}{k^3} \mathcal P_{\delta \phi, 1\vx}(k,\eta),
\end{align}
where $\mathcal P_{\delta \phi, 1\vx}$ is defined in Eq.~(\ref{eq:1vx_final}).

%%%%%%%%%%%%%%%%%%%%%%%%%%%%%%%%
\section{Perturbative expansion of the tree-level power spectrum}
\label{app:P_tree_pertb}
\begin{figure}%[t] 
        \centering \includegraphics[width=0.7\columnwidth]{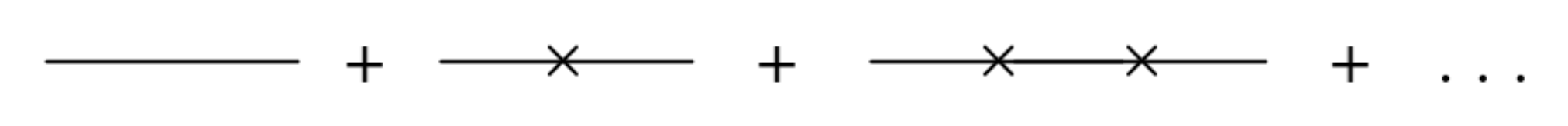}
        \caption{ Diagrammatic expansion of $\mathcal{P}_{\delta \phi}$.
        }
        \label{fig:tree_diagrams}
\end{figure}

%%%%%%%%%%%%%%%%%%%%%%%%%%%%%%%%
In the main text,  we computed the tree-level power spectrum by solving the equation of motion of the mode function. This amounts to a non-perturbative computation of the tree-level power spectrum \cite{Chen:2015dga,Werth:2023pfl}. On the other hand, when dealing with primordial features, it is sometimes useful to treat the quadratic interaction in the Hamiltonian itself as a perturbation on top of the free Hamiltonian, which consists only in the slow-roll background. Then the power spectrum can be computed as the sum of Feynmann diagrams as those shown in Fig.~\ref{fig:tree_diagrams}, where a cross corresponds to the insertion of the quadratic interaction $H_{\text{int},2}$. When the amplitude of the features is small, as for example for features that fit the CMB residuals, the series are perturbative and the non-perturbative result obtained using the equation of motion matches well the sum of the first two diagrams in Fig.~\ref{fig:tree_diagrams} (see e.g. Refs.~\cite{Chen:2015dga,Braglia:2022ftm}). In this Appendix, we explicitly show that the perturbative expansion is no longer adequate for the oscillatory feature model considered in this paper and that we have to resort to the  equation of motion to get the correct shape of the tree power spectrum. To do so, we compute the diagrams in Fig.~\ref{fig:tree_diagrams} and show that the one with two insertions of $H_{{\rm int},2}$ is larger than the one with one insertion.

We recall the expression for the second order Hamiltonian density,
\begin{equation}
	\mathcal H_2 = \frac{1}{a^2} \left[ \frac{1}{2}\delta {\phi'}^2 + \frac{1}{2}(\partial_i \delta \phi)^2 + \frac{a^2}{2} V^{(2)}(\phi) \delta \phi^2 \right].
\end{equation}
Writing $V^{(2)}\equiv \overline{V^{(2)}}+\Delta V^{(2)}$, we can separate the contributions into the featureless part of the potential, which we define to be the free Hamiltonian, and the feature part of the potential, which we define as a quadratic interaction: 
\begin{align}
	\overline{\mathcal H_{2}} &= \frac{1}{a^2} \left[ \frac{1}{2}\delta {\phi'}^2 + \frac{1}{2}(\partial_i \delta \phi)^2 + \frac{a^2}{2} \overline{V^{(2)}}(\phi) \delta \phi^2 \right], \\
	\mathcal H_{{\rm int},2} &= \frac{1}{2} \Delta V^{(2)}(\phi) \delta \phi^2,
\end{align}
where $\overline{V^{(2)}} = V_0 (n_s-1)/(2 M_\Pl^2)$.
Note that we again neglect the modification of the potential around the end of inflation.

We define the effects of the feature on the power spectrum as $\Delta \mathcal{P} \equiv \mathcal{P}_{f}-\mathcal{P}_0$, where $\mathcal P_f$ and $\mathcal{P}_0$ are the power spectra with and without the feature, respectively.
Note that $\mathcal P_f$ can be a power spectrum with perturbative features, as we will see below.
We compare $\Delta \mathcal P/\mathcal P_0$ for the non-perturbative result obtained by solving the equation of motion and that for the perturbative expansion results.
For the non-perturbative case, $\mathcal P_f$ is $\mathcal P_{\delta \phi,\tre}$ and we numerically derive $\mathcal{P}_{\delta\phi,{\rm tr}}$ and $\mathcal P_0$. 
Although both $\mathcal{P}_{\delta\phi,{\rm tr}}$ and $\mathcal{P}_0$ weakly evolve in time even on superhorizon because of the growth of $\epsilon$, the fractional change $\Delta \mathcal{P}/\mathcal{P}_0$ is constant on superhorizon scales as the same time evolution appears in the numerator and denominator.

On the other hand, for the perturbative expansion, we use the analytical mode function.
The $\delta\phi$'s are operators in the interaction picture, which are evolved using $\overline{\mathcal H_{2}}$, unlike in the main text, where they are evolved using the full $\mathcal H_{2}$. 
Imposing the Bunch-Davies conditions when the modes are well inside the horizon, we obtain the mode function in the featureless potential as
\begin{align}
    &\overline{U}_k(\eta)\simeq\frac{H}{\sqrt{2\, k^3}}(1+i k \eta)\ee^{-i k \eta}, 
	\label{eq:overline_u}
\end{align}
where the overline denotes the quantity in the featureless potential, which is different from $U_k$ in the main text.
Note that we do not take into account the superhorizon evolution because we focus on $\Delta \mathcal P/\mathcal P_0$. In the fractional change, the superhorizon evolution is canceled out even if we take into account the superhorizon evolution of $\overline U_k$. 
Given this, we use Eq.~(\ref{eq:overline_u}) instead of the mode function obtained by solving the equation of motion, for simplicity.
The first correction to the featureless power spectrum, i.e. that represented by the second diagram in Fig.~\ref{fig:tree_diagrams}, is given by: 
\begin{align}
    (2\pi)^3\delta(\mathbf{k}+\mathbf{k}')\Delta\mathcal{P}_1&=\frac{k^3}{2\pi^2}(-i)\int_{-\infty}^0{\rm d}\eta \,\langle0\lvert\left[\delta\phi_{\mathbf k}(0)\delta\phi_{\mathbf k'}(0),\,H_{{\rm int},2}(\eta)\right]\rvert0\rangle\notag\\
    &=\frac{k^3}{2\pi^2}(-i)\int_{-\infty}^0 {\rm d}\eta\, a^4\,\int {\rm d}^3 x\,\int\frac{{\rm d}^3 p}{(2\pi)^3}\frac{{\rm d}^3 q}{(2\pi)^3}\ee^{i(\mathbf{p}+\mathbf{q})\cdot\mathbf{x}}\,\frac{\Delta V^{(2)}}{2}\,\langle0\lvert\left[\delta\phi_{\mathbf k}(0)\delta\phi_{\mathbf k'}(0),\, \delta\phi_{\mathbf p}(\eta)\delta\phi_{\mathbf q}(\eta) \right]\rvert0\rangle\notag \\
    &=\frac{k^3}{2\pi^2}(2\pi)^3\delta(\mathbf{k}+\mathbf{k}')\overline{U}_k(0)^2\,4\,{\rm Im}\left[ \int_{-\infty}^0 {\rm d}\eta\, a^4\,\,\frac{\Delta V^{(2)}}{2}\,\overline{U}^*_k(\eta)^{2}\right]\notag\\
    &=(2\pi)^3\delta(\mathbf{k}+\mathbf{k}')\mathcal{P}_0\,3\,{\rm Im} \left[\int_0^\infty{\rm d}x\,\frac{\Delta V^{(2)} M_\Pl^2}{V_0}\frac{\ee^{-2 i x y }}{x^4 y^3}\left(-y^2x^2 + 2 i x y +1\right) \right],
\end{align}
where $H_{\text{int},2} \equiv \int \dd^3 x\, a^4 \mathcal H_{\text{int},2}$, $\mathcal P_0 \equiv k^3 |\overline{U}_k(0)|^2/(2\pi^2) = H^2/(2\pi)^2$, and we have used $\overline{U}^2_k(0) = |\overline{U}_k(0)|^2$, $a(\eta)\simeq-1/H\eta$, and $H \simeq \sqrt{V_0/(3M_\Pl^2)}$. 
We have also changed variables to $x=\eta/\eta_0$ and $y= - k\eta_0$ from the third to the fourth line. Note that we are calculating the correlator at $\eta\to0$, i.e. long after the feature.

The contribution to the second order diagram is:
\begin{align}
    (2\pi)^3\delta(\mathbf{k}+\mathbf{k}')\Delta\mathcal{P}_2&=\frac{k^3}{2\pi^2}\Biggl\{i^2(-1) \int_{-\infty}^0{\rm d}\eta_1 \,\int_{-\infty}^0{\rm d}\eta_2  \,\langle0\lvert H_{{\rm int},2}(\eta_1)\delta\phi_{\mathbf k}(0)\delta\phi_{\mathbf k'}(0)\,H_{{\rm int},2}(\eta_2)\rvert0\rangle\notag\\
    &+2\,{\rm Re} \left[i (-1)^2 \int_{-\infty}^0{\rm d}\eta_1 \,\int_{-\infty}^{\eta_1}{\rm d}\eta_2  \,\langle0\lvert \delta\phi_{\mathbf k}(0)\delta\phi_{\mathbf k'}(0)\,H_{{\rm int},2}(\eta_1)\,H_{{\rm int},2}(\eta_2) \rvert0\rangle \right]\Biggr\}\notag\\
    &\equiv (2\pi)^3 \delta(\mathbf{k}+\mathbf{k}')\left(\Delta\mathcal{P}_{2,\,1} + \Delta\mathcal{P}_{2,\,2}\right).
\end{align}
After performing the contractions, we get the following expressions for $\Delta\mathcal{P}_{2,\,1}$ and $\Delta\mathcal{P}_{2,\,2}$:
\begin{align}
  (2\pi)^3\delta(\mathbf{k}+\mathbf{k}')\Delta\mathcal{P}_{2,\,1}  &=\frac{k^3}{2\pi^2}  (2\pi)^3 \delta(\mathbf{k}+\mathbf{k}')\,8\,\lvert \overline{U}_k(0)\rvert^2\, \int_{-\infty}^0{\rm d}\eta_1 \,a^4(\eta_1)\,\frac{\Delta V^{(2)}(\eta_1)}{2 }\,\overline{U}_k(\eta_1)^2\,\notag \\
  & \qquad \times \int_{-\infty}^0{\rm d}\eta_2 \,a^4(\eta_2)\,\frac{\Delta V^{(2)}(\eta_2)}{2}\,\overline{U}^*_k(\eta_2)^2\notag\\
  &=(2\pi)^3\delta(\mathbf{k}+\mathbf{k}')\mathcal{P}_0\,\frac{1}{2}\Biggl\lvert3\int_0^\infty \frac{{\rm d}x}{y^3 x^4}\ee^{2 i x y}\frac{\Delta V^{(2)}(x)M_\Pl^2}{V_0}\left(-y^2x^2-2 i x y +1\right)\Biggr\rvert^2,\\
   (2\pi)^3\delta(\mathbf{k}+\mathbf{k}')\Delta\mathcal{P}_{2,\,2}  &=-\frac{k^3}{2\pi^2}  (2\pi)^3 \delta(\mathbf{k}+\mathbf{k}')\,16\, {\rm Im} \left[\overline{U}_k(0)^2\, \int_{-\infty}^0{\rm d}\eta_1 \,a^4(\eta_1)\,\frac{\Delta V^{(2)}(\eta_1)}{2 }\,\lvert \overline{U}_k(\eta_1)\rvert^2 \right.\,\notag \\
  &\qquad \left.\times \int_{-\infty}^{\eta_1}{\rm d}\eta_2 \,a^4(\eta_2)\,\frac{\Delta V^{(2)}(\eta_2)}{2 }\,\overline{U}^*_k(\eta_2)^2 \right]\notag\\
  &=-(2\pi)^3\delta(\mathbf{k}+\mathbf{k}')\mathcal{P}_0\, {\rm Im} \left[\int_0^\infty \frac{{\rm d}x_1}{y^3 x_1^4}\,3\,\frac{\Delta V^{(2)}(x_1)M_\Pl^2}{V_0 }\,(1+x_1^2 y^2) \right.\,\notag\\
  &\qquad \left.\times\int_{x_1}^\infty \frac{{\rm d}x_2}{y^3 x_2^4}e^{-2 i x_2 y}\,3\,\frac{\Delta V^{(2)}(x_2)\,M_\Pl^2}{V_0 }\left(-x_2^2y^2+2 i x_2 y +1\right) \right], 
\end{align}
where $x_1 = \eta_1/\eta_0$ and $x_2 = \eta_2/\eta_0$.

\begin{figure}[t] 
        \centering \includegraphics[width=0.495\columnwidth]{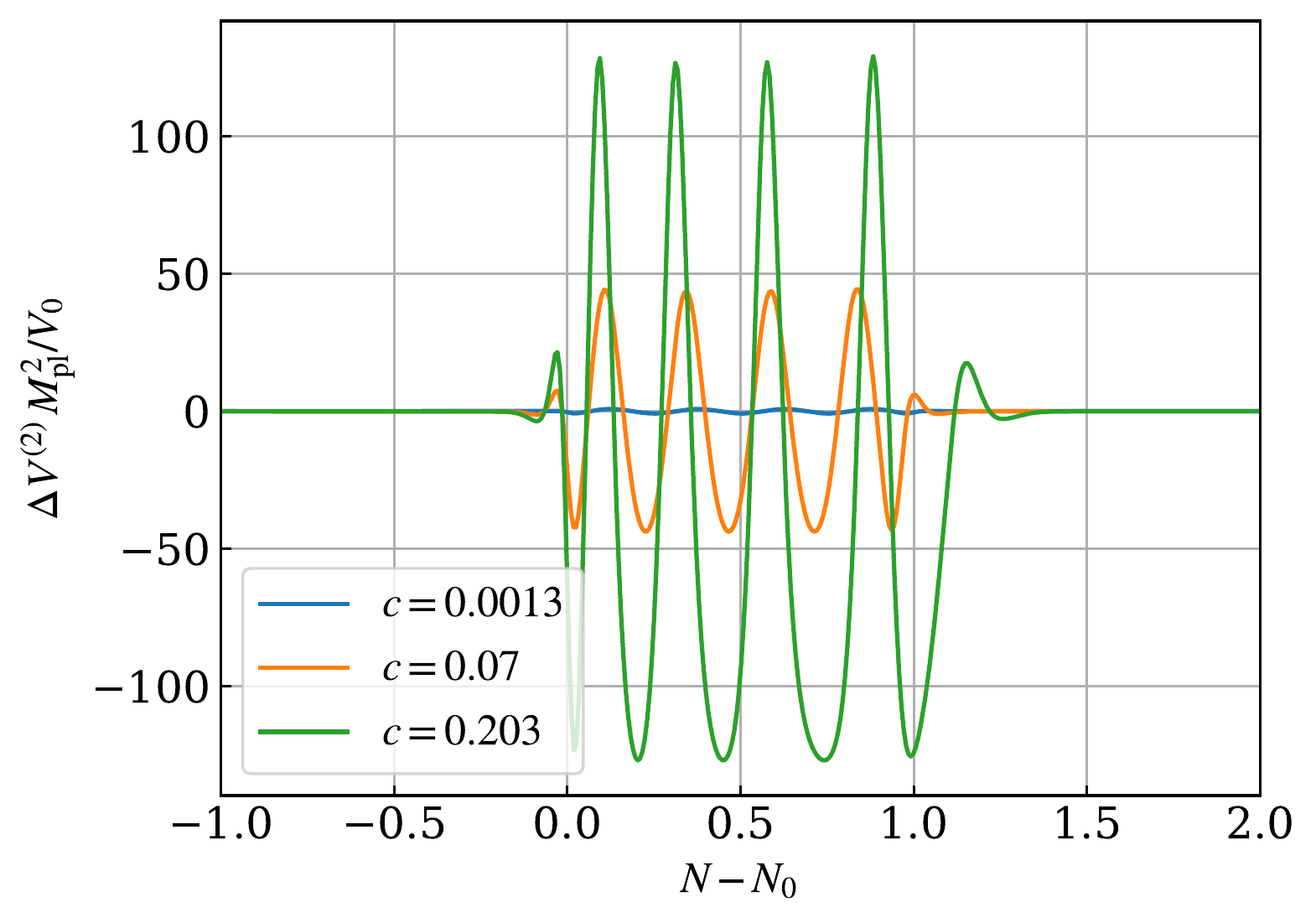}
        \includegraphics[width=0.495\columnwidth]{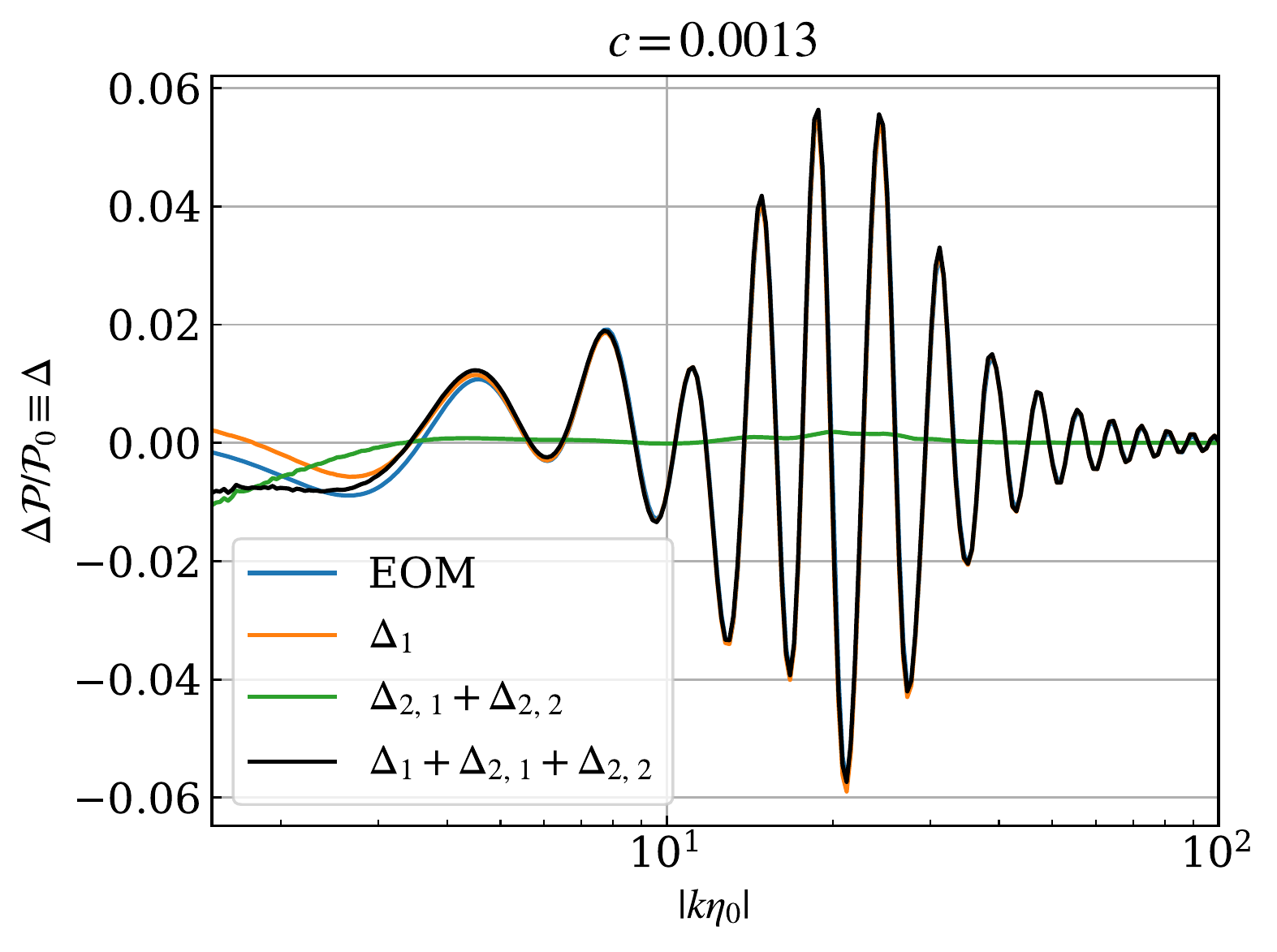}
        \includegraphics[width=0.495\columnwidth]{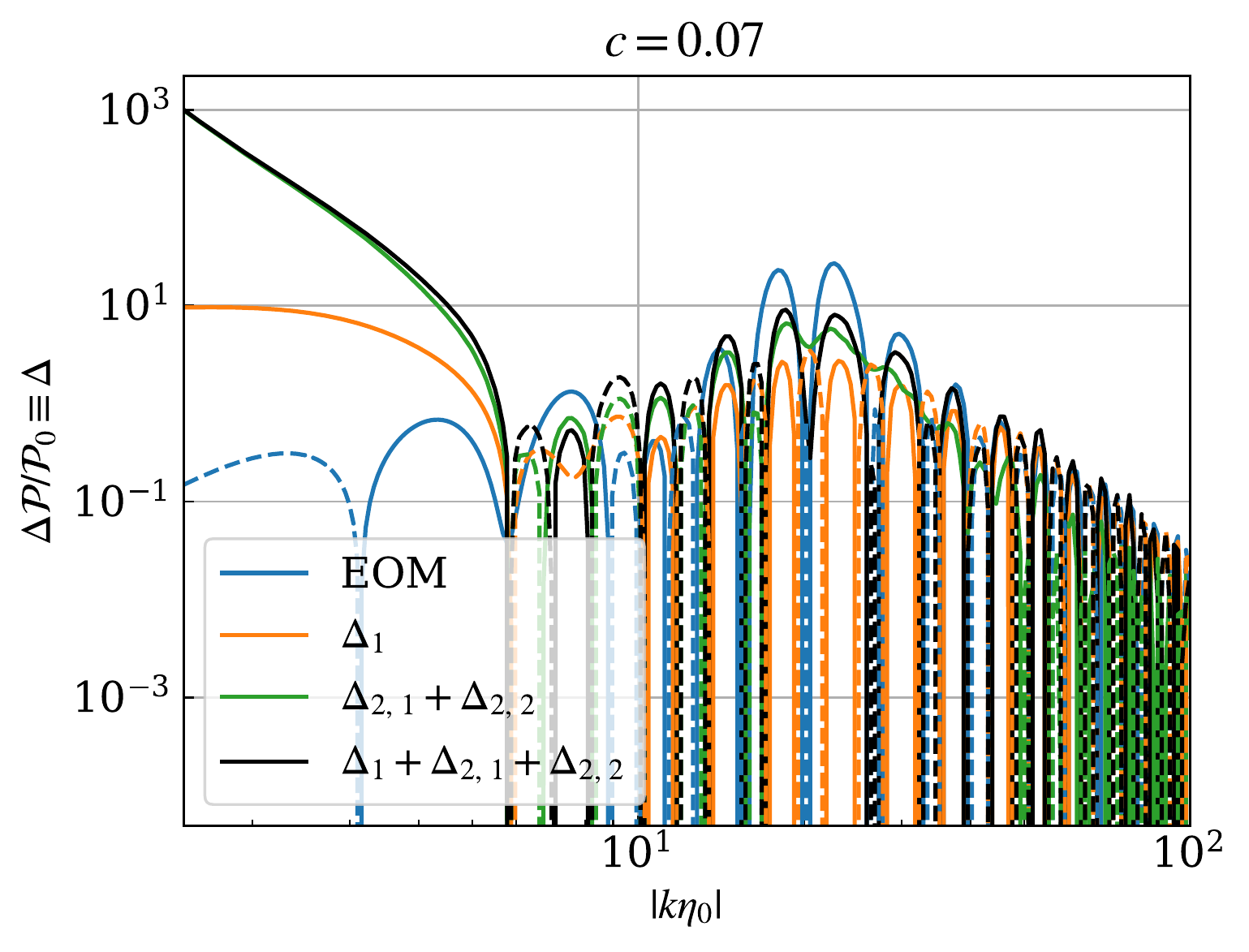}
        \includegraphics[width=0.495\columnwidth]{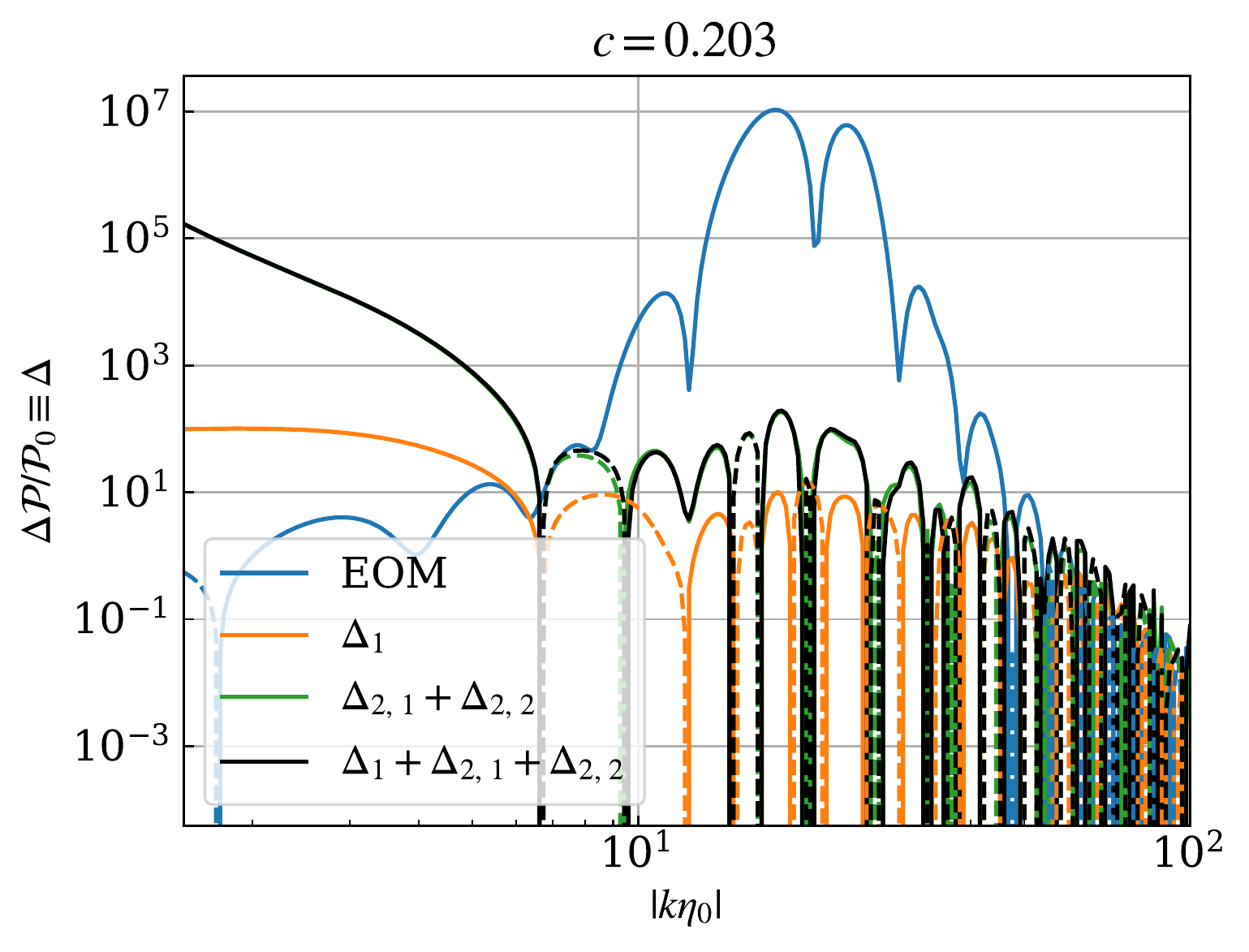}
        \caption{ Perturbative corrections to the tee-level power spectrum of $\delta\phi$ compared to the non-perturbative result, obtained by solving the equation of motion (``EOM'' in blue). We use the parameters for case A in Fig.~\ref{fig:power_evol_pbh_pl} and change the amplitude of the feature as $c=0.0013,\,0.07,\,0.203$. 
        The y-axes in the bottom panels are in log scale and the dashed lines show the absolute values of negative values.
         In the top-left panel, we also show the interacting term $\Delta V^{(2)}M_\Pl^2/V_0$.
        }
        \label{fig:P_tree}
\end{figure}

 The equations above can be integrated if some analytical approximation for the interaction $\Delta V^{(2)}M_\Pl^2/V_0$ is provided. While it is possible to do that, we evaluate the integrals numerically, as we did throughout this paper. We adopt the parameters of case A in Fig.~\ref{fig:power_evol_pbh_pl} and vary $c=0.0013,\,0.07,\,0.203$. The first value of $c$ corresponds to a feature with a maximum amplitude of $\Delta \mathcal{P}/\mathcal{P}_0\simeq0.05$, which is the typical upper limit on primordial features at large scales from CMB data~\cite{Planck:2018jri}. The third value of $c$ is the one used in Fig.~\ref{fig:power_evol_pbh_pl} and we chose the second value of $c$ just as an intermediate case. The results are shown in Fig.~\ref{fig:P_tree}.
 As anticipated, our results show that higher order diagrams become increasingly important for larger values of $c$. 
 In particular, for $\Delta\mathcal{P}/\mathcal{P}_0 > \mathcal{O}(0.1)$, the higher order term in the in-in formalism becomes more important than the lower one.
 Our result is consistent with our findings in the main text, i.e. for a sufficiently large feature amplitude, the computation of the power spectrum is not perturbative {\rm already} at the tree level. Fortunately, at the tree level, a simple non-perturbative way to compute it is to numerically solve the equation of motion to get the mode functions as the desired (non-perturbative) result. 
 This method can also be generalized to multifield models with a prescribed procedure that is consistent with the first-principle in-in formalism \cite{Chen:2015dga,Werth:2023pfl}.
 
On the other hand, Fig.~\ref{fig:P_tree} also shows that, for feature amplitudes $\Delta \mathcal{P}/\mathcal{P}_0\lesssim \mathcal{O}(0.1)$, the correction to the power spectrum is well approximated by the lowest order diagram. This is reassuring in that it shows that, for primordial features of immediate observational interest, i.e. those that can be tested with large scale data such as CMB and LSS, the in-in expansion is well controlled and the theoretical tools that have been used in the literature to compute features are more than adequate.

%%%%%%%%%%%%%%%%%
\section{Cutoff scale dependence}
\label{app:cutoff}
%%%%%%%%%%%%%%%%%

In the calculation of the loop power spectrum, we introduced the cutoff scales to avoid the UV and IR divergences by assuming that such divergences are appropriately regularized.
However, if we take very small/large cutoff scales, the results can be changed by the contributions from the very small/large scale perturbations that are not amplified and should be regularized.
In this appendix, we obtain analytical estimates on the cutoff scale dependence of the results and discuss how large or small the cutoff scales can be without changing the loop power spectrum associated with the amplified perturbations.
Note that we focus on the power spectrum at $\eta_\lmax$ and ignore $\mathcal O(1)$ factors for simplicity throughout this appendix.

\subsection{UV cutoff}
Let us begin with the UV contributions.
If the scales of the UV modes are much smaller than the peak scale, the UV modes are not amplified by the parametric resonance because of $(k/a)^2 \gg |V^{(2)}|$.
Hereafter, we consider the UV modes as the small-scale perturbations not amplified by the oscillatory feature. 
From Eq.~(\ref{eq:u_sol}), we can get the $U_k(\eta)$ for the UV modes:
\begin{align}
	 U_{k_\UV}(\eta) 
	 &\simeq \frac{-i}{a\sqrt{2k_\UV} } \ee^{-i k_\UV \eta},
\end{align}
where we have used $aH \simeq -1/\eta$, valid during the slow-roll inflation.
Using this expression and Eq.~(\ref{eq:mathcal_p_delta_sym}), we can approximate the UV contribution in $\mathcal P^a_{\delta \phi, 2\vx}(k_\pe,\eta_\lmax)$ as 
\begin{align}
 \mathcal P^a_{\delta \phi, 2\vx, \text{UV}}(k_\pe,\eta_\lmax) &\simeq \int^{v_\UV}_{v_*} \dd v \int^{\text{min}[1+v,v_\UV]}_{\text{max}[|1-v|,v_*]} \dd u \frac{uv}{4\pi^4} k^6_\pe \nonumber \\
 & \qquad \times \left|\int^{\eta_\lmax}_{\eta_0} \dd \eta' \lambda(\eta') 2\, \text{Im} \left[ U_{k_\pe}(\eta_\lmax) U^*_{k_\pe}(\eta')\right] U_{k_\pe u}(\eta') U_{k_\pe v}(\eta')\right|^2 \nonumber \\
 &\simeq  \int^{v_\UV}_{v_*} \dd v \int^{\text{min}[1+v,v_\UV]}_{\text{max}[|1-v|,v_*]} \dd u \frac{uv}{4\pi^4} k^6_\pe\nonumber \\
 & \qquad \times
  \left|\int^{\eta_\lmax}_{\eta_0} \dd \eta' \lambda(\eta') 2\, \text{Im} \left[ U_{k_\pe}(\eta_\lmax) U^*_{k_\pe}(\eta')\right] \frac{\ee^{-i k_\pe(u+v)\eta'}}{2 a^2(\eta') k_\pe \sqrt{uv}}\right|^2 \nonumber \\
 &\sim  \int^{v_\UV}_{v_*} \dd v \int^{\text{min}[1+v,v_\UV]}_{\text{max}[|1-v|,v_*]} \dd u \frac{uv}{4\pi^4} k^6_\pe \left| \frac{\Delta \eta_{\text{osc}}(\eta_\lmax) \lambda(\eta_\lmax)}{a^2(\eta_\lmax) k_\pe(u+v)} \frac{1}{2 a^2(\eta_\lmax) k_\pe \sqrt{uv}}\right|^2 \nonumber \\
 &\sim  \lambda^2(\eta_\lmax) \frac{a_0^2}{a^{10}(\eta_\lmax)} \left[\frac{1}{v_*} - \frac{1}{v_\UV} \right],
 \label{eq:p_2vx_s_uv}
\end{align}
where we have used the fact that the dominant contribution comes from $\eta' \sim \eta_\lmax$.
$v_\UV (\gg 1)$ is the UV cutoff and $v_*$ is defined such that $k_\pe v_*$ is the smallest wavenumber in $k>k_\pe$ where the perturbations are not amplified during the oscillatory phase.
From the second to the third approximate equality, we have used Eq.~\eqref{eq:im_uu_approx} and $\int^{\eta_\lmax}_{\eta_0} \dd\eta' f(\eta') \ee^{-ik_\pe(u+v)\eta'} \sim  f(\eta_\lmax)/(k_\pe(u+v))$ because the shortest oscillation timescale is determined by $\ee^{-ik_\pe(u+v)\eta'}$ in the integrand with $u, v \gg 1$.
In the final line, we have used $\Delta \eta_{\text{osc}}(\eta_\lmax) \sim a_0/(a(\eta_\lmax) k_\pe)$ and $\int^{\text{min}[1+v,v_\UV]}_{\text{max}[|1-v|,v_*]} \dd u \sim \mathcal O(1)$.
Note that the UV contributions come from two UV modes with $v \gg 1$ and $u \gg 1$ due to the momentum conservation.
Combining Eqs.~(\ref{eq:p_2vx_s_uv}) and (\ref{eq:p_2vx_sym_est}), we find 
\begin{align}
	\frac{\mathcal P^a_{\delta \phi, 2\vx, \text{UV}}(k_\pe,\eta_\lmax)}{\mathcal P^a_{\delta \phi, 2\vx}(k_\pe,\eta_\lmax)} \sim \frac{a^2_0  \mathcal P^2_{\delta \phi, \tr}(k_\pe,\eta_0)}{a^2(\eta_\lmax) \mathcal P^2_{\delta \phi, \tr}(k_\pe,\eta_\lmax)} \left[\frac{1}{v_*} - \frac{1}{v_\UV} \right] \ll 1,
\end{align}
where we have used $\mathcal P_{\delta \phi, \tr}(k_\pe,\eta_0) \simeq k_\pe^2/(4 \pi^2 a_0^2)$.
The UV divergence does not appear in $\mathcal P^a_{\delta \phi, 2\vx, \text{UV}}$.

Using Eq.~(\ref{eq:asym_final}), we can approximate the UV contribution in $\mathcal P^b_{\delta \phi, 2\vx}$ as 
\begin{align}
&\left|\mathcal P^b_{\delta \phi, 2\vx, \text{UV}}(k_\pe,\eta_\lmax)\right| \nonumber \\
&\simeq \left|8 \int^{v_\UV}_{v_*} \dd v \int^{\text{min}[1+v,v_\UV]}_{\text{max}[|1-v|,v_*]} \dd u \frac{uv}{4\pi^4} k_\pe^6 \int^{\eta_\lmax}_{\eta_0} \dd \eta' \int^{\eta'}_{\eta_0} \dd \eta'' \lambda(\eta') \lambda(\eta'')\text{Im} \left[U_{k_\pe}(\eta_\lmax) U^*_{k_\pe}(\eta')\right]  \right.  \nonumber \\
  & \left. \phantom{\int^\infty_0 } 
  \times \Re\left[U_{k_\pe}(\eta_\lmax)U^*_{k_\pe}(\eta'')\right]\left(\text{Im}\left[ U_{k_\pe v}(\eta')U^*_{k_\pe v}(\eta'')\right]\Re\left[U_{k_\pe u}(\eta') U^*_{k_\pe u}(\eta'')\right] + (u \leftrightarrow v)\right) \right| \nonumber \\
  &\simeq \left|8 \int^{v_\UV}_{v_*} \dd v \int^{\text{min}[1+v,v_\UV]}_{\text{max}[|1-v|,v_*]} \dd u \frac{uv}{4\pi^4} k_\pe^6 \int^{\eta_\lmax}_{\eta_0} \dd \eta' \int^{\eta'}_{\eta_0} \dd \eta'' \lambda(\eta') \lambda(\eta'')\text{Im} \left[U_{k_\pe}(\eta_\lmax) U^*_{k_\pe}(\eta')\right]  \right.  \nonumber \\
  & \left. \phantom{\int^\infty_0 } 
  \times \Re\left[U_{k_\pe}(\eta_\lmax)U^*_{k_\pe}(\eta'')\right]\text{Im}\left[ U_{k_\pe v}(\eta')U^*_{k_\pe v}(\eta'') U_{k_\pe u}(\eta') U^*_{k_\pe u}(\eta'')\right]  \right| \nonumber \\
  &\sim \left|8 \int^{v_\UV}_{v_*} \dd v \int^{\text{min}[1+v,v_\UV]}_{\text{max}[|1-v|,v_*]} \dd u \frac{uv}{4\pi^4} k_\pe^6 \int^{\eta_\lmax}_{\eta_0} \dd \eta' \int^{\eta'}_{\eta_0} \dd \eta'' \lambda(\eta') \lambda(\eta'')\text{Im} \left[U_{k_\pe}(\eta_\lmax) U^*_{k_\pe}(\eta')\right]  \right.  \nonumber \\
  & \left. \phantom{\int^\infty_0 } 
  \times \Re\left[U_{k_\pe}(\eta_\lmax)U^*_{k_\pe}(\eta'')\right]\frac{\sin[k_\pe(u+v)(\eta''-\eta')]}{4 a^2(\eta') a^2(\eta'') k_\pe^2 u v}   \right| \nonumber \\  
  &\sim \left|8 \int^{v_\UV}_{v_*} \dd v \int^{\text{min}[1+v,v_\UV]}_{\text{max}[|1-v|,v_*]} \dd u \frac{uv}{4\pi^4} k_\pe^6 \lambda^2(\eta_\lmax) \right. \nonumber \\
  &\left. \phantom{\int^\infty_0 } 
  \times |U_{k_\pe}(\eta_\lmax)|^2 \frac{\Delta \eta_{\text{osc}}(\eta_\lmax)}{a^2(\eta_\lmax)} \frac{\Delta \eta_{\text{osc}}(\eta_\lmax)}{ k_\pe (u+v)} \frac{1}{4 a^4(\eta_\lmax) k_\pe^2 u v}  \right| \nonumber \\
  &\sim \left||U_{k_\pe}(\eta_\lmax)|^2 \frac{a_0^2 \lambda^2(\eta_\lmax) k_\pe}{a^8(\eta_\lmax)} \ln(v_\UV/v_*) \right|,
 \label{eq:p_2vx_b_uv}
\end{align}
where we have used $\int^{\eta_\lmax}_{\eta_0} \dd \eta' \int^{\eta'}_{\eta_0} \dd \eta''f(\eta')g(\eta'') \sin[k_\pe(u+v)(\eta'' - \eta')] \sim \frac{\Delta \eta_{\text{osc}}(\eta_\lmax) f(\eta_\lmax)g(\eta_\lmax)}{k_\pe(u+v)} + \frac{f(\eta_\lmax)g(\eta_0)}{(k_\pe(u+v))^2}$. 
In addition, we have also used $\text{Im} \left[U_{k_\pe}(\eta_\lmax) U^*_{k_\pe}(\eta')\right]\Re\left[U_{k_\pe}(\eta_\lmax)U^*_{k_\pe}(\eta'')\right] \sim |U_{k_\pe}(\eta_\lmax)|^2\frac{\Delta \eta_{\text{osc}}(\eta_\lmax)}{a^2(\eta_\lmax)}$ for $\eta' \sim \eta_\lmax$ and $\eta''\sim \eta_\lmax$.
Note that the contribution from $\eta' \sim \eta_\lmax$ and $\eta'' \sim \eta_\lmax$ is dominant because we are considering the case in $k_\pe(u+v) \gg \Delta \eta_{\text{osc}}(\eta_\lmax)$ with $u, v \gg 1$.
Combining Eqs.~(\ref{eq:p_2vx_b_uv}) and (\ref{eq:b_approx_2vx_s}), we find 
\begin{align}
  \left|\frac{\mathcal P^b_{\delta \phi, 2\vx,\text{UV}}(k_\pe,\eta_\lmax)}{\mathcal P^b_{\delta \phi, 2\vx}(k_\pe,\eta_\lmax)}\right| \sim \frac{\mathcal P_{\delta \phi, \tr}(k_\pe,\eta_0)}{\mathcal P_{\delta \phi, \tr}(k_\pe,\eta_\lmax)} \ln(v_\UV/v_*).
\end{align}
Although a log divergence appears in the limit of $v_\UV \rightarrow \infty$, the cutoff scale dependence is much weaker than Eq.~(\ref{eq:uv_cut_dep_1vx}).
For example, at $N-N_0 = 1.18$ in Fig.~\ref{fig:loop_pbh_pl} (case A), the right hand side becomes $\sim \mathcal O(10^{-6}) \ln(v_\UV/v_*)$.
Then, we need to set $\ln(v_\UV/v_*) < \mathcal O(10^{6})$ to cut the UV contribution, though this is automatically satisfied if we assume the scale for $v_\UV$ is larger than the Planck length.

Using Eq.~(\ref{eq:1vx_ana}), we can approximate the UV contribution in $\mathcal P_{\delta \phi, 1\vx}(k_\pe,\eta_\lmax)$ as 
\begin{align}
	|\mathcal P_{\delta \phi, 1\vx, \text{UV}}(k_\pe,\eta_\lmax)| &\sim \Delta \eta_{\text{osc}}^2(\eta_\lmax) \left|\frac{\mu(\eta_\lmax)}{a^2(\eta_\lmax)}\right| 6\, \frac{k^3}{2\pi^2}|U_k(\eta_\lmax)|^2 \int^{k_\pe v_\UV}_{k_\pe v_*} \frac{\dd p}{p} \mathcal P_{\delta \phi,\tre}(p,\eta_\lmax),
\end{align}
where we have used the fact that the dominant contribution comes from $\eta'\sim \eta_\lmax$.
Combining this equation and Eq.~(\ref{eq:1vx_ana2}), we find 
\begin{align}
	\left|\frac{\mathcal P_{\delta \phi, 1\vx,\text{UV}}(k_\pe,\eta_\lmax)}{\mathcal P_{\delta \phi,1\vx}(k_\pe,\eta_\lmax)} \right|\sim \frac{\int^{k_\pe v_\UV}_{k_\pe v_*} \frac{\dd p}{p} \mathcal P_{\delta \phi,\tre}(p,\eta_\lmax)}{\mathcal P_{\delta \phi,\tre}(k_\pe,\eta_\lmax)} \sim \frac{a_0^2 \mathcal P_{\delta \phi,\tre}(k_\pe,\eta_0) (v_\UV^2 - v_*^2)}{a^2(\eta_\lmax) \mathcal P_{\delta \phi,\tre}(k_\pe,\eta_\lmax)}.
	\label{eq:uv_cut_dep_1vx}
\end{align}
For example, at $N-N_0 = 1.18$ in Fig.~\ref{fig:loop_pbh_pl} (case A), the right hand side becomes $\sim \mathcal O(10^{-8}) (v_\UV^2-v_*^2)$.
This means that, if we set $v_\UV^2 < \mathcal O(10^8)$ in that case, we can safely calculate the one-vertex loop power spectrum without being dominated by the UV contributions that should be regularized and subtracted.

\subsection{IR cutoff}

Next, let us discuss the IR contributions. 
Hereafter, we consider the IR modes as the superhorizon perturbations at the beginning of the resonance, which are not amplified by the oscillatory features. 
Similar to Eq.~(\ref{eq:u_q_approx}), we simply approximate $U_k(\eta)$ for the IR mode as 
\begin{align}
	U_{k_\IR}(\eta) \simeq \frac{H }{\sqrt{2k_\IR^3}}.
	\label{eq:u_k_ir_approx}
\end{align}
Note again that this is a good approximation at $\eta_\lmax$ (see Sec.~\ref{subsec:tadpole}).
Using this approximation and Eq.~(\ref{eq:mathcal_p_delta_sym}), we can approximate the IR contribution in $\mathcal P^a_{\delta \phi,2\vx}$ as 
\begin{align}
 \mathcal P^a_{\delta \phi, 2\vx, \IR}(k_\pe,\eta_\lmax) &\simeq \int^{v_h}_{v_\IR} \dd v \int^{1+v}_{|1-v|} \dd u \frac{uv}{4\pi^4} k^6_\pe \left|\int^{\eta_\lmax}_{\eta_0} \dd \eta' \lambda(\eta') 2\, \text{Im} \left[ U_{k_\pe}(\eta_\lmax) U^*_{k_\pe}(\eta')\right] U_{k_\pe u}(\eta') U_{k_\pe v}(\eta')\right|^2 \nonumber \\
 &\simeq  \int^{v_h}_{v_\IR} \dd v \int^{1+v}_{|1-v|} \dd u \frac{uv}{4\pi^4} k^6_\pe \left|\int^{\eta_\lmax}_{\eta_0} \dd \eta' \lambda(\eta') 2\, \text{Im} \left[ U_{k_\pe}(\eta_\lmax) U^*_{k_\pe}(\eta')\right] U_{k_\pe u}(\eta') \frac{H}{\sqrt{2 (k_\pe v)^3}} \right|^2 \nonumber \\
 &\sim  \int^{v_h}_{v_\IR} \dd v \int^{1+v}_{|1-v|} \dd u \frac{uv}{4\pi^4} k^6_\pe \frac{H^2}{2(k_\pe v)^3} \left| \frac{\Delta \eta_{\text{osc}}^2(\eta_\lmax) \lambda(\eta_\lmax)}{a^2(\eta_\lmax)} U_{k_\pe u}(\eta_\lmax) \right|^2 \nonumber \\
 &\sim  \lambda^2(\eta_\lmax) \frac{a_0^4 H^2}{a^{8}(\eta_\lmax) k_\pe^4} \mathcal P_{\delta \phi, \tre}(k_\pe,\eta_\lmax) \ln(v_h/v_\IR),
 \label{eq:p_a_IR_cut}
\end{align}
where we have used the fact that the dominant contribution comes from $\eta' \sim \eta_\lmax$.
$v_\IR (\ll 1)$ is the IR cutoff and $k_\pe v_h$ is the horizon scale at the beginning of the resonance.
Note that the IR contributions come from the mixing of the IR mode and the peak-scale mode with $v \ll 1$ and $u \simeq 1$ due to the momentum conservation.
Combining this equation and Eq.~(\ref{eq:p_2vx_sym_est}), we find 
\begin{align}
	\frac{\mathcal P^a_{\delta \phi, 2\vx, \IR}(k_\pe,\eta_\lmax)}{\mathcal P^a_{\delta \phi, 2\vx}(k_\pe,\eta_\lmax)} \sim \frac{\mathcal P_{\delta \phi, \tr,\IR}}{ \mathcal P_{\delta \phi, \tr}(k_\pe,\eta_\lmax)} \ln(v_h/v_\IR),
	\label{eq:2vx_sym_ir_cond}
\end{align}
where we have defined $\mathcal P_{\delta \phi, \tr,\IR} \equiv H^2/(2\pi)^2$ as the IR tree power spectrum.
For example, at $N-N_0 = 1.18$ in Fig.~\ref{fig:loop_pbh_pl} (case A), the right hand side becomes $\sim \mathcal O(10^{-9}) \ln(v_h/v_\IR)$, which means that we need to set $\ln(v_h/v_\IR) < \mathcal O(10^9)$ to cut the IR contribution.

Using Eq.~(\ref{eq:asym_final}), we can approximate the IR contribution in $\mathcal P^b_{\delta \phi, 2\vx}$ as 
\begin{align}
  &\left|\mathcal P^b_{\delta \phi, 2\vx, \IR}(k_\pe,\eta_\lmax)\right| \nonumber \\
  &\simeq \left|8 \int^{v_h}_{v_\IR} \dd v \int^{1+v}_{|1-v|} \dd u \frac{uv}{4\pi^4} k_\pe^6 \int^{\eta_\lmax}_{\eta_0} \dd \eta' \int^{\eta'}_{\eta_0} \dd \eta'' \lambda(\eta') \lambda(\eta'')\text{Im} \left[U_{k_\pe}(\eta_\lmax) U^*_{k_\pe}(\eta')\right]  \right.  \nonumber \\
  & \left. \phantom{\int^\infty_0 } 
  \times \Re\left[U_{k_\pe}(\eta_\lmax)U^*_{k_\pe}(\eta'')\right]\text{Im}\left[ U_{k_\pe v}(\eta')U^*_{k_\pe v}(\eta'')U_{k_\pe u}(\eta') U^*_{k_\pe u}(\eta'')\right] \right| \nonumber \\
  &\simeq \left|8 \int^{v_h}_{v_\IR} \dd v \int^{1+v}_{|1-v|} \dd u \frac{uv}{4\pi^4} k_\pe^6 \int^{\eta_\lmax}_{\eta_0} \dd \eta' \int^{\eta'}_{\eta_0} \dd \eta'' \lambda(\eta') \lambda(\eta'')\text{Im} \left[U_{k_\pe}(\eta_\lmax) U^*_{k_\pe}(\eta')\right]  \right.  \nonumber \\
  & \left. \phantom{\int^\infty_0 } 
  \times \Re\left[U_{k_\pe}(\eta_\lmax)U^*_{k_\pe}(\eta'')\right] \frac{H^2}{2 k_\pe^3 v^3} \Im[U_{k_\pe u}(\eta') U^*_{k_\pe u}(\eta'')]  \right| \nonumber \\  
  &\sim  \lambda^2(\eta_\lmax) \frac{a_0^4 H^2}{a^{8}(\eta_\lmax) k_\pe^4} \mathcal P_{\delta \phi, \tre}(k_\pe,\eta_\lmax) \ln(v_h/v_\IR),
\end{align}
where we have used the fact that the dominant contribution comes from $\eta'' \sim \eta_\lmax$ and $\eta' \sim \eta_\lmax$.
We can see that the order of the IR contribution is the same as Eq.~(\ref{eq:p_a_IR_cut}).
Then, the order of the ratio $\mathcal P^b_{\delta \phi, 2\vx, \IR}/\mathcal P^b_{\delta \phi, 2\vx}$ also becomes the same as Eq.~(\ref{eq:2vx_sym_ir_cond}).

Using Eq.~(\ref{eq:1vx_ana}), we can approximate the IR contribution in the one-vertex contribution as 
\begin{align}
	\left|\mathcal P_{\delta \phi, 1\vx, \IR}(k_\pe,\eta_\lmax)\right| &\sim \Delta \eta_{\text{osc}}^2(\eta_\lmax) \left|\frac{\mu(\eta_\lmax)}{a^2(\eta_\lmax)}\right| 6\, \frac{k^3}{2\pi^2}|U_k(\eta_\lmax)|^2 \int^{k_\pe v_h}_{k_\pe v_\IR} \frac{\dd p}{p} \mathcal P_{\delta \phi,\tre}(p,\eta_\lmax),
\end{align}
where we have used the fact that the dominant contribution comes from $\eta' \sim \eta_\lmax$.
Combining this equation and Eq.~(\ref{eq:1vx_ana2}), we find 
\begin{align}
	\left|\frac{\mathcal P_{\delta \phi, 1\vx,\IR}(k_\pe,\eta_\lmax)}{\mathcal P_{\delta \phi,1\vx}(k_\pe,\eta_\lmax)} \right|\sim \frac{\int^{k_\pe v_h}_{k_\pe v_\IR} \frac{\dd p}{p} \mathcal P_{\delta \phi,\tre}(p,\eta_\lmax)}{\mathcal P_{\delta \phi,\tre}(k_\pe,\eta_\lmax)} \sim \frac{\mathcal P_{\delta \phi,\tre,\IR}}{\mathcal P_{\delta \phi,\tre}(k_\pe,\eta_\lmax)} \ln(v_h/v_\IR).
	\label{eq:ir_cut_dep_1vx}
\end{align}
The right hand side is the same as the case of the two-vertex contribution, Eq.~(\ref{eq:2vx_sym_ir_cond}).

%%%%%%%%%%%%%%%%%%%%%%%%%%%%%%%%%%%
%%%%%%%%%%%%%%%%%%%%%%%%%%%%%%%%%%%
%%%%%%%%%%%%%%%%%%%%%%%%%%%%%%%%%%%
\small
\bibliographystyle{apsrev4-1}
\bibliography{draft_one_loop}

\end{document}